\documentclass[hyper]{JHEP3}

\pdfoutput=1
\usepackage{latexsym}
\usepackage{graphicx}
\usepackage{mathrsfs}
\usepackage{amsmath}
\usepackage{amsfonts} 
\usepackage{amssymb}  
\usepackage[bf, sf, FIGTOPCAP, nooneline, tight]{subfigure}
\usepackage{multirow}
\usepackage{multicol}
\usepackage[numbers,sort&compress]{natbib}
\usepackage{color}
\let\normalcolor\relax

\def\urltilda{\kern -.15em\lower .7ex\hbox{\~{}}\kern .04em}

\def\tab#1{Table~\ref{#1}}
\def\tabs#1#2{Tables~\ref{#1} and \ref{#2}}
\def\fig#1{Fig.~\ref{#1}}
\def\figs#1#2{Figs.~\ref{#1} and \ref{#2}}
\def\subfigs#1#2#3#4{Figs.~\ref{#1}{#3},{#4} and \ref{#2}{#3},{#4}}
\def\sec#1{Sec.~\ref{#1}}
\def\eq#1{Eq.~\ref{#1}}

\newcommand{\newc}{\newcommand}

\newc{\be}{\begin{equation}}
\newc{\ee}{\end{equation}}
\newc{\nn}{\nonumber}
\newc\ps{\mbox{ ps}}
\newc{\mev}{\mbox{ MeV}}
\newc{\gev}{\mbox{ GeV}}
\newc{\tev}{\mbox{ TeV}}
\newc{\GeV}{\gev}
\newc{\MeV}{\mev}
\newc{\TeV}{\tev}
\newc{\cl}{\text{CL}}
\newc\BR{BR}
\newc{\alphaemmz}{\alpha_{\text{em}}(m_Z)^{\overline{MS}}}
\newc{\alphas}{\alpha_s(m_Z)^{\overline{MS}}}
\newc\zetah{\zeta_h}
\newc\eg{{\rm {e.g.}}}
\newc\etal{{\rm {et al.}}}
\newc\ie{{\rm i.e.}}
\newc\etc{{\rm {etc}}}
\newc{\mhalf}{m_{1/2}}
\newc{\mzero}{m_0}
\newc{\tanb}{\tan\beta}
\newc{\azero}{A_0}
\newc{\sgn}{{\rm sgn}}
\newc{\deltaamususy}{\delta a_{\mu}^{\text{SUSY}}}
\newc{\bsg}{\bsgamma}
\newc\gmtwo{(g-2)_{\mu}}
\newc\deltaamu{\Delta a_{\mu}}
\newc{\abundchi}{\Omega_{\chi} h^2}
\newc{\msbar}{\overline{MS}}
\newc{\mtop}{m_t}
\newc{\mtpole}{m_t}
\newc{\hl}{h}
\newc{\mhl}{m_{\hl}}  
\newc{\mgut}{M_{\rm GUT}}
\newc{\mplanck}{M_{\rm P}}
\newc{\mpl}{M_{\text{Pl}}}
\newc{\msusy}{M_{\rm SUSY}}
\newc{\ms}{M_{\text{S}}}
\newc{\VEV}[1]{\langle #1 \rangle}
\newc{\sineff}{\sin^2 \theta_{\rm{eff}}}
\newc\MN{{\sf {MultiNest}}} 

\newcommand\pb{\,\mbox{pb}}

\newc\bsgamma{b\rightarrow s \gamma }
\newc\brbsgamma{\BR(\overline{B}\rightarrow X_s\gamma)}
\newc\bsmumu{\overline{B}_s\to\mu^+\mu^-}
\newc\brbsmumu{\BR(\overline{B}_s\to\mu^+\mu^-)}
\newc\bdmmumu{\overline{B}_d\to\mu^+\mu^-}
\newc\bbbarmix{\overline{B}_s\mbox{--}B_s}
\newc\delmbs{\Delta M_{B_s}}
\newc\brbtaunu{\BR(\overline{B}_u\to \tau \nu)}
\newc{\mbmbmsbar}{m_b(m_b)^{\msbar} }

\newc\AIPCP[3] {{\em AIP Conf. Proc.} {\bf #1} (#2) #3}
\newc\AJ[3] {{\em Astrophys. J.} {\bf #1} (#2) #3}
\newc\AMS[3] {{\em Ann. Math. Statist.} {\bf #1} (#2) #3}                
\newc\AP[3] {{\em Ann. Phys.} {\bf #1} (#2) #3}
\newc\APJ[3] {{\em Astropart. J.} {\bf #1} (#2) #3}
\newc\APP[3] {{\em Astropart. Phys.} {\bf #1} (#2) #3}
\newc\APS[3] {{\em Astrophys. J. Suppl.} {\bf #1} (#2) #3}
\newc\ARNPS[3] {{\em Ann. Rev. Nucl. Part. Sci.} {\bf C#1} (#2) #3}
\newc\BA[3] {{\em Bayesian Anal.} {\bf C#1} (#2) #3}              
\newc\CPC[3] {{\em Comput. Phys. Commun.} {\bf C#1} (#2) #3}
\newc\CP[3] {{\em Contemp. Phys.} {\bf #1} (#2) #3}                     
\newc\EPJ[3] {{\em Euro. Phys. Journ.} {\bf C#1} (#2) #3}
\newc\JCAP[3] {{\em JCAP} {\bf #1} (#2) #3}
\newc\JHEP[3] {{\em JHEP} {\bf #1} (#2) #3}
\newc\JPG[3] {{\em J. Phys.} {\bf G #1} (#2) #3}
\newc\IJMP[3] {{\em Int. J. Mod. Phys.} {\bf A #1} (#2) #3}
\newc\MNRAS[3] {{\em Mon. Not. Roy. Astron. Soc.} {\bf #1} (#2) #3}
\newc\MPL[3] {{\em Mod. Phys. Lett.} {\bf A #1} (#2) #3}
\newc\NAR[3] {{\em New Astron. Rev.} {\bf #1} (#2) #3}                  
\newc\NCA[3] {{\em Nuovo Cimento} {\bf #1} (#2) #3}
\newc\NIM[3] {{\em Nucl. Instrum. Methods} {\bf #1} (#2) #3}
\newc\NIMA[3] {{\em Nucl. Instrum. Methods} {\bf A #1} (#2) #3}
\newc\NAT[3] {{\em Nature} {\bf #1} (#2) #3}
\newc\NPB[3] {{\em Nucl. Phys.} {\bf B #1} (#2) #3}
\newc\NPA[3] {{\em Nucl. Phys.} {\bf A #1} (#2) #3}
\newc\NPPS[3] {{\em Nucl. Phys. Proc. Suppl.} {\bf #1} (#2) #3}                
\newc\PLB[3] {{\em Phys. Lett.} {\bf B #1} (#2) #3}
\newc\PR[3] {{\em Phys. Rep.} {\bf #1} (#2) #3}
\newc\PRL[3] {{\em Phys. Rev. Lett.} {\bf #1} (#2) #3}
\newc\PRD[3] {{\em Phys. Rev.} {\bf D #1} (#2) #3}
\newc\PRC[3] {{\em Phys. Rev.} {\bf C #1} (#2) #3}
\newc\PTP[3] {{\em Prog. Theor. Phys.} {\bf #1} (#2) #3}
\newc\RMP[3] {{\em Rev. Mod. Phys.} {\bf #1} (#2) #3 }
\newc\RPP[3] {{\em Rept. Prog. Phys.} {\bf #1} (#2) #3 }
\newc\SC[3] {{\em Science} {\bf #1} (#2) #3 }
\newc\ZPC[3] {{\em Z. Phys.} {\bf C #1} (#2) #3}
\newc\Err[3] {{\em Erratum-ibid.} {\bf #1} (#2) #3 }

\title{A Profile Likelihood Analysis of the Constrained MSSM with Genetic Algorithms}

\author{Yashar Akrami, Pat Scott, Joakim Edsj\"o, Jan Conrad and Lars Bergstr\"om\\
       Oskar Klein Centre for Cosmoparticle Physics\\
       Department of Physics, Stockholm University\\
       AlbaNova, SE-10691 Stockholm, Sweden \\
       E-mails: \email{yashar, pat, edsjo, conrad, lbe@fysik.su.se}}

\preprint{}	

\abstract{The Constrained Minimal Supersymmetric Standard Model (CMSSM) is one of the simplest and most widely-studied supersymmetric extensions to the standard model of particle physics.  Nevertheless, current data do not sufficiently constrain the model parameters in a way completely independent of priors, statistical measures and scanning techniques.  We present a new technique for scanning supersymmetric parameter spaces, optimised for frequentist profile likelihood analyses and based on Genetic Algorithms.  We apply this technique to the CMSSM, taking into account existing collider and cosmological data in our global fit.  We compare our method to the \MN~algorithm, an efficient Bayesian technique, paying particular attention to the best-fit points and implications for particle masses at the LHC and dark matter searches.  Our global best-fit point lies in the focus point region.  We find many high-likelihood points in both the stau co-annihilation and focus point regions, including a previously neglected section of the co-annihilation region at large $\mzero$.  We show that there are many high-likelihood points in the CMSSM parameter space commonly missed by existing scanning techniques, especially at high masses.  This has a significant influence on the derived confidence regions for parameters and observables, and can dramatically change the entire statistical inference of such scans.}


\begin{document}

\section{Introduction} \label{sec:intro}

New physics beyond the Standard Model (SM) is broadly conjectured to appear at TeV energy scales.  Particular attention has been paid to supersymmetric (SUSY) extensions of the SM, widely hoped to show up at the Large Hadron Collider (LHC).  One of the strongest motivations for physics at the new scale is the absence of any SM mechanism for protecting the Higgs mass against radiative corrections; this is known as the hierarchy or fine-tuning problem~\cite{Martin:9709356}.  Softly-broken weak-scale supersymmetry (for an introduction, see Ref.~\citealp{EWSUSY}) provides a natural solution to this problem via the cancellation of quadratic divergences in radiative corrections to the Higgs mass.  The natural connection between SUSY and grand unified
theories (GUTs) also offers extensive scope for achieving gauge-coupling unification in this framework~\cite{Ellis:1990}.  Supersymmetry has even turned out to be a natural component of many string theories, so it may be worth incorporating into extensions of the SM anyway (though in these models it is not at all necessary for SUSY to be detectable at low energies).

Another major theoretical motivation for supersymmetry is that most weak-scale versions contain a viable dark matter (DM) candidate~\cite{DMreview}.  Its stability is typically achieved via a conserved discrete symmetry ($R$-parity) which arises naturally in some GUTs, and makes the lightest supersymmetric particle (LSP) stable.  Its `darkness' is achieved by having the LSP be a neutral particle, such as the lightest neutralino or sneutrino.  These are both weakly-interacting massive particles (WIMPs), making them prime dark matter candidates~\cite{DMreview}.  The sneutrino is strongly constrained due to its large nuclear-scattering cross-section, but the neutralino remains arguably the leading candidate for DM.  

Describing the low-energy behaviour of a supersymmetric model typically requires adding many new parameters to the SM.  This makes phenomenological
analyses highly complicated.  Even upgrading the SM to its most minimal SUSY form, the Minimal Supersymmetric Standard Model (MSSM; for a recent review, see Ref.~\citealp{Chung:0312378}), introduces more than a hundred free parameters.  All but one of these come from the soft terms in the SUSY-breaking sector.  Fortunately, extensive regions of the full MSSM parameter space are ruled out phenomenologically, as generic values of many of the new
parameters allow flavour changing neutral currents (FCNCs) or CP violation at levels excluded by experiment.

One might be able to relate many of these seemingly-free parameters theoretically, dramatically reducing their number.  This requires specification of either the underlying SUSY-breaking mechanism itself, or a mediation mechanism by which SUSY-breaking would be conducted from some undiscovered particle sector to the known particle spectrum and its SUSY counterparts.  Several mediation mechanisms (for recent reviews, see e.g. Refs.~\citealp{Chung:0312378} and~\citealp{Luty:0509029}) have been proposed which relate the MSSM parameters in very different ways, but so far no clear preference has been established for one mechanism over another.  For a comparison of some mediation models using current data, see Ref.~\citealp{AbdusSalam:09060957}.  Gravity-mediated SUSY breaking, based on supergravity unification, naturally leads to the suppression of many of the dangerous FCNC and CP-violating terms.  Its simplest version is known as minimal supergravity (mSUGRA)~\cite{mSUGRA1,mSUGRA2}.

An alternative approach is to directly specify a phenomenological MSSM reduction at low energy.  Here one sets troublesome CP-violating and FCNC-generating terms to zero by hand, and further reduces the number of parameters by assuming high degrees of symmetry in e.g.~mass and mixing matrices.

A hybrid approach is to construct a phenomenological GUT-scale model, broadly motivated by the connection between SUSY and GUTs.  Here one imposes boundary conditions at the GUT scale ($\sim$10$^{16}$\,GeV) and then explores the low-energy phenomenology by means of the Renormalisation Group Equations (RGEs).  One of the most popular schemes is the Constrained MSSM (CMSSM)~\cite{CMSSM}, which incorporates the phenomenologically-interesting parts of mSUGRA.  The CMSSM includes four continuous parameters: the ratio of the two Higgs vacuum expectation values ($\tanb$), and the GUT-scale values of the SUSY-breaking scalar, gaugino and trilinear mass parameters ($\mzero$, $\mhalf$ and $\azero$).  The sign of $\mu$ (the MSSM Higgs/higgsino mass parameter) makes for one additional discrete parameter; its magnitude is determined by requiring that SUSY-breaking induces electroweak symmetry-breaking.  Despite greatly curbing the range of possible phenomenological consequences, the small number of parameters in the CMSSM has made it a tractable way to explore basic low-energy SUSY phenomenology.

Before drawing conclusions about model selection or parameter values from experimental data, one must choose a statistical framework to work in.  There are two very different fundamental interpretations of probability, resulting in two approaches to statistics (for a detailed discussion, see e.g. Ref.~\citealp{Cowan:1998}).  Frequentism deals with relative frequencies, interpreting probability as the fraction of times an outcome would occur if a measurement were repeated an infinite number of times.  Bayesianism deals with subjective probabilities assigned to different hypotheses, interpreting probability as a measure of the degree of belief that a hypothesis is true.  The former is used for assigning statistical errors to measurements, whilst the latter can be used to quantify both statistical and systematic uncertainties.  In the Bayesian approach one is interested in the probability of a set of model parameters given some data, whereas in the frequentist approach the only quantity one can discuss is the probability of some dataset given a specific set of model parameters, i.e. a likelihood function.

In a frequentist framework, one simply maps a model's likelihood as a function of the model parameters.  The point with the highest likelihood is the best fit, and uncertainties upon the parameter values can be given by e.g. iso-likelihood contours in the model parameter space.  To obtain joint confidence intervals on a subset of parameters, the full likelihood is reduced to a lower-dimensional function by maximising it along the unwanted directions in the parameter space.  This is the profile likelihood procedure~\cite[and references therein]{profilelike}.  In the Bayesian picture~\cite{BayesRevs}, probabilities are directly assigned to different volumes in the parameter space.  One must therefore also consider the state of subjective knowledge about the relative probabilities of different parameter values, independent of the actual data; this is a prior.  In this case, the statistical measure is not the likelihood itself, but a prior-weighted likelihood known as the posterior probability density function (PDF).  Because this posterior is nothing but the joint PDF of all the parameters, constraints on a subset of model parameters are obtained by marginalising (i.e.~integrating) it over the unwanted parameters.  This marginalised posterior is then the Bayesian counterpart to the profile likelihood.  Bayesians report the posterior mean as the most-favoured point (given by the expectation values of the parameters according to the marginalised posterior), with uncertainties defined by surfaces containing set percentages of the total marginalised posterior, or `probability mass'.

One practically interesting consequence of including priors is that they are a powerful tool for estimating how robust a fit is.  If the posterior is strongly dependent on the prior, the data are not sufficient to constrain the model parameters.  It has been shown that the prior still plays a large role in Bayesian inferences in the CMSSM~\cite{Trotta:08093792}.  If an actual detection occurs at the LHC, this dependency should disappear~\cite{Roszkowski:09070594}. 

Clearly the results of frequentist and Bayesian inferences will not coincide in general.  This is especially true if the model likelihood has a complex dependence on the parameters (i.e. not just a simple Gaussian form), and if insufficient experimental data is available.  Note that this is true even if the prior is taken to be flat; a flat prior in one parameter basis is certainly not flat in every such basis.  In the case of large sample limits both approaches give similar results, because the likelihood and posterior PDF both become almost Gaussian; this is why both are commonly used in scientific data analysis.

The first CMSSM parameter scans were performed on fixed grids in parameter space~\cite{Drees:9207234,grid-cmssm}.  Predictions of e.g.~the relic density of the neutralino as a cold dark matter candidate or the Higgs/superpartner masses were computed for each point on the grid, and compared with experimental data.  In these earliest papers, points for which the predicted quantities were within an arbitrary confidence level (e.g.~1$\sigma$, 2$\sigma$) were deemed ``good''.  Because all accepted points are considered equivalently good, this method provides no way to determine points' relative goodnesses-of-fit, and precludes any deeper statistical interpretation of results.

The first attempts at statistical interpretation were to perform (frequentist) $\chi^2$ analyses with grid scans~\cite{chisquare-cmssm}.  Limited random scans were also done in some of these cases.  Despite some advantages of grid scans, their main drawback is that the number of points sampled in an $N$-dimensional space with $k$ points for each parameter grows as $k^{N}$, making the method extremely inefficient.  This is even true for spaces of moderate dimension like the CMSSM.  The lack of efficiency is mainly due to the complexity and highly non-linear nature of the mapping of the CMSSM parameters to physical observables; many important features of the parameter space can be missed by not using a high enough grid resolution.

Another class of techniques that has become popular in SUSY analyses is based on more sophisticated scanning algorithms.  These are techniques designed around the Bayesian requirement that a probability surface be mapped in such a way that the density of the resultant points is proportional to the actual probability.  However, the points they return can also be used in frequentist analyses.  Foremost amongst these techniques is the Markov Chain Monte Carlo (MCMC) method~\cite{Baltz:0407039,Allanach:0507283,Allanach:0601089,Austri:0602028,Allanach:0609295,Roszkowski:0611173,Roszkowski:07052012,Allanach:07050487,Trotta:0609126,Roszkowski:07070622,Allanach:08061184,Allanach:08061923,Buchmueller:08084128,Roszkowski:09031279,Martinez:09024715,Balazs:09065012,Belanger:09065048,Buchmueller:09075568,Bechtle:09072589,Lafaye:07093985},
which has also been widely used in other branches of science, in particular cosmological data analysis~\cite{MCMCCosmo}.  The MCMC method provides a greatly improved scanning efficiency in comparison to traditional grid searches, scaling as $kN$ instead of $k^{N}$ for an $N$-dimensional parameter space.  More recently, the framework of nested sampling~\cite{SkillingNS} has come to prominence, particularly via the publicly-available implementation \MN~\cite{MultiNest}.  A handful of recent papers have explored the CMSSM
parameter space or its observables using this technique~\cite{Trotta:08093792,Martinez:09024715,Feroz:08074512,Feroz:09032487,Trotta:09060366,Roszkowski:09070594,Scott:FermiLAT,Scott:09084082},
as well as the higher-dimensional spaces of the Constrained Next-to-MSSM (CNMSSM)~\cite{LopezFogliani:09064911} and phenomenological MSSM (pMSSM)~\cite{pMSSM}. \MN~was also the technique of choice in the supersymmetry-breaking study of Ref.~\citealp{AbdusSalam:09060957}. A CMSSM scan with \MN~takes roughly a factor of $\sim$200 less computational effort than a full MCMC scan, whilst results obtained with both algorithms are identical (up to numerical noise)~\cite{Trotta:08093792}.

Besides improved computational efficiency, MCMCs and nested sampling offer other convenient features for both frequentist and Bayesian analyses.  In a fully-defined statistical framework, all significant sources of uncertainty can be included, including theoretical uncertainties and our imperfect knowledge of the relevant SM parameters.  These can be introduced as additional `nuisance' parameters in the scans, and resultant profile likelihoods and posterior PDFs profiled/marginalised over them.  In a similar way, one can profile and marginalise over all parameters at once to make straightforward statistical inferences about any arbitrary function of the model parameters, like neutralino annihilation fluxes~\cite{Roszkowski:07070622,Martinez:09024715,Scott:09084082} or cross-sections~\cite{Scott:FermiLAT}.  The profiling/marginalisation takes almost no additional computational effort: profiling simply requires finding the sample with the highest likelihood in a list, whereas marginalisation, given the design of MCMCs and nested sampling, merely requires tallying the number of samples in the list.  Finally, it is straightforward to take into account all priors when using these techniques for Bayesian analyses.

Although the prior-dependence of Bayesian inference can be useful for determining the robustness of a fit, it may be considered undesirable when trying to draw concrete conclusions from the fitting procedure.  This is because the prior is a subjective quantity, and most researchers intuitively prefer their conclusions not to depend on subjective assessments.  In this case, the natural preference would be to rely on a profile likelihood analysis rather than one based on the posterior PDF.  The question then becomes: how does one effectively and efficiently explore a parameter space like the CMSSM, with its many finely-tuned regions, in the context of the profile likelihood?

As a first attempt to answer this question, the profile likelihood of the CMSSM was recently mapped with MCMCs~\cite{Allanach:07050487} and \MN~\cite{Trotta:08093792}.  Despite the improved efficiency of these Bayesian methods with respect to grid searches by several orders of magnitude, they are not optimised to look for isolated points with large likelihoods.  They are thus very likely to entirely miss high-likelihood regions occupying very tiny volumes in the parameter space.  Such regions might have a strong impact on the final results of the profile likelihood scan\footnote{Missing fine-tuned regions could even modify the posterior PDF if they are numerous or large enough.}, since the profile likelihood is normalised to the best-fit point and places all regions, no matter how small, on an equal footing.  It appears that in the case of the CMSSM there are many such fine-tuned regions.  This is seen in e.g.~CMSSM profile likelihood maps with different \MN~scanning priors~\cite{Trotta:08093792}.  Given that the profile likelihood is independent of the prior by definition, these results demonstrate that many high-likelihood regions are missed when using a scanning algorithm optimised for Bayesian statistics.  In order to make valid statistical inferences in the context of the profile likelihood, the first (and perhaps most crucial) step is to correctly locate the best-fit points.  Setting confidence limits and describing other statistical characteristics of the parameter space makes sense only if this first step is performed correctly.

If one wishes to work confidently in a frequentist framework, some alternative scanning method is clearly needed.  The method should be optimised for calculating the profile likelihood, rather than the Bayesian evidence or posterior.  Even if the results obtained with such a method turn out to be consistent with those of MCMCs and nested sampling, the exercise would greatly increase the utility and trustworthiness of those techniques.

In this paper, we employ a particular class of optimisation techniques known as Genetic Algorithms (GAs) to scan the CMSSM parameter space, performing a global frequentist fit to current collider and cosmological data.  GAs and other evolutionary algorithms have not yet been widely used in high energy physics or cosmology; to our knowledge, their only prior use in SUSY phenomenology~\cite{Allanach:0406277} has been for purposes completely different to ours\footnote{See e.g. Refs.~\citealp{GenExpHepRef,GenPheHepRef,Teodorescu:08040369} for their use in high energy physics, Ref.~\citealp{GenCosRef}~for uses in cosmology and general relativity, and Ref.~\citealp{GenNucRef}~for applications to nuclear physics.}.  There are two main reasons GAs should perform well in profile likelihood scans.  Firstly, the sole design purpose of GAs is to maximise or minimise a function.  This is exactly what is required by the profile likelihood; in the absence of any need to marginalise (i.e.~integrate) over dimensions in the parameter space, it is more important that a search algorithm finds the best-fit peaks than accurately maps the likelihood surface at lower elevations.  Secondly, the ability of GAs to probe global extrema excels most clearly over other techniques when the parameter space is very large, complex or poorly understood; this is precisely the situation for SUSY models.  We focus exclusively on the CMSSM as our test-bed model, but the algorithms could easily be employed in higher-dimensional SUSY parameter spaces without considerable change in the scanning efficiency (as they scale as $kN$ for an $N$-dimensional parameter space).  We compare our profile likelihood results directly with those of the \MN~scanning algorithm.  This means that we also compare indirectly with MCMC scans, because \MN~and MCMCs give essentially identical results~\cite{Trotta:08093792}. We find that GAs uncover many better-fit points than previous \MN~scans.

This paper proceeds as follows: in~\sec{sec:analysis} we briefly review the parameters of the CMSSM, the predicted observables, constraints and bounds on them from collider and cosmological observations.  We then introduce GAs as our specific scanning technique of choice.  We present and discuss results in~\sec{sec:results}, comparing those obtained with GAs to those produced by \MN.  We include the best-fit points and the highest-likelihood regions of the parameter space, as well as the implications for particle discovery at the LHC and in direct and indirect dark matter searches.  We draw conclusions and comment on future prospects in~\sec{sec:concl}.

\section{Model and analysis} \label{sec:analysis}

\subsection{CMSSM likelihood} \label{sec:CMSSMlike}

Our goal is to compare the results of a profile likelihood analysis of the CMSSM performed with
GAs to those obtained using Bayesian scanning techniques, in particular \MN.  For this reason, we work with the same set of free parameters and ranges, the same observables (measurable physical quantities predicted by the model) and the same constraints (the observed values of observables, as well as physicality requirements) as in Ref.~\citealp{Trotta:08093792}.  We also perform the theoretical calculations and construct the full likelihood function of the model based on these variables and data in the same way as in Ref.~\citealp{Trotta:08093792}.  We limit ourselves to just a brief review of these quantities and constraints here.  For a detailed discussion, the reader is referred to previous papers~\cite{Austri:0602028,Trotta:08093792,Roszkowski:0611173}.

\subsubsection{Parameters and ranges}

As pointed out in~\sec{sec:intro}, there are four new continuous parameters $\mhalf$, $\mzero$, $\azero$ and $\tanb$, which are the main free parameters of the model to be fit to the data.  There is also a new discrete parameter, the sign of $\mu$, which we fix to be positive.\footnote{The choice of positive $\mu$ is motivated largely by constraints on the CMSSM from the anomalous magnetic moment of the muon $\deltaamususy$.  The branching fraction $\brbsgamma$ actually prefers negative $\mu$ (see, for example, Ref.~\citealp{Feroz:08074512} and the references therein).  $\mu > 0$ was apparently chosen in earlier work~\cite{Trotta:08093792} to stress the seeming inconsistency between the SM predictions for $\deltaamususy$ and experimental data.  We employ the same fixed sign in the present work for consistency, although it is of course possible to leave this a free discrete parameter in any global fit.}

We add four additional nuisance parameters to the set of free parameters in our scans.  These are the SM parameters with the largest uncertainties and strongest impacts upon CMSSM predictions: $\mtpole$, the pole top quark mass, $\mbmbmsbar$, the bottom quark mass evaluated at $m_b$, $\alphaemmz$, the electromagnetic coupling constant evaluated at the $Z$-boson pole mass $m_Z$, and $\alphas$, the strong coupling constant, also evaluated at $m_Z$.  These last three quantities are computed in the modified minimal subtraction renormalisation scheme $\msbar$.

The set of CMSSM and SM parameters together constitute an 8-dimensional parameter space
\be \label{allparams} \Theta = (\mzero, \mhalf, \azero, \tanb,
\mtpole, \mbmbmsbar, \alphaemmz, \alphas ), \ee
to be scanned and constrained according to the available experimental data.  The ranges over which we scan the parameters are $\mzero,\mhalf\in(50\gev,4\tev)$, $\azero\in(-7\tev,7\tev)$, ~$\tanb\in(2,62)$, ~$\mtpole\in[167.0\gev,~178.2\gev]$, ~$\mbmbmsbar\in[3.92\gev,~4.48\gev]$, ~$1/\alphaemmz\in[127.835,~128.075]$ and $\alphas\in[0.1096,~0.1256]$.

\subsubsection{Constraints: physicality, observables, and uncertainties}

In order to compare the predictions of each point in the parameter space with data, one has to first derive some quantities which are experimentally measurable.  For a global fit, one needs to take into account all existing (and upcoming) data, such as collider and cosmological observations, and direct and indirect dark matter detection experiments.  This is indeed the ultimate
goal in any attempt to constrain a specific theoretical model like the CMSSM.  Because our main goal in this paper is to assess how powerful GAs are compared to conventional methods (in particular the~\MN~algorithm), we restrict our analysis to the
same set of observables and constraints as in the comparison paper~\cite{Trotta:08093792}.  These include the collider limits on Higgs and superpartner masses, electroweak precision measurements, $B$-physics quantities and the cosmologically-measured abundance of dark matter.  These quantities and constraints are given in \tab{tab:obs}.

The observables are of three types:
\begin{itemize}
    \setlength{\itemsep}{1pt}
    \setlength{\parskip}{0pt}
    \setlength{\parsep}{0pt}
    \item The SM nuisance parameters.  Although they are considered free parameters of the model along with the CMSSM parameters, these are rather well-constrained by the data, and therefore used in constructing the full likelihood function.
    \item Observables for which positive measurements have been made.  These are the $W$-boson pole mass ($m_W$), the effective leptonic weak mixing
angle ($\sineff{}$), anomalous magnetic moment of the muon ($\deltaamususy$), branching fraction $\brbsgamma$, $\bbbarmix$ mass difference ($\delmbs$), branching fraction $\brbtaunu$, and dark matter relic density ($\abundchi$) assuming the neutralino is the only constituent of dark matter.
    \item Observables for which at the moment only (upper or
lower) limits exist, i.e. the branching fraction $\brbsmumu$, the lightest MSSM Higgs boson mass $\mhl$ (assuming its coupling to the $Z$-boson is
SM-like), and the superpartner masses.
\end{itemize}

Our sources for these experimental data are indicated in the table.  Note that there are no theoretical uncertainties associated with the SM parameters, because they are simultaneously observables and free input parameters.  For details about these quantities, experimental values and errors, particularly the reasoning behind theoretical uncertainties, see Refs.~\citealp{Trotta:08093792,Austri:0602028} and~\citealp{Roszkowski:0611173}.
\TABLE[h!]{ \centering
{\footnotesize
\begin{tabular}{|l | l l l | l|} \hline
Observable &   Mean value & \multicolumn{2}{c|}{Uncertainties} & Reference \\
           &         &    \multicolumn{2}{c|}{(standard deviations)} &  \\
&         & experimental  & theoretical &
 \\\hline\hline
\multicolumn{5}{|c|}{SM nuisance parameters}
\\ \hline\hline
$\mtpole$           &  $172.6\gev$    & $1.4\gev$&  - & \cite{topmass:mar08} \\
$m_b (m_b)^{\overline{MS}}$ & $4.20\gev$  & $0.07\gev$ & - &  \cite{pdg07} \\
$\alphas$       &   $0.1176$   & $0.002$ & - & \cite{pdg07}\\
$1/\alphaemmz$  & $127.955$ & $0.03$ & - & \cite{Hagiwara:2006jt}
\\ \hline\hline
\multicolumn{5}{|c|}{measured}
\\ \hline\hline
$m_W$     &  $80.398\gev$   & $25\mev$ & $15\mev$ & \cite{lepwwg} \\
$\sineff{}$    &  $0.23153$      & $16\times10^{-5}$
               & $15\times10^{-5}$ &  \cite{lepwwg}  \\
$\deltaamususy \times 10^{10}$       &  29.5 & 8.8 &  1.0 &
\cite{gm2} \\
$\brbsgamma \times 10^{4}$ &
3.55 & 0.26 & 0.21 & \cite{hfag}   \\
$\delmbs$     &  $17.77\ps^{-1}$  & $0.12\ps^{-1}$  &
$2.4\ps^{-1}$
& \cite{cdf-deltambs} \\
$\brbtaunu \times 10^{4}$ &  $1.32$  & $0.49$  & $0.38$
& \cite{hfag} \\
$\abundchi$ &  0.1099 & 0.0062 & $0.1\,\abundchi$& \cite{wmap5yr}
\\ \hline\hline
\multicolumn{5}{|c|}{limits only (95\%~\cl)}\\\hline\hline
$\brbsmumu$ & $ <5.8\times 10^{-8}$& & 14\% & \cite{cdf-bsmumu}\\
$\mhl$  & $>114.4\gev$\ (SM-like Higgs)  & & $3\gev$ & \cite{lhwg} \\
$\zetah^2$ & $f(m_h)$\ (see Ref.~\citealp{Austri:0602028}) & & negligible & \cite{lhwg} \\
$m_{\tilde\chi_1^0}$   & $>50\gev$\ & & 5\% & \cite{lsp} \\
$m_{\tilde\chi^\pm_1}$& $>103.5\gev$ ($>92.4\gev$)\ & & 5\% & \cite{lepsusy} (\cite{aleph,l3})\\
$m_{\tilde{e}_R}$ & $>100\gev$ ($>73\gev$)\ & & 5\% & \cite{lepsusy} (\cite{aleph,l3})\\
$m_{\tilde{\mu}_R}$ & $>95\gev$ ($>73\gev$)\ & & 5\% & \cite{lepsusy} (\cite{aleph,l3})\\
$m_{\tilde{\tau}_1}$ & $>87\gev$ ($>73\gev$)\ & & 5\% & \cite{lepsusy} (\cite{aleph,l3})\\
$m_{\tilde{\nu}}$ & $>94\gev$ ($>43\gev$)\ & & 5\% & \cite{delphi} (\cite{aleph,l3})\\
$m_{\tilde{t}_1}$ & $>95\gev$ ($>65\gev$)\ & & 5\% & \cite{lepsusy} (\cite{aleph,l3})\\
$m_{\tilde{b}_1}$ & $>95\gev$ ($>59\gev$)\ & & 5\% & \cite{lepsusy} (\cite{aleph,l3})\\
$m_{\tilde{q}}$ & $>375\gev$& & 5\% & \cite{pdg07} \\
$m_{\tilde{g}}$ & $>289\gev$& & 5\% & \cite{pdg07} \\ \hline
\end{tabular}
} \caption[aa]{\footnotesize{List of all physical observables used in the analysis.  These are: collider limits on Higgs and superpartner masses, electroweak precision measurements, $B$-physics quantities and the dark matter relic density.  For the sake of comparability, these are the same quantities and values as used in Ref.~\citealp{Trotta:08093792}.  The upper sub-table gives measurements of SM nuisance parameters.  The central panel consists of observables for which a positive measurement has been made, and the lower panel shows observables for which only limits exist at the moment.  The numbers in parenthesis are more conservative bounds applicable only under specific conditions.  For details and arguments, see Refs.~\citealp{Trotta:08093792,Austri:0602028} and~\citealp{Roszkowski:0611173}.
Table adapted mostly from Ref.~\citealp{Feroz:09032487}.}} \label{tab:obs}
}

In order to calculate all observables and likelihoods for different points in the CMSSM
parameter space, we have used \textsf{SuperBayeS~1.35}~\cite{superbayes}, a publicly available package which combines
\textsf{SoftSusy}~\cite{softsusy},
\textsf{DarkSusy}~\cite{darksusy},
\textsf{FeynHiggs}~\cite{feynhiggs}, \textsf{Bdecay} and
\textsf{MicrOMEGAs}~\cite{micromegas} in a statistically consistent way.  The public version of the package offers three different scanning algorithms: MCMCs,~\MN, and fixed-grid scanning.  We have modified the code to also include GAs.  Other global-fit packages are also available: \textsf{Fittino}~\cite{Bechtle:09072589}, for MCMC scans of the CMSSM, \textsf{SFitter}~\cite{Lafaye:07093985}, for MCMC scans of the CMSSM and also weak-scale MSSM, and \textsf{Gfitter}~\cite{Flacher:08110009}, for Standard Model fits to electroweak precision data (SUSY fits will soon be included as well). Amongst other search strategies, \textsf{Gfitter} can make use of GAs.

In \textsf{SuperBayeS}, the likelihoods of observables for which positive measurements exist (indicated in the upper and central panels of~\tab{tab:obs}), are modeled by a multi-dimensional Gaussian function.  The variance of this Gaussian is given by the sum of the experimental and theoretical variances associated with each observable; the corresponding standard deviations are shown in~\tab{tab:obs}.  For observables where only upper or lower limits exist (indicated in the lower panel of~\tab{tab:obs}), a smeared step-function likelihood is used, constructed taking into account estimated theoretical errors in calculating the predicted values of the observables.  For details on the exact mathematical forms of these likelihood functions, see Ref.~\citealp{Austri:0602028}.  

In addition to experimental constraints from collider and cosmological observations, one must also ensure that each point is physically self-consistent; those that are not should be discarded or assigned a very low likelihood value.  Unphysical points are ones where no self-consistent solutions to the RGEs exist, the conditions of electroweak symmetry-breaking are not satisfied, one or more masses become tachyonic, or the theoretical assumption that the neutralino is the LSP is violated.  This is done in \textsf{SuperBayeS} by assigning an extremely small (almost
zero) likelihood to the points that do not fulfil the physicality conditions.

\subsection{Genetic Algorithms and profile likelihoods of the CMSSM} \label{sec:GA}

GAs~\cite{Goldberg:1989,GArecrev,Goldberg:2010} are a class of adaptive heuristic search techniques that draw on the evolutionary ideas of natural selection and survival of the fittest to solve optimisation problems.  According to these principles, individuals in a breeding population which are better adapted to their environment generate more offspring than others.

GAs were invented in early 1970s, primarily by John Holland and colleagues for solving optimisation problems~\cite{Holland}, although the idea of evolutionary computing had been introduced as early as the 1950s.  Holland also introduced a formal mathematical framework, known as \textit{Holland's Schema Theorem}, which is commonly considered to be the theoretical explanation for the success of GAs.  Now, about thirty years after their invention, GAs have amply demonstrated their practical usefulness (and robustness) in a variety of complex optimisation problems in computational science, economics, medicine and engineering~\cite{GArecrev}.

The idea is very simple:  in general, solving a problem means nothing more than finding the one solution most compatible with the conditions of the problem amongst many candidate solutions, in an efficient way.  In most cases, the quality of different candidate solutions can be formulated in terms of a mathematical `fitness function', to be maximised by some algorithm employed to solve the problem.  With a GA, one repeatedly modifies a population of individual candidate solutions in such a way that after several iterations, the fittest point in the population evolves towards an optimal solution to the problem.

These iterative modifications are designed to imitate the reproductive behaviour of living organisms.  At each stage, some individuals are selected randomly or semi-randomly from the current population to be parents.  These parents produce children, which become the next generation of candidate solutions.  If parents with larger fitness values have more chance to recombine and produce children, over successive generations the best individual in the population should approach an optimal solution.

Because GAs are based only on fitness values at each point, and are insensitive to the function's gradients, they can also be applied to problems for which the fitness function has many discontinuities, is stochastic, highly non-linear or non-differentiable, or possesses any other special features which make the optimisation process extremely difficult.  GAs are generally prescribed when the traditional optimisation algorithms either fail entirely or give substandard results.  These properties make GAs ideal for our particular problem, as the CMSSM has exactly those unruly properties.

In what follows, we describe the general algorithmic strategy employed in a GA, followed by our particular implementation for a profile likelihood analysis of the CMSSM.

\subsubsection{Main strategy}

Denote the full model likelihood by $\mathcal{L}(\Theta)$, where $\Theta$ is the set of free parameters introduced in~\eq{allparams}.  This function, as a natural proxy for the goodness-of-fit given a fixed number of degrees of freedom, indicates how fit each particular $\Theta$, or individual, is.  It is thus a good choice for the genetic fitness function.  We now want to find a specific individual, say $\Theta_{max}$, for which the fitness function $\mathcal{L}(\Theta)$ is globally maximised.

Consider a population of $I$ candidate individuals $\Theta_i$ ($i=1,...,I$).  Denote this entire population by $P$.  This population is operated on $K$ times by a semi-random \emph{genetic operator} $\mathbb{G}$ to produce a series of new populations $P^k$, where ($k=1,...,K$) and $P^k = \mathbb{G}P^{k-1}$.  The $i$th individual in the $k$th generation is $\Theta_i^k$.  For the general fitness of individuals to improve from one generation to the next, $\mathbb{G}$ must clearly depend upon $\mathcal{L}(\Theta)$.

The search must be initialised with some starting population $P^0$, which is then evolved under the action of $\mathbb{G}$ until some convergence criterion $\mathbb{T}$ is met.  At this stage, the algorithm returns its best estimate of $\Theta_{max}$ as the individual $\Theta$ where $\mathcal{L}(\Theta_i^k)$ is maximised.  If $\mathbb{G}$ and $\mathbb{T}$ have been chosen appropriately, this should occur at $k=K$, i.e. in the last generation.  This algorithm can be summarised as follows:\\

\small{
\begin{array}[t]{llllr}
 &\multicolumn{2}{l}{\textcolor{red}{\mbox{\texttt{initialisation:}}}} & & \\
 &\multicolumn{2}{l}{P^0:=\{\Theta_i^0\},~\forall i\in[1,I]} & & \\
 &\multicolumn{2}{l}{k:=0} & & \\
 &\multicolumn{2}{l}{\textcolor{red}{\mbox{\texttt{reproduction~loop:}}}} & & \\
 &\multicolumn{2}{l}{\mbox{do~while~not}~\mathbb{T}} & & \\
 & &k:=k+1 & & \\
 & &\multicolumn{2}{l}{\textcolor{red}{\mbox{\texttt{generating~new~population~through~genetic~operators:}}}} & \\
 & &P^k:=\mathbb{G}P^{k-1} & \\
 &\mbox{end~do} & & & \\
 &\multicolumn{2}{l}{\textcolor{red}{\mbox{\texttt{reading~the~best-fit~point:}}}} & & \\
 &\multicolumn{3}{l}{\Theta_{max}:=\Theta_l^m~\mbox{where}~\mathcal{L}(\Theta_l^m)=\max{\{\mathcal{L}(\Theta_i^k)\}},~\forall i\in[1,I],~\forall k\in[1,K]}&  \\ 
 & & & &
\end{array}}
\normalsize

Three main properties define a GA.  Firstly, $\mathbb{G}$ operates on a population of points rather than a single individual.  This makes GAs rather different from conventional Monte Carlo search techniques such as the MCMC, though the nested sampling algorithm does also act on a population of points.  The parallelism of a GA means that if appropriate measures of population diversity are incorporated into the algorithm, local maxima in the likelihood surface can be dealt with quite effectively; if a population is required to maintain a certain level of diversity, concentrations of individuals clustering too strongly around local maxima will be avoided by the remaining members of the population.  This parallelism increases the convergence rate of the algorithm remarkably.

Secondly, $\mathbb{G}$ does not operate directly on the real values of the parameters $\Theta_i$ (the
phenotype), but acts on their encoded versions (the chromosomes, or genotype).  Depending on the problem, individuals can be encoded as a string of binary or decimal digits, or even more complex data structures.  $\mathbb{G}$ then acts on the chromosomes in the current generation based only on their fitnesses, i.e. the likelihood function in our case.  No further information is required for the GA to work; this means discontinuities in the likelihood function or its derivatives have virtually no affect on the performance of the algorithm.

Finally, the transition rules used in $\mathbb{G}$ are probabilistic, not deterministic.  The constituent genetic operators contained within $\mathbb{G}$ are the key elements of the algorithm, and how they act on different populations defines different types of GAs.  In our case
\be \mathbb{G} = \mathbb{RMCS},\ee
where $\mathbb{S}$ is the selection procedure (how parents are selected for breeding from a source population), $\mathbb{C}$ is the crossover (how offspring are to inherit properties from their parents), $\mathbb{M}$ is the mutation process (random changes made to the properties of newly-created offspring), and $\mathbb{R}$ is the reproduction scheme used to place offspring into a broader population.  $\mathbb{C}$ and $\mathbb{M}$ are stochastic processes defined at the genotype level, whereas $\mathbb{S}$ and $\mathbb{R}$ are phenotype-level processes, semi-random and essentially deterministic in nature, respectively.

The randomised operators of GAs are strongly distinct from simple random walks.  This is because every new generation of individuals inherits some desirable characteristics from the present generation.  Crossover (or recombination) rules play a crucial role in determining how the parents create the children of the next generation.  The children are not copied directly to the next population; mutation rules specify that a certain degree of random modification should be applied to the newly-produced offsprings' chromosomes.  Mutation is very important at this stage, since it is often the only mechanism preventing the algorithm from getting stuck in local maxima; its strength is typically linked dynamically to some measure of population diversity.

The reproduction loop is terminated whenever $\mathbb{T}$ is fulfilled.  In our case, whenever a predefined number of generations $N$ have been produced (i.e. $K\equiv N$).  The termination parameter (i.e. the number of generations $N$) and the chosen termination criterion itself depend upon the particular problem at hand and how accurate a solution is required.  The fittest individual in the final population is then accepted as the best-fit solution to the problem.  If one is also interested in mapping the likelihood function in the vicinity of the best-fit points, so as to be able to plot e.g. 1 and 2$\sigma$ confidence regions for different CMSSM parameters, then it is useful to also retain all the individuals produced during the iterations of the algorithm.

\subsubsection{Our specific implementation}

Although any algorithm with the basic features described above ensures progressive improvement over successive generations, to guarantee absolute maximisation one usually needs to implement
additional strategies and techniques, depending on the particular problem at hand. To implement a GA for analysing the CMSSM parameter space, we have taken advantage of the public GA package \textsf{PIKAIA}~\cite{pikaia,pikaiaweb}.  Here we briefly describe the options and ingredients from \textsf{PIKAIA~1.2} that we used in our GA implementation; the majority of these were the default choices.

$\bullet$ \textbf{Fitness function:}  A natural fitness function to choose is the log-likelihood of the model, $\ln \mathcal{L}(\Theta)$. This function is however always negative (or zero).  Since \textsf{PIKAIA} internally seeks to maximise a function, and this function must be positive definite, we chose the inverse chi-square (i.e. $\frac{1}{\chi^2}$) as the simplest appropriate fitness function.  Except for the way we adjust the mutation rate for different generational iterations (see below), all the other genetic operators implemented in our algorithm are functions only of the ranking of individuals in the population; the actual values of the fitness function at these points do not matter as long as the ranking is preserved.

$\bullet$ \textbf{Encoding:}  We encode individuals in the population (i.e. points in the CMSSM parameter space) using a decimal alphabet.  That is, a string of base 10 integers, such that every normalised parameter $\theta_i$ is encoded into a string $d_1d_2...d_{n_d}$, where the $d_i\in[0,9]$ are positive integers.  This is different from many other public-domain GAs which usually make use of binary encoding.  We use 5 digits to encode each parameter.  This means that every individual chromosome is a decimal string of length $m\times n_d=8\times5=40$.

$\bullet$ \textbf{Initialisation and population size:}  We use completely random points in the parameter space for the initial population.  We choose a population size of $n_p=100$ (the typical number usually used in GAs), keeping it fixed throughout the entire evolutionary process.

$\bullet$ \textbf{Selection:}  In order to select parents in any given iteration, a stochastic mechanism is used.  The probability of an individual to be selected for breeding is determined based on its fitness in the following way:  first, we assign to each individual $\Theta_i$ a rank $r_i$ based on its fitness $f_i$ such that $r=1$ corresponds to the fittest individual and $r=n_p$ to the most unfit.  Then a ranking fitness $f'_i$ is defined in terms of this rank:
\be f'_i=n_p-r_i+1.\nn\ee
The sum of all ranking fitness values in the population is computed as
\be F=\sum_{i=1}^{n_p} f'_i,\nn\ee
and $n_p$ running sums are defined as
\be S_j=\sum_{i=1}^{j}f'_i,~j=1,...,n_p.\nn\ee
Obviously $S_{j+1}\geq S_{j}$ (since $f'_i$ are all positive), and $S_{n_p}=F$.  As the next step, a random number $R\in[0,F]$ is generated and the element $S_j$ is located for which $S_{j-1}\leq R <S_j$. The corresponding individual is one of the parents selected for breeding; the other one is also chosen in the same manner.  This selection procedure is called the Roulette Wheel Algorithm (see Ref.~\citealp{pikaia} and references therein for more on this procedure and the motivation for using the ranking fitness in place of the true fitness).

$\bullet$ \textbf{Crossover:}  When a pair of parent chromosomes have been chosen, a pair of offspring are produced via the crossover operator $\mathbb{C}$.  We use two different types of operators for this purpose, namely, uniform one-point and two-point crossovers (see Ref.~\citealp{Tomasz:2006} for a comprehensive review of existing crossover operators).  \tab{crossover} illustrates how these two processes work.  In our case, parents are encoded as $40$-digit strings.  The one-point crossover begins by randomly selecting a cutting point along the chromosomes' length, and dividing each parent string into two sub-strings.  This is done by generating a random integer $K\in[1,40]$.  According to the crossover scheme, the strings located in identical parts of the two strings are then swapped to give the two children chromosomes.  It is clear that even though the information encoded in the parent chromosomes is transfered to the offspring chromosomes, the latter are in general different from the former, corresponding to a different set of model parameters in the parameter space.  In the two-point crossover scheme, two slicing points are selected randomly by generating two random integers $K_1,K_2\in[1,40]$, and the string segments between these two cutting points are exchanged.

\TABLE{
\centering
{\small
\begin{tabular}{|l | l | l|} \hline
\multicolumn{3}{|c|}{uniform one-point crossover} \\ \hline\hline
initial parent chromosomes & \textsf{\textcolor{blue}{6739...8451}}  & \textsf{\textcolor{red}{4394...0570}} \\ \hline
selecting a random cutting point &  \textsf{\textcolor{blue}{6739...84}}$|$\textsf{\textcolor{blue}{51}}    & \textsf{\textcolor{red}{4394...05}}$|$\textsf{\textcolor{red}{70}} \\ \hline
swapping the sub-strings &  \textsf{\textcolor{blue}{6739...84}}$|$\textsf{\textcolor{red}{70}} & \textsf{\textcolor{red}{4394...05}}$|$\textsf{\textcolor{blue}{51}} \\ \hline
final offspring&  \textsf{\textcolor{blue}{6739...84}\textcolor{red}{70}} & \textsf{\textcolor{red}{4394...05}\textcolor{blue}{51}} \\ \hline \hline
\multicolumn{3}{|c|}{uniform two-point crossover} \\ \hline\hline
initial parent chromosomes & \textsf{\textcolor{blue}{6739...8451}}  & \textsf{\textcolor{red}{4394...0570}} \\ \hline
selecting two random cutting points &  \textsf{\textcolor{blue}{67}}$|$\textsf{\textcolor{blue}{39...845}}$|$\textsf{\textcolor{blue}{1}}    & \textsf{\textcolor{red}{43}}$|$\textsf{\textcolor{red}{94...057}}$|$\textsf{\textcolor{red}{0}} \\ \hline
swapping the sub-strings &  \textsf{\textcolor{blue}{67}}$|$\textsf{\textcolor{red}{94...057}}$|$\textsf{\textcolor{blue}{1}} & \textsf{\textcolor{red}{43}}$|$\textsf{\textcolor{blue}{39...845}}$|$\textsf{\textcolor{red}{0}} \\ \hline
final offspring&  \textsf{\textcolor{blue}{67}\textcolor{red}{94...057}\textcolor{blue}{1}} & \textsf{\textcolor{red}{43}\textcolor{blue}{39...845}\textcolor{red}{0}} \\ \hline
\end{tabular}
} \caption[aa]{\footnotesize{Schematic description of the uniform one-point and two-point crossover operators employed in our analysis.}} \label{crossover}
} 

In our algorithm, for each pair of parent chromosomes, either one-point or two-point crossover is chosen with equal probability.  This combination of the one-point and two-point crossovers is proposed to avoid the so-called ``end-point bias'' problem produced by using only the one-point crossover.  This happens when, for example, a combination of two sub-strings situated at opposite ends of a parent string are advantageous when decoded back to the phenotypic level (in the sense that they give a high fitness), \emph{but only when they are expressed simultaneously}.  Such a feature is impossible to pass on to offspring when using a uniform one-point crossover, as cutting the string at any single point always destroys this combination.  This is much less of a problem for sub-strings located more centrally in the parent string (see Ref.~\citealp{pikaia} again for more details on this issue).

Once two parents have been selected for breeding, the crossover operation is applied only with a preset probability.  This probability is usually taken to be large but not $100\%$.  We use $85\%$ in our analysis, meaning that there is a $15\%$ chance that any given breeding pair will be passed on intact to the next stage, where they will be affected by mutation.

$\bullet$ \textbf{Mutation:}  We employ the so-called uniform one-point mutation operator. Different genes in the offspring's chromosomes (i.e. decimal digits in the 40-digit strings) are replaced with a predefined probability (the `mutation rate'), by a random integer in the interval $[0,9]$.  Like the crossover operator, mutation preserves the parameter bounds.  The choice of mutation rate is highly important, and very problem-dependent; in general it cannot be chosen {\it a priori}.  If too large, it can destroy a potentially excellent offspring and, in the extreme case, make the whole algorithm behave effectively as an entirely random search.  If too small, it can endanger the variability in the population and cause the whole population to become trapped in local maxima; a large enough mutation rate is often the only mechanism to escape premature convergence at local maxima.  Therefore, instead of using a fixed mutation rate we allow it to vary dynamically throughout the run, such that the degree of `biodiversity' is monitored and the mutation rate is adjusted accordingly.  When the majority of individuals in a population (as estimated by the median individual) are very similar to the best individual, the population is probably clustered around a local maximum and the mutation rate should increase.  The converse is also true: a high degree of diversity indicates that the mutation rate should be kept low.  The degree of clustering and the subsequent amount of adjustment in our algorithm are assessed based on the difference between the actual fitness values of the best and median points.  This is done by defining the quantity $\Delta f=(f_{r=1}-f_{r=n_p/2})/(f_{r=1}+f_{r=n_p/2})$ in which $f_{r=1}$ and $f_{r=n_p/2}$ correspond to the fitnesses of the best and median individuals, respectively.  The mutation rate is then increased (decreased) by a fixed multiplicative factor whenever $\Delta f$ is smaller (larger) than a preset lower (upper) bound.  We choose the lower and upper critical values of $\Delta f$ to be $0.05$ and $0.25$ respectively, and the multiplicative factor to be $1.5$.  We bound the mutation rate to lie between the typical values of $0.0005$ and $0.25$.  We use an initial mutation rate of $0.005$.

$\bullet$ \textbf{Reproduction plans:}  After selecting parents and producing offspring by acting on them with the crossover and mutation operators, the newly bred individuals must somehow be incorporated into the population.  The strategy which controls this process is called a reproduction plan.  Although there are several advanced reproduction plans on the market, we utilise the simplest off-the-shelf option: full generational replacement. This means that the whole parent population is replaced by the newly-created children in each iteration, all in a single step.

$\bullet$ \textbf{Elitism:}  There is always a possibility that the fittest individual is not passed on to the next generation, since it may be destroyed under the action of the crossover and mutation operations on the parents.  To guarantee survival of this individual, we use an elitism feature in our reproduction plan.  Under full generational replacement, elitism simply means that the fittest individual in the offspring population is replaced by the fittest parent, if the latter has a higher fitness value.

$\bullet$ \textbf{Termination and number of generations:}  There are several termination criteria one could use for a GA.  Rather than evolving the population generation after generation until some tolerance criterion is met, we perform the evolution over a fixed and predetermined number of generations.  This is because the former strategy is claimed to be potentially dangerous when approaching a new problem, in view of the usual convergence trends exhibited by GA-based optimisers (see Ref.~\citealp{pikaia} for more details).  In our analysis, we used 10 separate runs of the algorithm with different initial random seeds, and a fixed number of likelihood evaluations ($\sim 3\times10^{5}$) for each.  We then compiled all points into a single list, and used it to map the profile likelihood of the CMSSM.

\section{Results and discussion} \label{sec:results}

We now present and analyse the results of our global profile likelihood fit of the CMSSM using GAs.  We compare results with a similar global fit using the state-of-the-art Bayesian algorithm \MN~\cite{MultiNest}.  In \sec{sec:BFP} we give the best-fit points and high-likelihood regions in the CMSSM parameter space.  In \sec{sec:LHC} we discuss and compare implications of both methods for the detection of supersymmetric and Higgs particles at the LHC.  \sec{sec:DM} is devoted to an investigation of dark matter in the CMSSM and the prospects for its direct and indirect detection.  Throughout this section we compare our GA profile likelihood results mainly with those of the \MN~algorithm implemented with linear (flat) priors.  The reasons for this are outlined in \sec{sec:TechComp}, along with a comparison of the two scanning techniques in terms of the computational speed and convergence.

\subsection{Best-fit points and high-likelihood regions} \label{sec:BFP}

\begin{figure}
\begin{center}
\subfigure[][]{\label{m0mhf:a}\includegraphics[width=0.49\linewidth, trim = 70 0 70 50, clip=true]{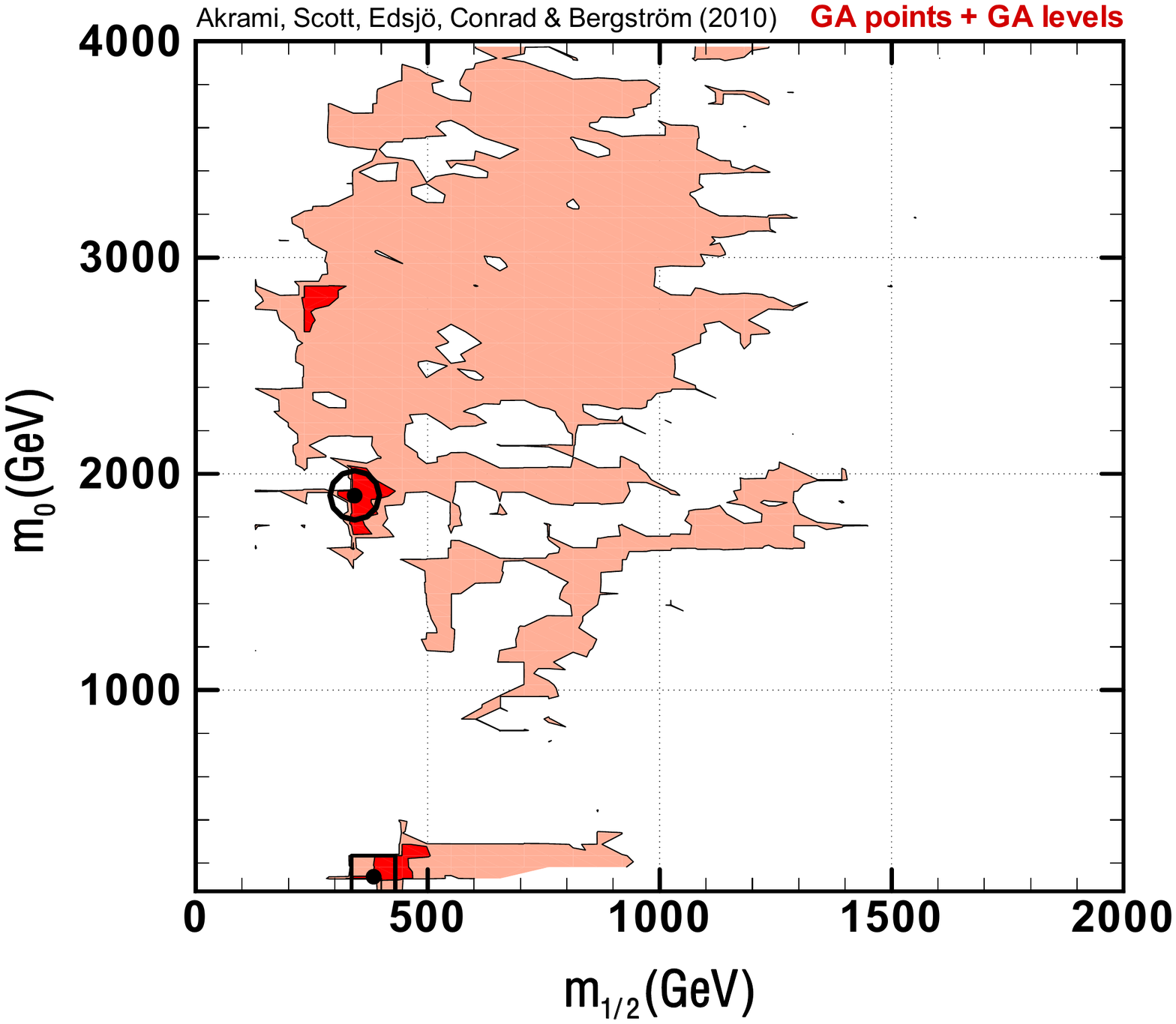}}
\subfigure[][]{\label{m0mhf:b}\includegraphics[width=0.49\linewidth, trim = 70 0 70 50, clip=true]{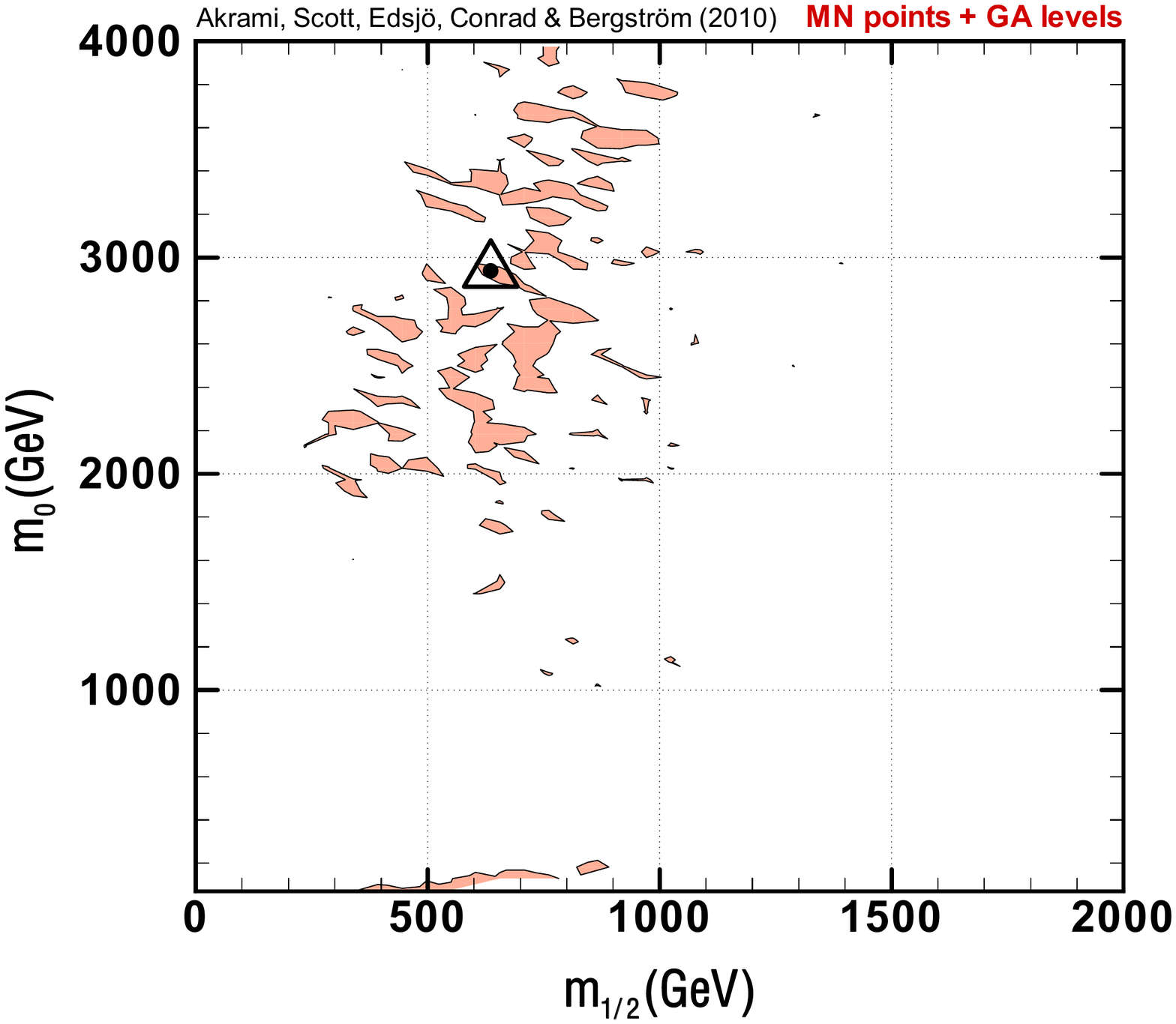}}\\
\subfigure[][]{\label{m0mhf:c}\includegraphics[width=0.49\linewidth, trim = 70 0 70 50, clip=true]{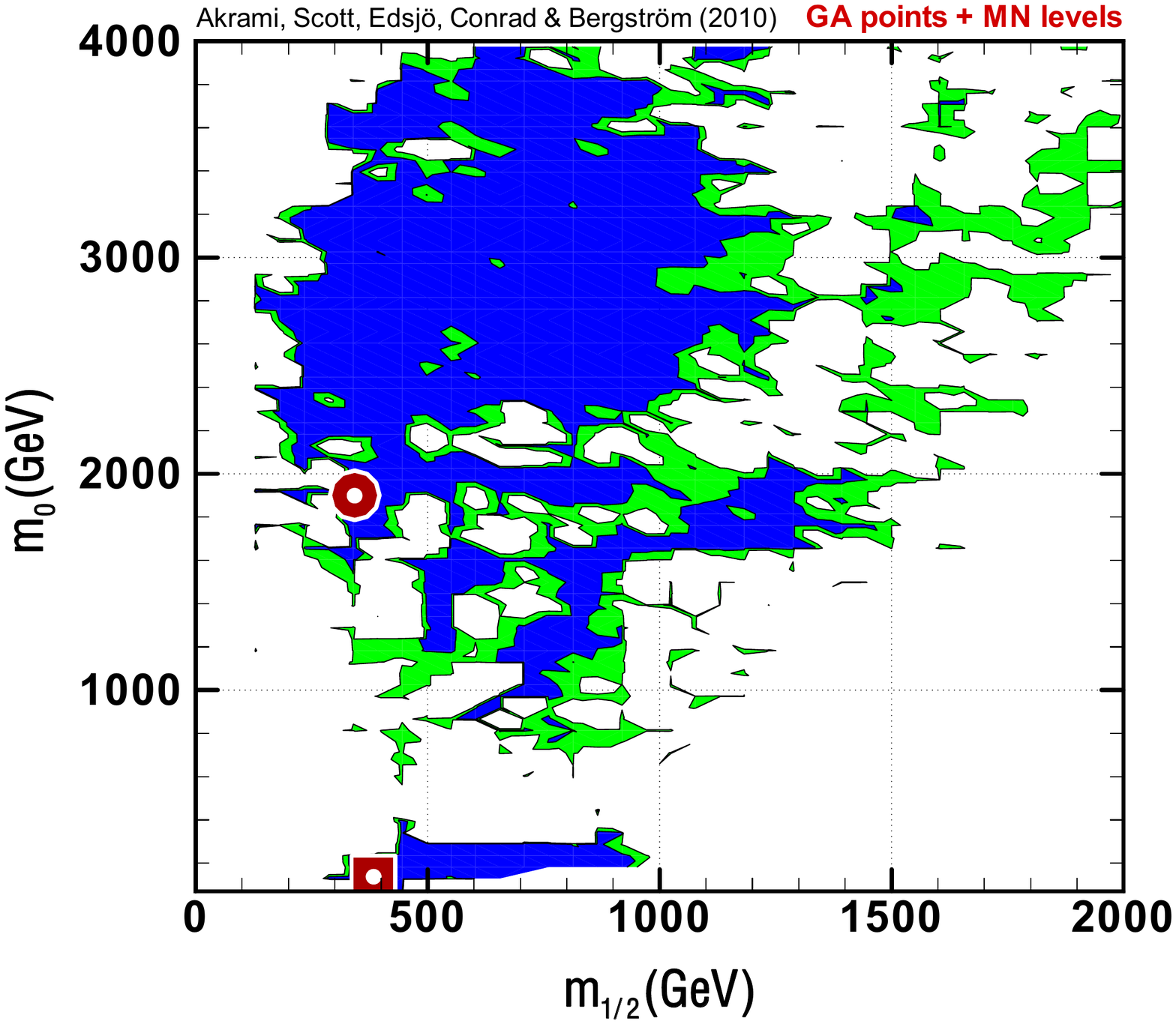}}
\subfigure[][]{\label{m0mhf:d}\includegraphics[width=0.49\linewidth, trim = 70 0 70 50, clip=true]{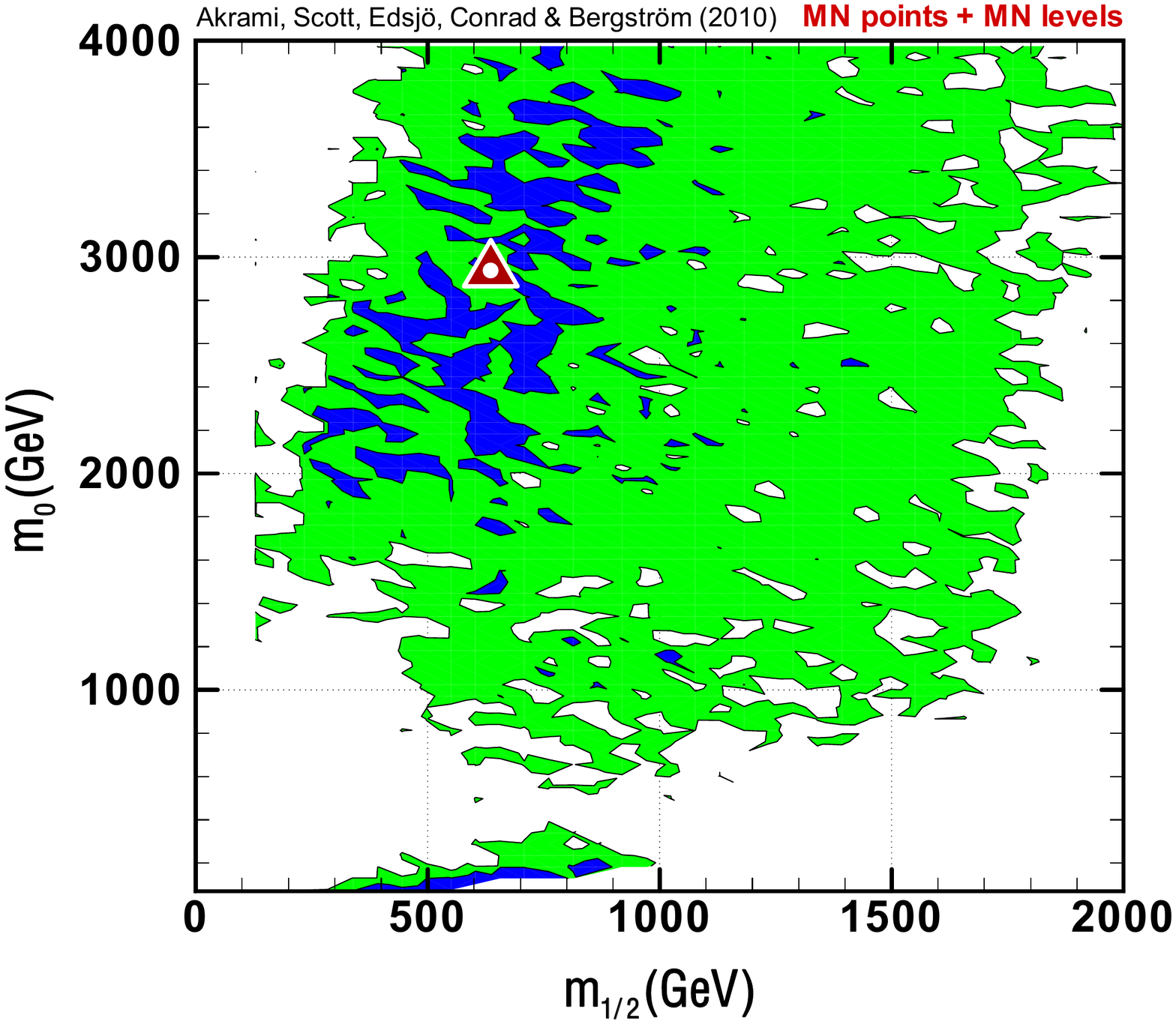}}
\caption[aa]{\footnotesize{Two-dimensional profile likelihoods in the $m_0$-$m_{1/2}$ plane for CMSSM scans with GAs (a) and \MN~(d).  These two panels show the statistically-consistent results of each scan.  The inner and outer contours represent $68.3\%$ ($1\sigma$) and $95.4\%$ ($2\sigma$) confidence regions respectively, for each scan.  The dotted circle, square and triangle show respectively the GA global best-fit point with a $\chi^2$ of $9.35$ (located in the focus point region), the GA best-fit point in the stau co-annihilation region ($\chi^2$ = $11.34$), and the best-fit point found by linear-prior \MN~($\chi^2$ = $13.51$).  Panels (b) and (c) are given for comparative purposes only.  Panel (b) shows the same \MN~sample points as in (d), but with iso-likelihood contour levels drawn as in (a), based on the GA best-fit likelihood value.  Panel (c) shows the same GA sample points as in (a), but with iso-likelihood contours as in (d), based on the \MN~best-fit likelihood value.  The sample points have been divided into $75\times75$ bins in all plots.  Here we see that the GA uncovers a large number of points with much higher likelihoods than \MN, across large sections of the $m_0$-$m_{1/2}$ plane.}}
  \label{fig:m0mhf}
\end{center}
\end{figure}

The $\chi^2$ at our best-fit point is $9.35$.  This is surprisingly better than the values of $13.51$ and $11.90$ found by \MN~with linear and logarithmic priors, respectively, for the same problem.  This improvement in the best fit can in principle have a drastic impact on the statistical inference drawn about the model parameters.

\begin{figure}
\begin{center}
\subfigure[][]{\label{A0tanbeta:a}\includegraphics[width=0.49\linewidth, trim = 70 0 70 50, clip=true]{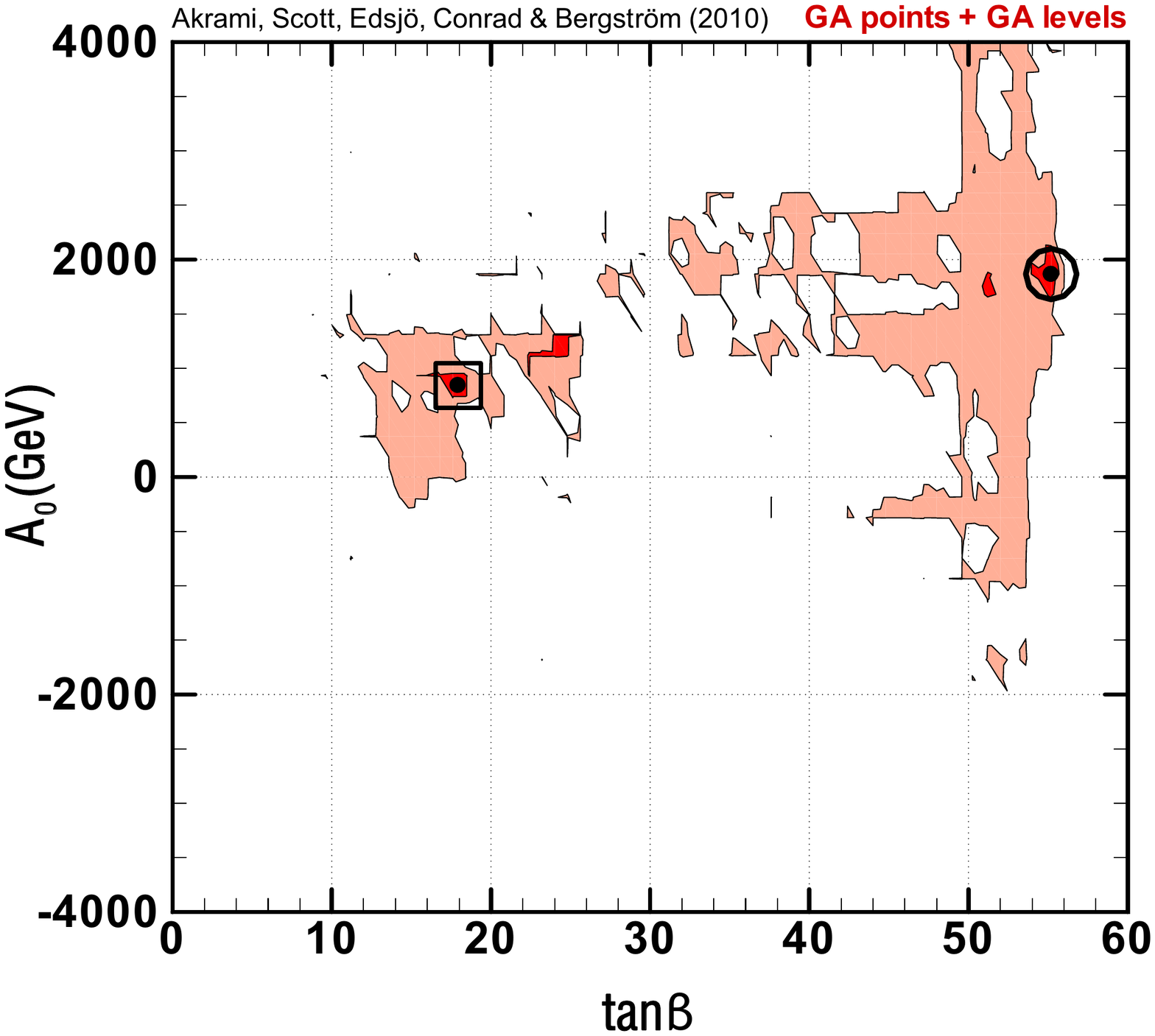}}
\subfigure[][]{\label{A0tanbeta:b}\includegraphics[width=0.49\linewidth, trim = 70 0 70 50, clip=true]{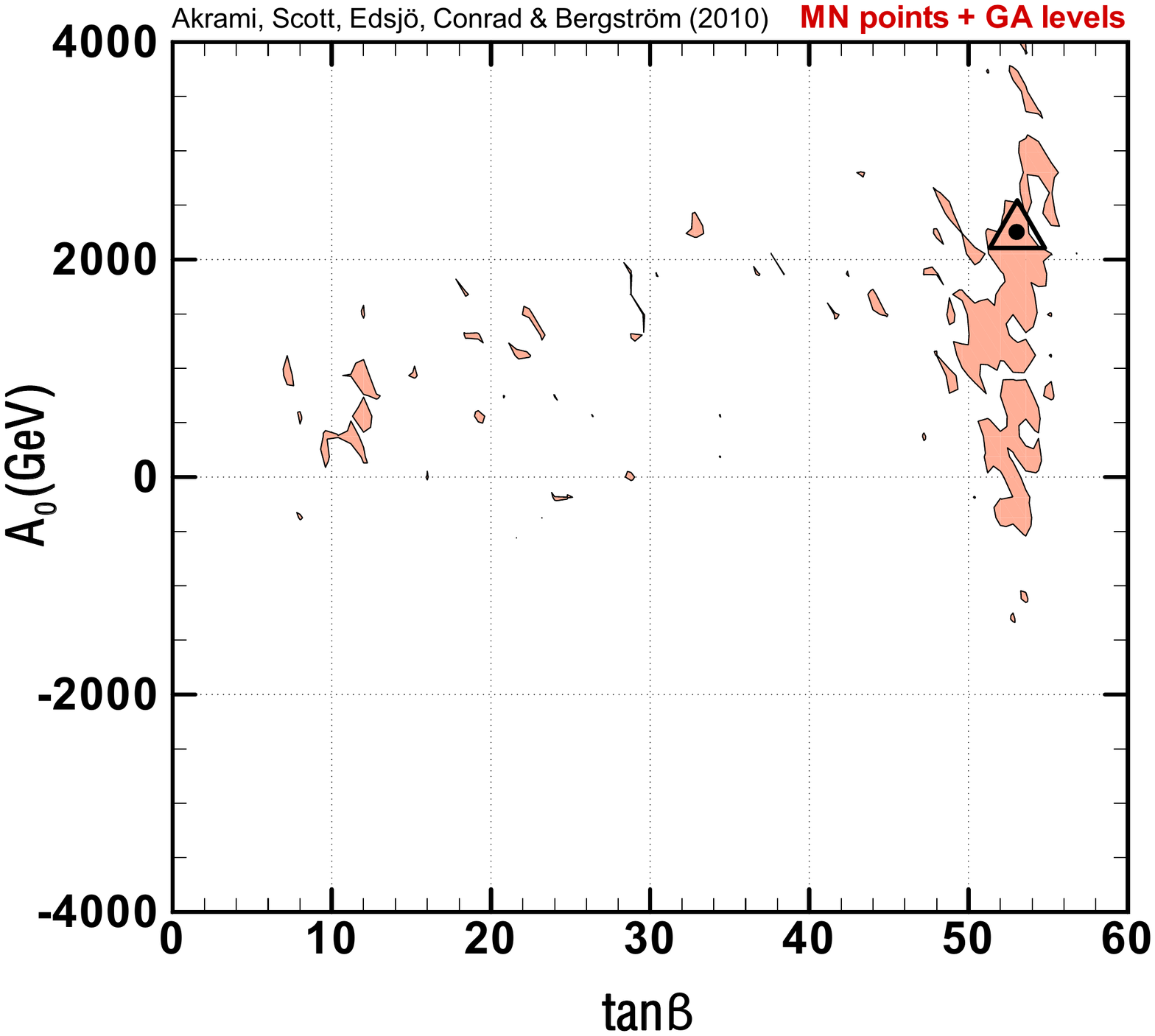}} \\
\subfigure[][]{\label{A0tanbeta:c}\includegraphics[width=0.49\linewidth, trim = 70 0 70 50, clip=true]{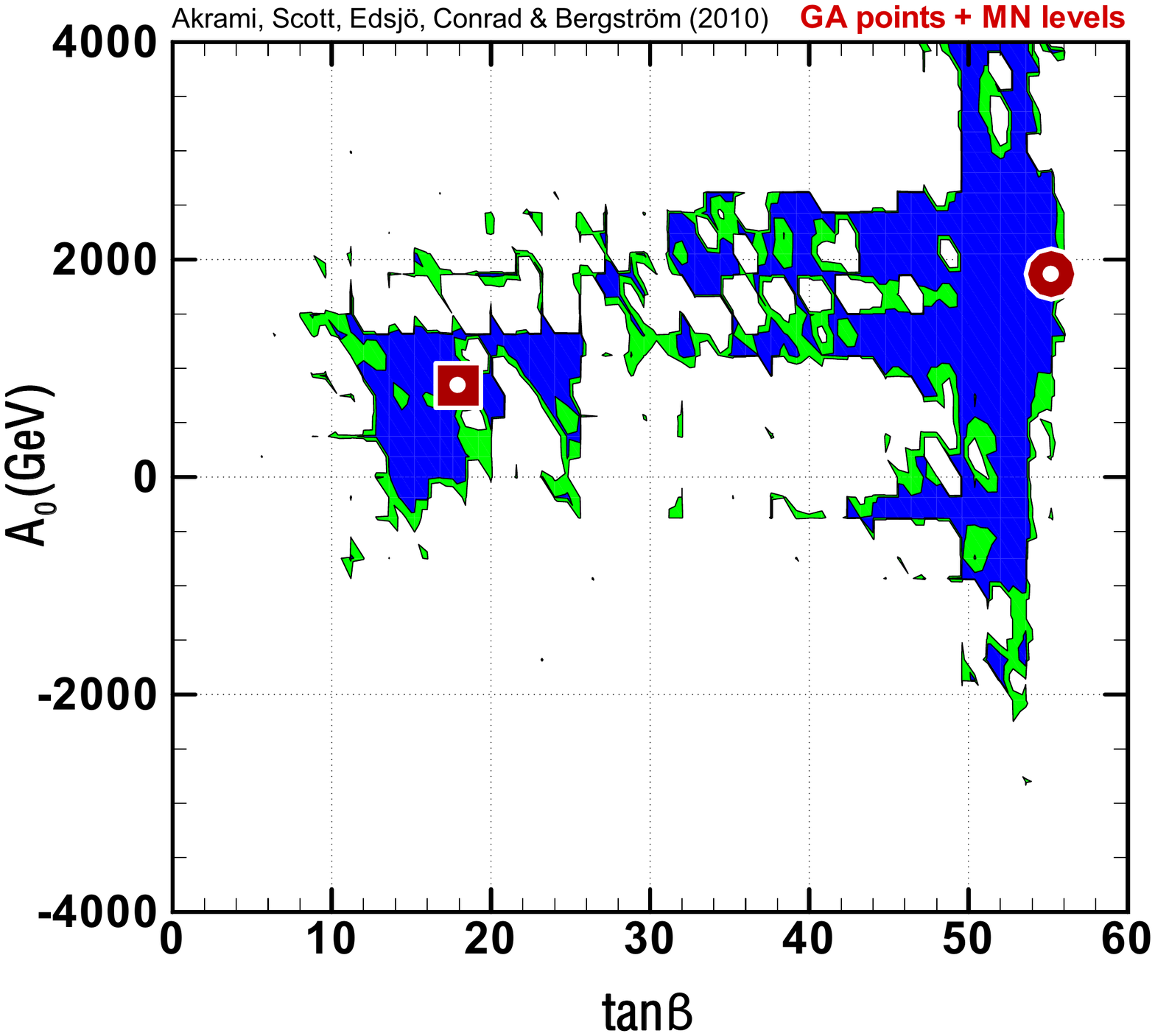}}
\subfigure[][]{\label{A0tanbeta:d}\includegraphics[width=0.49\linewidth, trim = 70 0 70 50, clip=true]{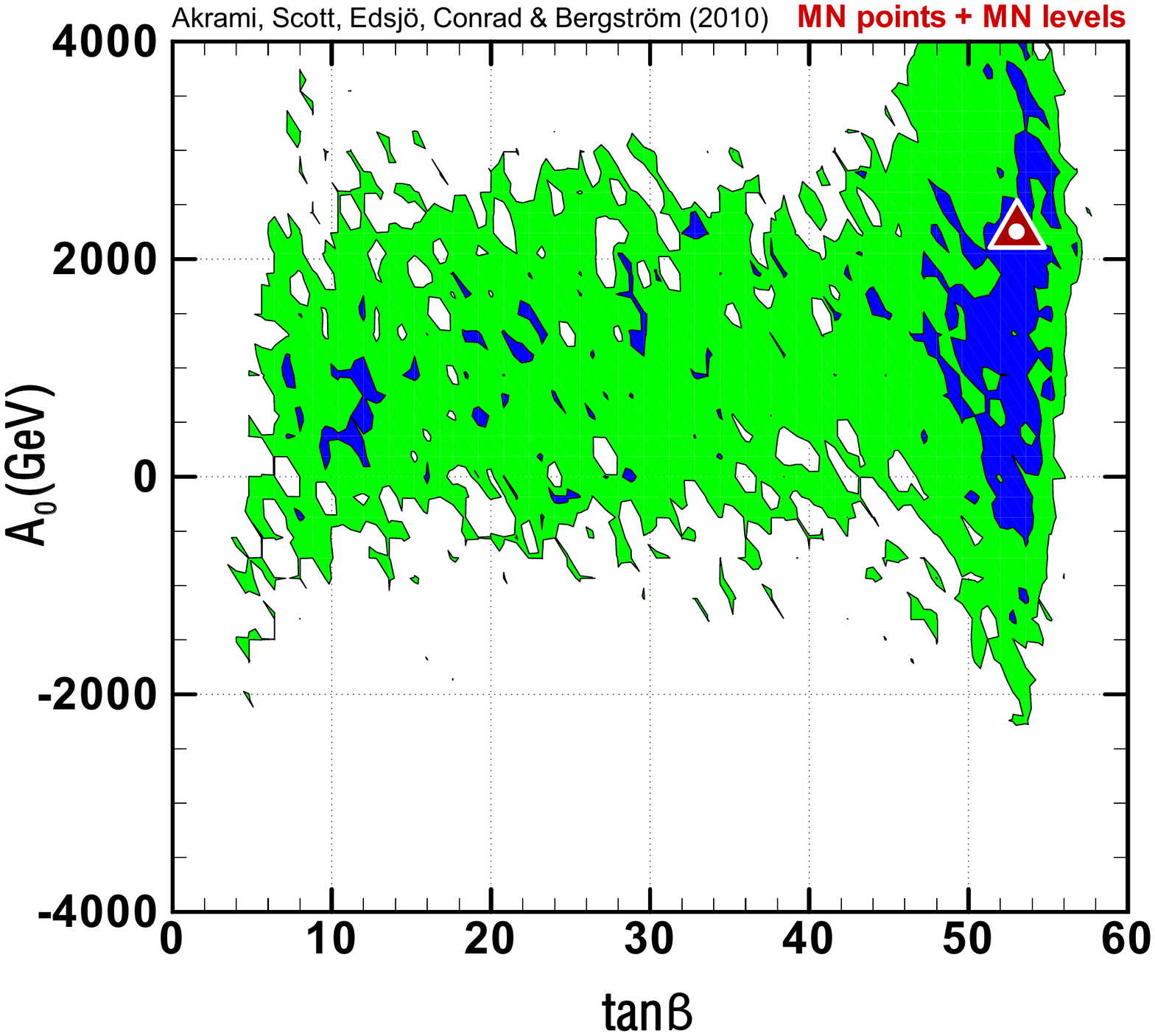}}
\caption[aa]{\footnotesize{As in \fig{fig:m0mhf}, but in the $A_0$-$\tanb$ plane.}}
\label{fig:A0tanbeta}
\end{center}
\end{figure}

To demonstrate these effects, let us start with two-dimensional (2D) plots for the principal parameters of the CMSSM, i.e. $\mzero$, $\mhalf$, $\azero$ and $\tanb$, shown in \figs{fig:m0mhf}{fig:A0tanbeta}.  These figures show 2D profile likelihood maps.  In the first figure, the full likelihood is maximised over all free (CMSSM plus SM nuisance) parameters except $\mzero$ and $\mhalf$.  Similar diagrams are shown in \fig{fig:A0tanbeta}, but now for the 2D profile likelihoods in terms of the parameters $\azero$ and $\tanb$.

Figs.~\ref{fig:m0mhf}a and \ref{fig:A0tanbeta}a show the 2D profile likelihood maps obtained by taking into account all the sample points in the parameter space resulting from the GA scan.  The inner and outer contours indicate $68.3\%$ ($1\sigma$) and $95.4\%$ ($2\sigma$) confidence regions based on the GA best-fit point of $\chi^2=9.35$.  That is, points with $\chi^2\leq11.65$ fall into the $1\sigma$ region, and points with $\chi^2\leq15.52$ fall into the $2\sigma$ region.  Similar plots are presented for the \MN~results in Figs.~\ref{fig:m0mhf}d and \ref{fig:A0tanbeta}d, where the $1\sigma$ and $2\sigma$ contours are drawn based on the \MN~best-fit of $\chi^2=13.51$ (1 and $2\sigma$ regions are given by $\chi^2\leq15.81$ and $\chi^2\leq19.68$, respectively).  These panels reflect the outcomes of each scanning algorithm in the absence of any information from the other.  The sample points have been divided into $75\times75$ bins in all plots and no smoothing is applied.

It is important to realise that although both the 1 and $2\sigma$ confidence regions in the GA results seem to be rather smaller in size than the corresponding regions in the \MN~scan (especially the $1\sigma$ region), this is by no means an indication that \MN~has found more high-likelihood points in the parameter space.  The situation is in fact the exact opposite.  This is clear when we recall that the best-fit likelihood values are very different in the two cases giving rise to two completely different sets of iso-likelihood contours.  To make this point clear, suppose for the sake of argument that the GA best-fit likelihood value is indeed the absolute global maximum we were looking for.  If so, we can now check how well the \MN~algorithm has sampled the parameter space, by looking at Figs.~\ref{fig:m0mhf}b and \ref{fig:A0tanbeta}b.  These show how many of the \MN~samples are located in the correct confidence regions set by the absolute maximum; the contours are drawn based on this best-fit value rather than the one found by \MN~itself.  The plots show that \MN~has discovered no points in the $1\sigma$ region and only a small fraction of the $2\sigma$ region.  In particular, it is interesting to notice that the \MN~best-fit point, i.e. the one with $\chi^2=13.51$ (marked as dotted triangles in \subfigs{fig:m0mhf}{fig:A0tanbeta}{b}{d}) now sits in the $2\sigma$ region.  These all come from the fact that only points with $\chi^2\leq11.65$ and $\chi^2\leq15.52$ fall in the $1\sigma$ and $2\sigma$ regions, respectively, and there are not many points found by \MN~with such low $\chi^2$s.  The same statement holds for the log-prior \MN~best-fit point with $\chi^2=11.90$.

It is obvious from these plots that the GA has found many points in the parameter space with rather high likelihood (or equivalently, low $\chi^2$) which were missed (or skipped) by \MN.  This indicates that the use of \MN~scans in the context of the frequentist profile likelihood is rather questionable.  This is not really surprising, given that \MN~is designed to sample the Bayesian posterior PDF, not map the profile likelihood.

We can also use the resultant GA samples in a different way to clarify this result.  In Figs.~\ref{fig:m0mhf}c and \ref{fig:A0tanbeta}c, the GA samples are plotted with the same contours as in Figs.~\ref{fig:m0mhf}d and \ref{fig:A0tanbeta}d, i.e.~based on the \MN~best-fit likelihood value.  Compared to \ref{fig:m0mhf}d and \ref{fig:A0tanbeta}d, we see that there are many high-likelihood points in the \MN~$1\sigma$ region found by the GA and missed by \MN.  In the sense of the profile likelihood, it appears that \MN~has converged prematurely; we see a much larger and more uniform pseudo-$1\sigma$ region with the GA data.  Here we see that most of the region labeled as being within the $2\sigma$ confidence level in the \MN~scan is actually part of its $1\sigma$ confidence region.

Our results confirm the complexity of the CMSSM parameter space, showing that much care should be taken in making any statistical statement about it.  This is especially true when using a frequentist approach, as this complication plays a crucial role in the final conclusions.  It is of course true that the convergence criterion for \MN~is defined on the basis of the Bayesian evidence, and the algorithm may have (indeed, probably has) converged properly in this context.  The point we want to emphasise is that even if \MN~is converged for a Bayesian posterior PDF analysis of the model, this convergence is far from acceptable for a profile likelihood analysis.  The same is also very likely to be true of other less sophisticated Bayesian methods, such as the MCMC;  this is the case at least for MCMC scans performed with the same physics and likelihood calculations as in our analysis (since MCMCs and \MN~give almost identical results in this case~\cite{Trotta:08093792}).

The aforementioned comparison does also suggest, however, that even the Bayesian posterior PDF obtained from \MN~and MCMC scans might not yet be quite properly mapped.  This is because in principle, a significant amount of probability mass could be contained in the regions found by the GA but missed or skipped by \MN.  Given the absence of any definition of a measure on the parameter space in profile likelihood analyses such as the one we perform, we are unfortunately not in a position to make any conclusive statement about the actual contribution of these regions to the Bayesian probability mass.  Nevertheless, the difference in size between the blue regions in Figs.~\ref{fig:m0mhf}c and \ref{fig:m0mhf}d is intriguing.  We discuss these convergence questions further in \sec{sec:TechComp}.

Our GA scan has found high-likelihood points in many of the CMSSM regions known to be consistent with data, in particular the relic abundance of dark matter~\cite{Allanach:0410091}.  These include the stau ($\tilde\tau$) {\it co-annihilation} (COA) region~\cite{Griest:1990kh} usually at small $\mzero$ where the lightest stau is close in mass to the neutralino, the {\it focus point} (FP) region~\cite{FocusP} at large $\mzero$ where a large Higgsino component causes neutralino annihilation into gauge boson pairs, and the light Higgs boson {\it funnel} region~\cite{Drees:9207234,Djouadi:0504090} at small $\mhalf$.  We have not found any high-likelihood points in the stop ($\tilde t$) co-annihilation region~\cite{Boehm:9911496,Arnowitt:0102181,Ellis:0112113} at large negative $A_0$, where the lightest stop is close in mass to the neutralino.  This could be interpreted as confirmation of the claim that this region, although compatible with the WMAP constraint on the relic density of dark matter, is highly disfavoured when other observables are also taken into account~\cite{Allanach:0507283}.

It is important to make the point that although our method does find some points in the funnel region, it does not spread out very well around those points to map the whole region.  Finding this very fine-tuned region is a known challenge for any scanning strategy, including nested sampling.  The failure of the GA to map other points in the funnel region can be understood.  We believe that this behaviour is caused by the specific crossover scheme employed in our analysis (i.e. one/two-point crossover), and could probably be cured by using a more advanced algorithm.  Alternatively, a different parameterisation of the model, such as a logarithmic scaling of the mass parameters (or equivalently, genetic encoding in terms of the logarithms of these parameters), would probably find the funnel region much more effectively (in the same way as it does when \MN~is implemented with logarithmic priors).  In any case, it is important to realise that these types of regions are findable by our method (although not very well), even without taking into account any ad hoc change in the model parameterisation (or choosing a non-linear prior such as the logarithmic one in the Bayesian language).  See \sec{sec:TechComp} for more discussions about the priors and parameterisation.

Returning to the best-fit points, it is visible from the plots that the global best-fit point is located in the FP region (dotted circles in Figs.~\ref{fig:m0mhf}a,c and \ref{fig:A0tanbeta}a,c).  This has a very interesting phenomenological implication, which we return to later in this section when we discuss contributions to the total likelihood of the best-fit model from different observables. For comparison, we have also marked the best-fit point located in the COA region, which has $\chi^2 = 11.34$ (dotted squares in Figs.~\ref{fig:m0mhf}a,c and \ref{fig:A0tanbeta}a,c).  This point is situated inside the $1\sigma$ confidence level contour (Figs.~\ref{fig:m0mhf}a and \ref{fig:A0tanbeta}a) and is well-favoured by our analysis.  It is interesting to notice that the $\chi^2$ for this point, although worse than the global best-fit $\chi^2$, is still better than the best value found by the \MN~scan, even when implemented with a log prior ($\chi^2=11.90$), which also corresponds to a point in the COA region.  This result is important as \MN~scans with logarithmic priors are usually considered a good way to probe low-mass regions such as the COA.  Our algorithm, even working with effectively linear priors (because the genome featured a linear encoding to the parameters), appears to have found a better point in this region as well.

\begin{figure}
\begin{center}
\includegraphics[width=0.49\linewidth, trim = 0 300 160 10, clip=true]{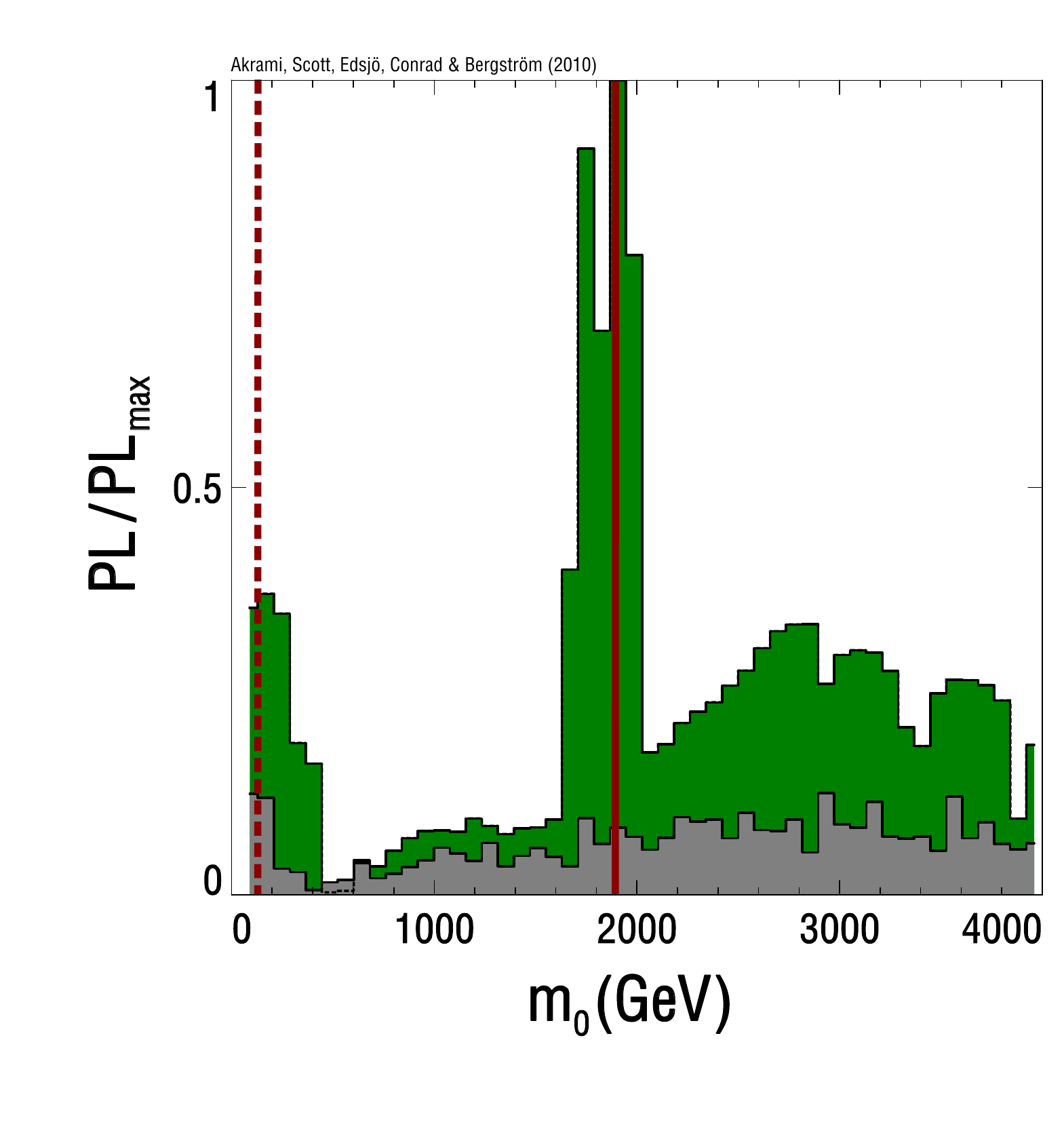}
\includegraphics[width=0.49\linewidth, trim = 0 300 160 10, clip=true]{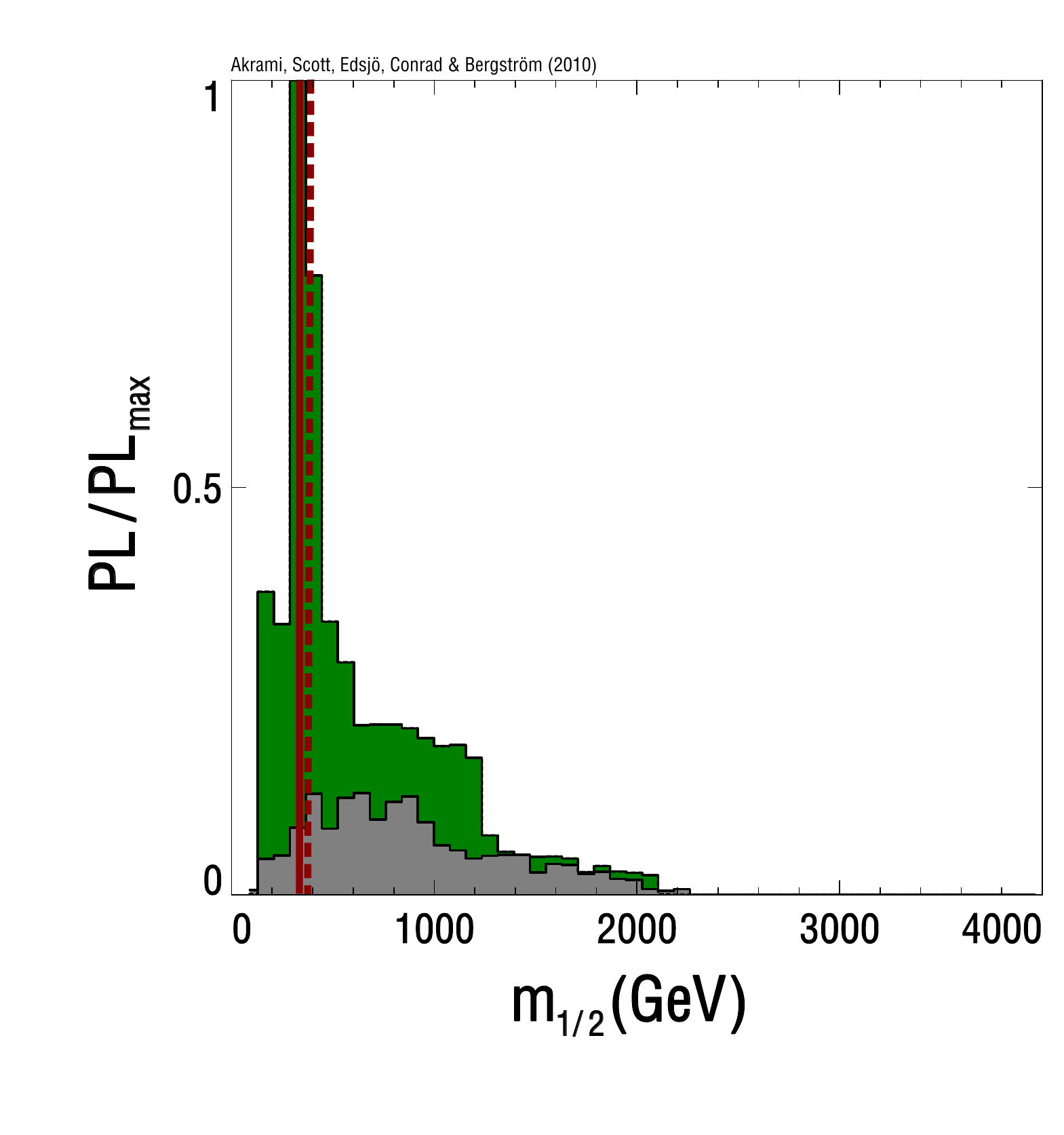} \\
\includegraphics[width=0.49\linewidth, trim = 0 340 160 10, clip=true]{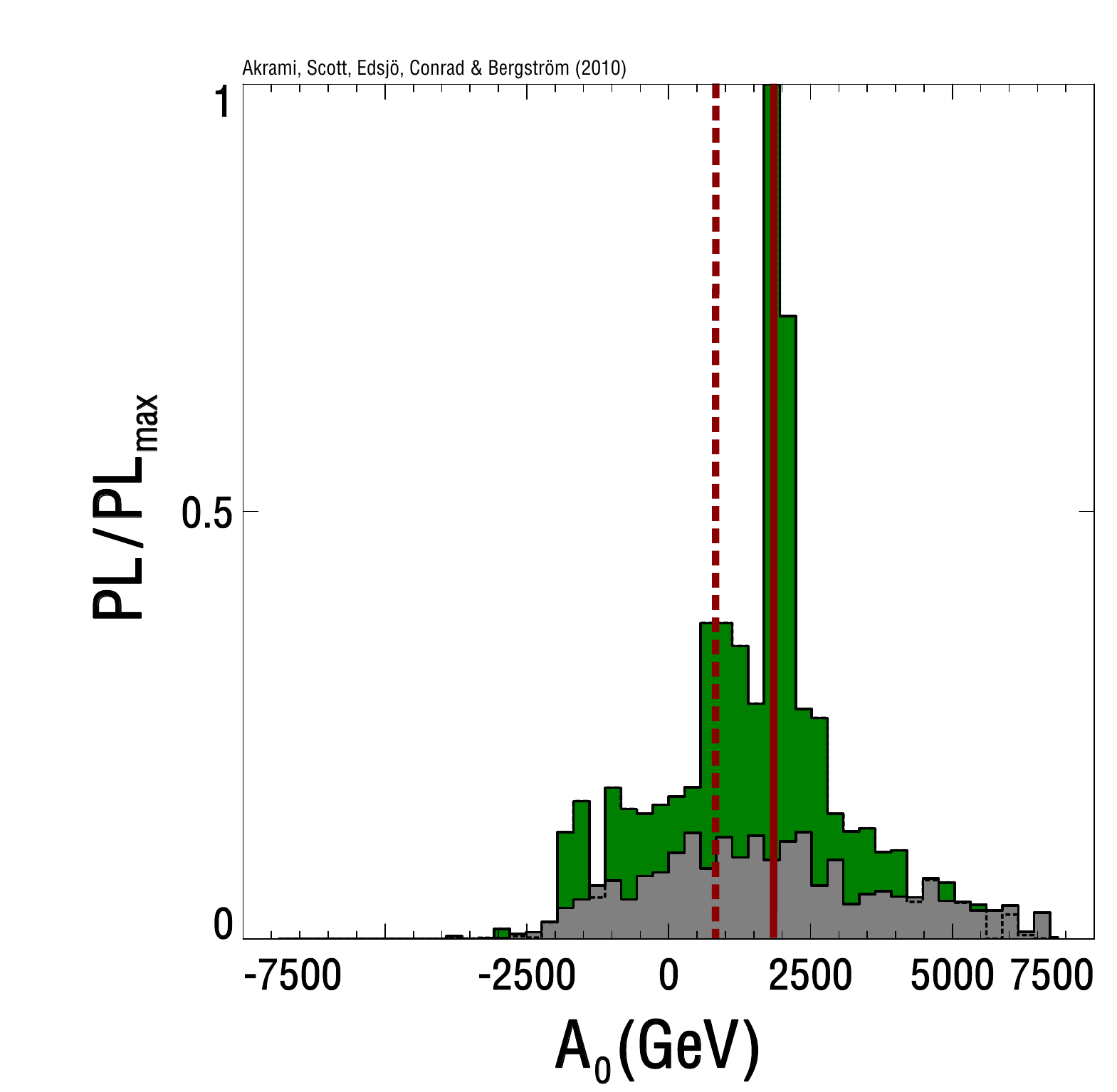}
\includegraphics[width=0.49\linewidth, trim = 0 340 160 10, clip=true]{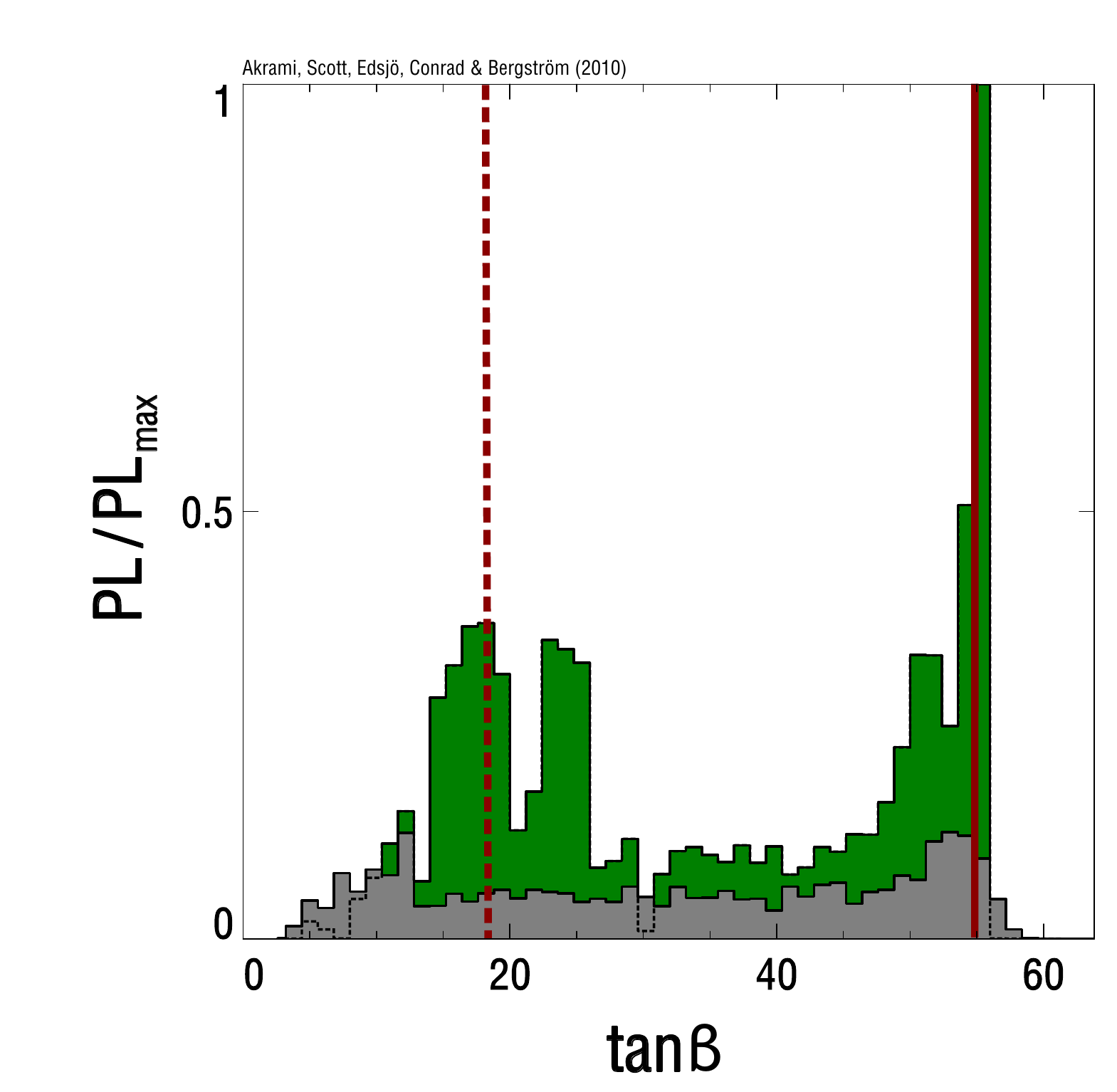}
\caption[aa]{\footnotesize{One-dimensional profile likelihoods ($PL$) of CMSSM parameters, normalised to the global GA best fit ($PL_{max}$).  The green and grey bars show results from GA and \MN~scans, respectively.  Solid and dashed red lines represent the GA global best-fit point ($\chi^2=9.35$, located in the focus point region) and the GA best-fit point in the stau co-annihilation region ($\chi^2=11.34$), respectively.  Samples are divided into $50$ bins.}} \label{fig:params1D}
\end{center}
\end{figure}

As another exhibition of the consequences of our results compared to the Bayesian nested sampling technique, it is interesting to look at the 1D profile likelihoods for the CMSSM parameters (\fig{fig:params1D}).  The horizontal axes in the plots indicate the CMSSM parameters and the vertical axes show the corresponding profile likelihoods, normalised to the best-fit GA value (i.e. with $\chi^2=9.35$).  Green and grey bars show the GA and \MN~1D profile likelihoods respectively.  We sorted points into $50$ bins for these plots.  We have also included the global (FP) and COA best-fit points in the plots, indicated by solid and dashed red lines, respectively.

The very different results in \fig{fig:params1D} from the two different algorithms are yet another confirmation that existing sampling techniques are probably still not sufficiently reliable for the exploration and mapping of SUSY likelihoods, at least in a frequentist framework.  These plots show that by employing a different scanning algorithm, it is quite possible to find many important new points in the parameter space.  This can in principle affect the whole inference about the model, especially when we are interested not only in drawing a general statement about the high-likelihood regions, but also in performing a more detailed exploration of the model likelihood around the best-fit points.  It also shows that the GA technology we made use of in this paper seems a better choice for frequentist analyses than conventional tools, which are typically optimised for Bayesian searches; we have found higher likelihood values for almost all the regions within the interesting range of model parameters.  Whilst this certainly favours this technique over others, it should also serve as a warning.  We can by no means be sure that the GA has actually found the true global best-fit point.  Clearly this concern should be taken much more seriously when one is dealing with more complicated models than the CMSSM, with more parameters and more complex theoretical structures.

Listed in \tabs{tab:BFPval}{tab:BFPlike} are all properties of the two GA best-fit points.  The upper part of the first table gives values of the CMSSM principal parameters and SM nuisance parameters, whereas the lower part shows all physical observables employed in our calculation of model likelihood.  These are the same quantities as given in \tab{tab:obs} except for the Higgs and sparticle masses, which will be presented in the upcoming section on implications for the LHC.  To make the differences between the properties of these two ``good'' points more clear, in the second table we have indicated the individual contributions from different observables to the total $\chi^2$ at each point.  These quantities are also given for \MN~best-fit points found using flat and logarithmic priors.

\TABLE[t]{
\centering
{\footnotesize
\begin{tabular}{|l | l | l|} \hline
\multicolumn{3}{|c|}{model (+nuisance) parameters} \\ \hline\hline
 & GA global BFP (located in FP region) & \multicolumn{1}{c|}{GA COA BFP} \\ \hline
$m_0$           &  $1900.5\gev$    & $133.9\gev$ \\
$m_{1/2}$       &  $342.8\gev$    & $383.1\gev$ \\
$A_0$           &  $1873.9\gev$    & $840.6\gev$ \\
$\tan\beta$     &  $55.0$    & $17.9$ \\
$\mtpole$       &  $172.9\gev$    & $173.3\gev$ \\
$m_b (m_b)^{\overline{MS}}$ & $4.19\gev$  & $4.20\gev$ \\
$\alphas$       &  $0.1172$ & $0.1183$ \\
$1/\alphaemmz$  & $127.955$ & $127.955$ \\ \hline\hline
\multicolumn{3}{|c|}{observables} \\ \hline\hline
 & GA global BFP (located in FP region)  & \multicolumn{1}{c|}{GA COA BFP}
\\ \hline
$m_W$     &  $80.366\gev$   & $80.371\gev$ \\
$\sineff{}$    &  $0.23156$      & $0.23153$ \\
$\deltaamususy \times 10^{10}$ & 5.9 & 14.5 \\
$\brbsgamma \times 10^{4}$ & 3.58 & 2.97 \\
$\delmbs$   &  $17.37\ps^{-1}$  & $19.0\ps^{-1}$ \\
$\brbtaunu \times 10^{4}$ &  $1.32$  & $1.46$ \\
$\abundchi$ &  0.10949 & 0.10985 \\
$\brbsmumu$ & $ 4.34\times 10^{-8}$ & $ 3.87\times 10^{-8}$\\ \hline
\end{tabular}
} \caption[aa]{\footnotesize{Parameter and observable values at the best-fit points (BFPs) found using Genetic Algorithms.  These quantities are shown for both the global best-fit point (located in the focus point (FP) region) and the best-fit point in the stau co-annihilation (COA) region. Higgs and sparticle masses will be given in \tab{tab:spectra}, when talking about implications for the LHC.}} \label{tab:BFPval}
}

One interesting fact seen in \tab{tab:BFPlike} is the apparent tension between $\deltaamususy$ and the other observables, in particular $\brbsgamma$.  This has been widely discussed in the past~\cite{Buchmueller:08084128,Buchmueller:09075568,Trotta:08093792,Feroz:09032487}.  While most of the discrepancy between the model and the experimental data at our global best-fit point (living in the FP region) comes from $\deltaamususy$ ($\sim 76\%$), and $\brbsgamma$ contributes only about $0.1\%$ to the total $\chi^2$, these two observables partially switch roles at the COA point.  This confirms that $\brbsgamma$ in general favours large $\mzero$ (the FP region), while $\deltaamususy$ favours smaller masses (the COA region).  A similar feature is also visible in the two \MN~best-fit points for flat and log priors, as they reside in the FP and COA regions, respectively.

\TABLE[t]{
\centering
{\footnotesize
\begin{tabular}{|l | l | l | l | l|} \hline
 & \multicolumn{4}{|c|}{partial $\chi^2$ (fractional contribution to the total $\chi^2$ in $\%$)} \\ \cline{2-5}
observable & GA global BFP & GA COA BFP & MN global BFP  & MN global BFP\\
 &  located in FP region &  & with flat priors  & with log priors \\ \hline
nuisance parameters & $0.12~(1.27\%)$ & $0.35~(3.10\%)$ & $0.48~(3.56\%)$ & $0.81~(6.78\%)$\\
$m_W$     &  $1.21~(12.95\%)$   & $0.83~(7.29\%)$ & $1.48~(10.92\%)$ & $0.69~(5.83\%)$\\
$\sineff{}$    &  $0.024~(0.26\%)$      & $\sim10^{-4}~(0.001\%)$ & $0.07~(0.49\%)$ & $0.0040~(0.034\%)$\\
$\deltaamususy$ & $7.09~(75.79\%)$ & $2.86~(25.21\%)$ & $9.21~(68.20\%)$ & $2.40~(20.18\%)$\\
$\brbsgamma$ & $0.010~(0.11\%)$ & $3.03~(26.76\%)$ & $0.10~(0.74\%)$ & $3.83~(32.20\%)$\\
$\delmbs$     & $0.028~(0.30\%)$ & $0.26~(2.31\%)$ & $0.09~(0.66\%)$ & $0.29~(2.41\%)$\\
$\brbtaunu$ &  $\sim10^{-5}~(10^{-4}\%)$  & $0.050~(0.44\%)$ & $1.91~(14.14\%)$ & $0.043~(0.36\%)$\\
$\abundchi$ &  $0.0011~(0.012\%)$ & $\sim10^{-5}~(10^{-4}\%)$ & $0.03~(0.2\%)$ & $0.13~(1.07\%)$\\
$\brbsmumu$ &  $0.016~(0.17\%)$ & $ 0.00~(0.00\%)$ & $0.00~(0.00\%)$ & $0.00~(0.00\%)$\\
$m_h$ & $0.85~(9.14\%)$ & $3.96~(34.88\%)$ & $0.15~(1.09\%)$ & $3.70~(31.13\%)$\\
sparticles & $ 0.00~(0.00\%)$ & $ 0.00~(0.00\%)$ & $0.00~(0.00\%)$ & $0.00~(0.00\%)$\\ \hline
\textbf{all} & $\textbf{9.35~(100~\%)}$ & $\textbf{11.34~(100~\%)}$ & $\textbf{13.51~(100~\%)}$ & $\textbf{11.90~(100~\%)}$\\ \hline
\end{tabular}
} \caption[aa]{\footnotesize{Contributions to the total $\chi^2$ by different observables employed in the scans (\tab{tab:obs}).  Contributions are shown for the GA global best-fit point (BFP) located in the focus point (FP) region and the GA best-fit point in the stau co-annihilation (COA) region.  Fractional contributions are also given in percent.  Similar quantities for both \MN~scans with flat (linear) and logarithmic priors are also listed for comparison, where the former is in the FP region and the latter is in the COA region.}} \label{tab:BFPlike}
}

In the case where our best-fit point is placed in the FP region, the total $\chi^2$ from all observables except $\deltaamususy$ and $\brbsgamma$ is a remarkably smaller fraction ($\sim 24\%$) of the total than in the COA region ($\sim 48\%$) (this is also the case if we compare the two \MN~points in the table, with contributions of $\sim 31\%$ and $\sim 47.5\%$).  This can be qualitatively interpreted as yet another reflection of the fact that in the absence of any constraint from $\deltaamususy$, the data is largely consistent with the global best-fit point being in the FP region.  That is, if one ignores $\deltaamususy$, it is much easier to fulfil all the experimental constraints on the CMSSM by moving towards larger $\mzero$.  If one wants to satisfy also the extra constraint coming from $\deltaamususy$, this might be possible by moving back towards lower masses (the COA region), but at the price of reducing the total likelihood.

It is important to stress that our global best-fit point is in fact part of the FP region, with high $\mzero$ (i.e. $\sim 1900\gev$).  This means that even taking into account the constraint from $\deltaamususy$, the FP is still favoured over the COA region in our analysis, in clear contradiction with some recent claims~\cite{Buchmueller:08084128,Buchmueller:09075568} that the latter is favoured by existing data.  These analyses were performed in a frequentist framework, and based on MCMC scans.  However, the large discrepancy with our findings probably comes more from differences in the likelihood functions themselves than the scanning algorithms, i.e.~in the calculations of physical observables and their contributions to the likelihood.  A direct comparison with these works would only be possible if we were to also work with exactly the same routines for the calculation of the likelihood as in Refs.~\citealp{Buchmueller:08084128} and \citealp{Buchmueller:09075568}, changing only the scanning algorithm (as we have here in comparing with Ref.~\citealp{Trotta:08093792}).  The difference we see in this case mostly reflects the discrepancy between the results of Ref.~\citealp{Trotta:08093792} and Refs.~\citealp{Buchmueller:08084128} and \citealp{Buchmueller:09075568}.

Nontheless, it is important to note that some differences could be due to the scanning technique.  We have shown in this paper that at least for the specific physics setup implemented in \textsf{SuperBayeS}, GAs find better-fit points than nested sampling, which in turn is known to find essentially the same points as MCMCs.  It is therefore quite reasonable to expect that GAs could find many points missed by MCMCs.  There are even some other FP points found by GAs with masses of about $2800\gev$ and located in the $1\sigma$ region (see \fig{fig:m0mhf}a), supporting the conclusion that although low masses are favoured over high masses in the previous \MN~and MCMC scans using \textsf{SuperBayeS}, the opposite holds in our GA scans.  This means that there exist many high-likelihood points in the FP region entirely missed by \MN~and MCMC scans performed in the \textsf{SuperBayeS} analyses.  It seems that those algorithms do not sample this region of the parameter space very well, at least when \textsf{SuperBayeS} routines are used for physics and likelihood calculations.  We see no reason why a similar situation could not also occur when different codes are used to evaluate the likelihood function.

Since we have not used exactly the same physics and likelihood setup, nor the same numerical routines for calculating different quantities as employed in Refs.~\citealp{Buchmueller:08084128} and~\citealp{Buchmueller:09075568} (and we cannot do that in a consistent way as the code employed in those studies is not publicly available), we cannot make a definitive statement as to the overall impact of the scanning algorithm in the discrepancy we see with their results.  One should however be very cautious in general when attempting to draw strong conclusions about e.g.~the FP being excluded by existing data.  The complex structure of the CMSSM parameter space makes the corresponding likelihood surface very sensitive to small changes in the codes and experimental data used to construct the full likelihood, which in turn can introduce a significant dependence upon the scanning algorithm.

\subsection{Implications for the LHC} \label{sec:LHC}

\begin{figure}
\begin{center}
\includegraphics[width=0.49\linewidth, trim = 0 330 160 10, clip=true]{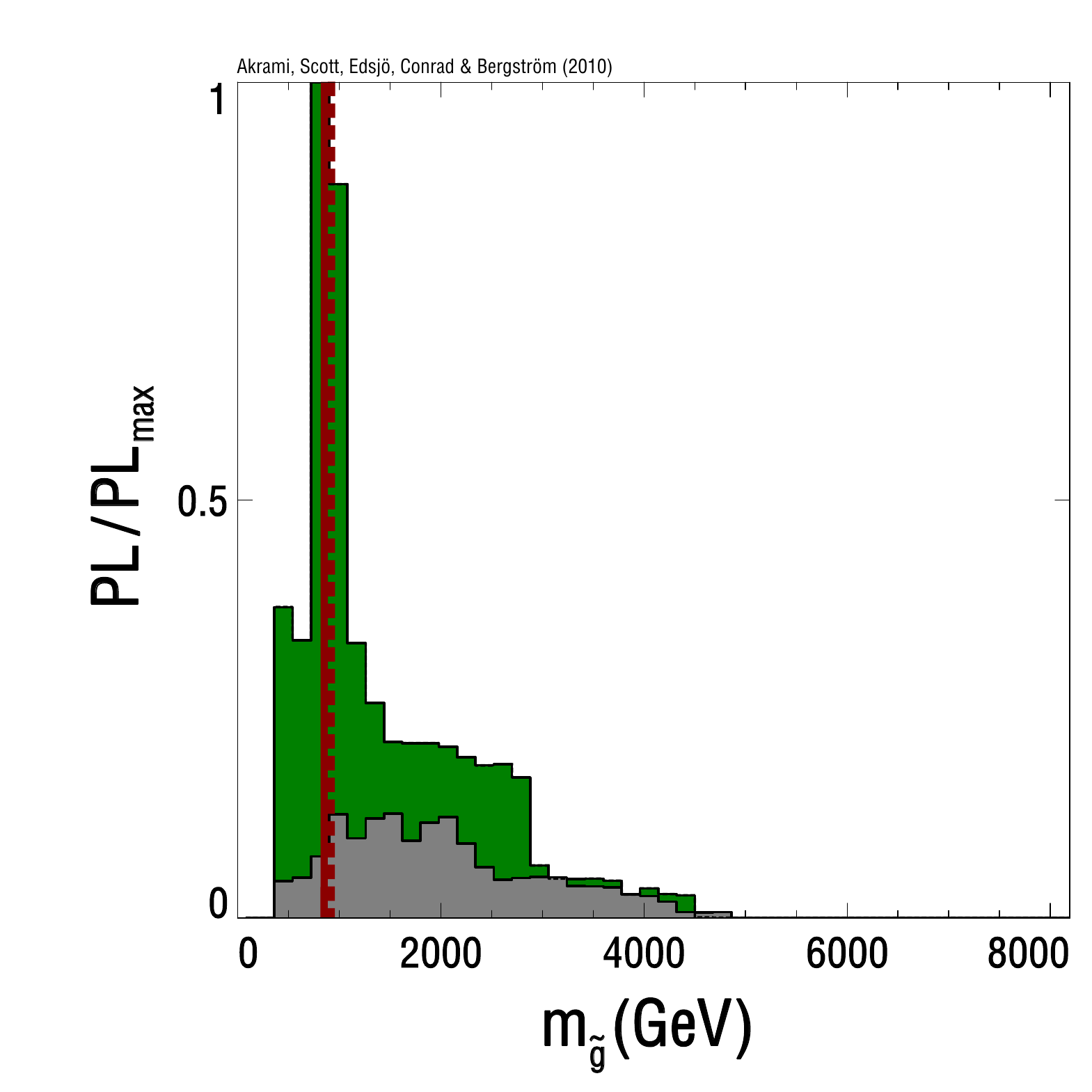}
\includegraphics[width=0.49\linewidth, trim = 0 330 160 10, clip=true]{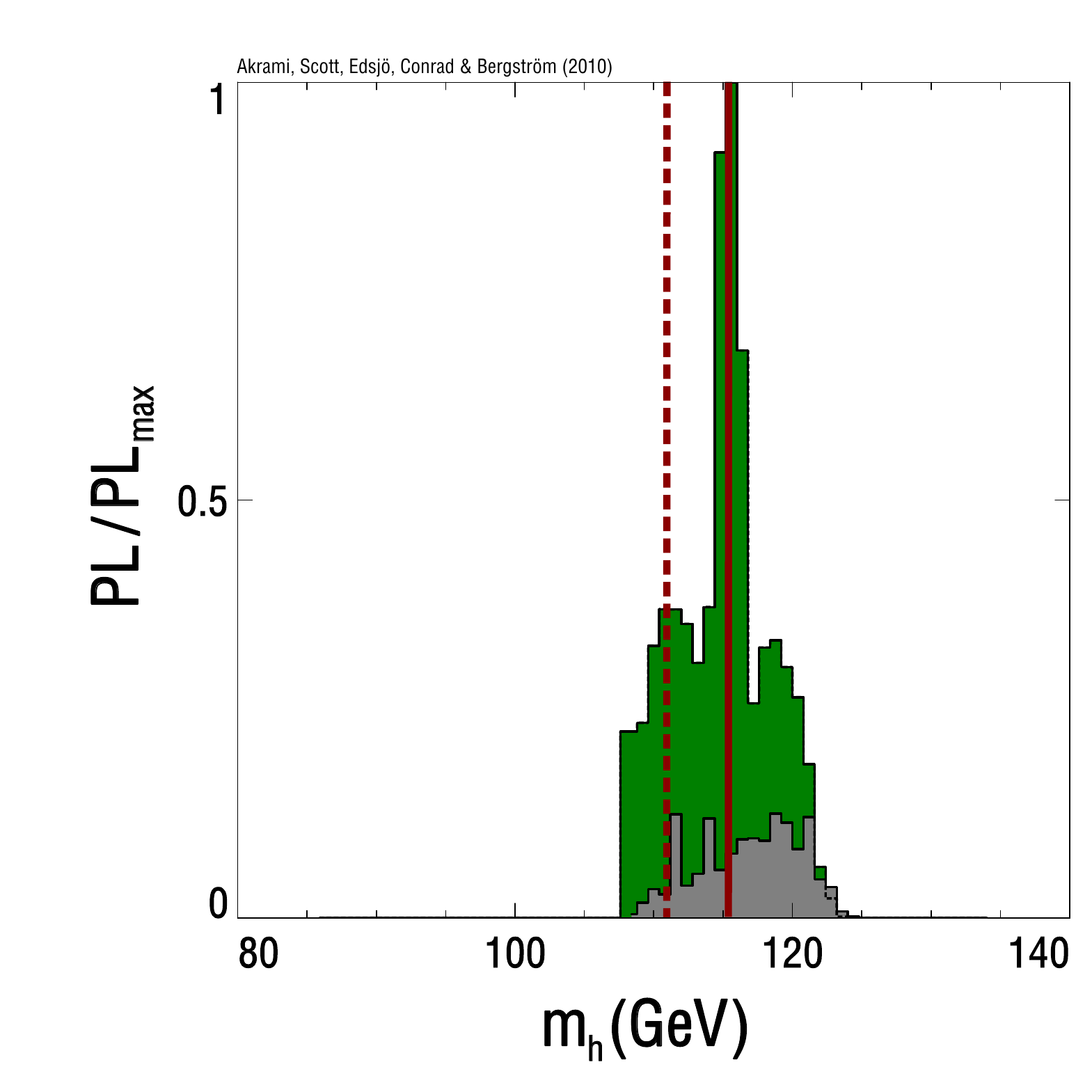}
\caption[aa]{\footnotesize{As in \fig{fig:params1D}, but for the gluino mass $m_{\tilde g}$ and the
lightest Higgs mass $m_h$.}}
\label{fig:LHC1D}
\end{center}
\end{figure}

We showed in the previous section that compared to the state-of-the-art Bayesian algorithm \MN, GAs are a very powerful tool for finding high-likelihood points in the CMSSM parameter space.  It is therefore interesting to examine how strongly these results impact predictions for future experimental tests at e.g.~the LHC.  We have calculated the 1D profile likelihoods corresponding to the gluino mass $m_{\tilde g}$ (as a popular representative of the sparticles) and the lightest Higgs mass $m_h$, both of which will be searched for at the LHC.  The resultant plots are given in \fig{fig:LHC1D}.  These plots are generated in the same way as those in \fig{fig:params1D}, indicating the differences between the two scanning strategies.  Here, we once again see that the GA has found much better fits in the mass ranges covered by \MN.

Looking first at the gluino mass prediction (left-hand plot of \fig{fig:LHC1D}), we confirm earlier findings~\cite{Trotta:08093792,Buchmueller:09075568} that the LHC will probe all likely CMSSM gluino masses if its reach extends beyond $\sim3\tev$.  The FP and COA best-fit points are both located at relatively low masses with quite similar values ($\sim 900\gev$).  These values are well within the reach of even the early operations of the LHC.  The detailed CMSSM mass spectra computed at both of these points are presented in \tab{tab:spectra}.  It can be seen from these spectra that our global best-fit point favours rather high masses for the sfermions, while the COA best-fit point favours low masses.

\TABLE[t]{
\centering
{\footnotesize
\begin{tabular}{|c|c|c||c|c|c|}\hline
& GA global BFP & GA COA BFP & & GA global BFP & GA COA BFP \\
(GeV) & located in FP region & & (GeV) & located in FP region & \\ \hline
$m_{\tilde{e}_L}$ & 1908& 294.1 & $m_{\tilde{d}_R}$ &  1994& 798.1 \\ \hline
$m_{\tilde{e}_R}$ & 1903& 202 & $m_{\tilde{s}_L}$ &  2000& 832.4\\ \hline
$m_{\tilde{\mu}_L}$ & 1907& 294.1 & $m_{\tilde{s}_R}$ &  1994& 798.1\\ \hline
$m_{\tilde{\mu}_R}$ & 1901& 201.9 & $m_{\tilde{b}_1}$ &  1354& 765\\ \hline
$m_{\tilde{\tau}_1}$ & 1100& 160.1 & $m_{\tilde{b}_2}$ &  1492& 793.4\\ \hline
$m_{\tilde{\tau}_2}$ & 1560& 289.2 & $m_{\tilde\chi_1^0}$ &  140.4& 152.6\\ \hline
$m_{\tilde{\nu}_{e}}$ & 1906 & 283.3 & $m_{\tilde\chi_2^0}$ &  269.9& 285.4\\ \hline
$m_{\tilde{\nu}_{\mu}}$ & 1905 & 283.3 & $m_{\tilde\chi_3^0}$ &  519.7& 451.1\\ \hline
$m_{\tilde{\nu}_{\tau}}$ & 1560& 272.5 & $m_{\tilde\chi_4^0}$ &  529.7& 469.6\\ \hline
$m_{\tilde{u}_L}$ & 1998& 826.1 & $m_{\tilde\chi_1^{\pm}}$ &  270.4& 286.9\\ \hline
$m_{\tilde{u}_R}$ & 1996& 805.4 & $m_{\tilde\chi_2^{\pm}}$ &  530.3& 468.5\\ \hline
$m_{\tilde{c}_L}$ & 1998& 826.1 & $m_{h}$ &  115.55& 111.11\\ \hline
$m_{\tilde{c}_R}$ & 1996& 805.4 & $m_{H}$ &  179.93& 504.24\\ \hline
$m_{\tilde{t}_1}$ & 1194& 672.8 & $m_{A}$ &  179.83& 504.04\\ \hline
$m_{\tilde{t}_2}$ & 1364& 803 & $m_{H^{\pm}}$  &  201.14& 510.67\\ \hline
$m_{\tilde{d}_L}$ &  2001& 832.4 & $m_{\tilde{g}}$ &  877.1& 898.8\\ \hline
\end{tabular}
} \caption[aa]{\footnotesize{Mass spectra of the GA global best-fit point (BFP) located in the focus point (FP) region and the GA best-fit point in the stau co-annihilation (COA) region.}} \label{tab:spectra}
}

Looking at the right-hand plot of \fig{fig:LHC1D}, corresponding to the likelihood of different values of $m_h$, we notice that although a large number of good points have Higgs masses higher than the SM limit from the Large Electron-Positron Collider (LEP; i.e. $m_h\geq114.4\gev$), including the global best-fit point (with $m_h=115.55\gev$), there are also many other important ones which violate this limit, including the best-fit COA point (with $m_h=111.11\gev$).  These points with low-mass Higgs bosons have been allowed by the smoothed likelihood function that we employed for the LEP limit (cf.~\sec{sec:CMSSMlike}).  Instead of this treatment of the Higgs sector, one could use a more sophisticated method, such as implemented by \textsf{HiggsBounds}~\cite{Bechtle:08114169}.  This would apply the collider bounds on the Higgs mass in a SUSY-appropriate manner, and give more accurate likelihoods at low masses around the $114.4\gev$ bound.

Looking again at \tab{tab:BFPlike}, the contribution from $m_h$ as a percentage of the total $\chi^2$ is considerably larger in the COA case than the FP.  This becomes clear when we compare their corresponding values for $m_h$ (i.e. $115.55\gev$ and $111.11\gev$, respectively) with the LEP limit.  The lower Higgs mass in the COA region is a reflection of the correlation between $m_0$ and $m_h$ in the CMSSM, confirming once more that moving to low $\mzero$ (i.e. approaching the COA region) causes models to become less compatible with all experimental data except $\deltaamususy$.

\subsection{Implications for dark matter searches} \label{sec:DM}

\begin{figure}
\begin{center}
\subfigure[][]{\label{sigmaSImChi:a}\includegraphics[width=0.49\linewidth, trim = 70 0 70 50, clip=true]{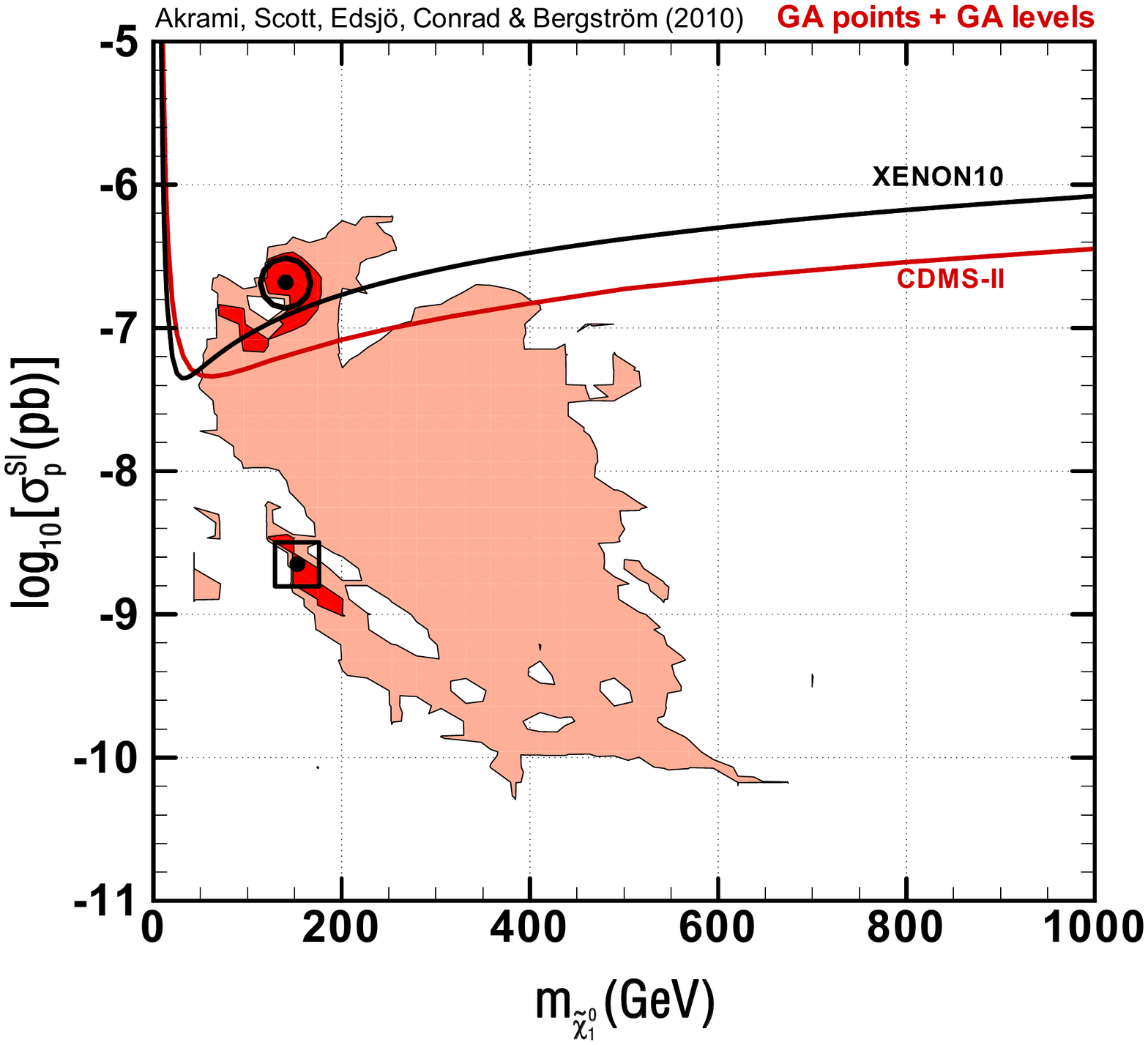}}
\subfigure[][]{\label{sigmaSImChi:b}\includegraphics[width=0.49\linewidth, trim = 70 0 70 50, clip=true]{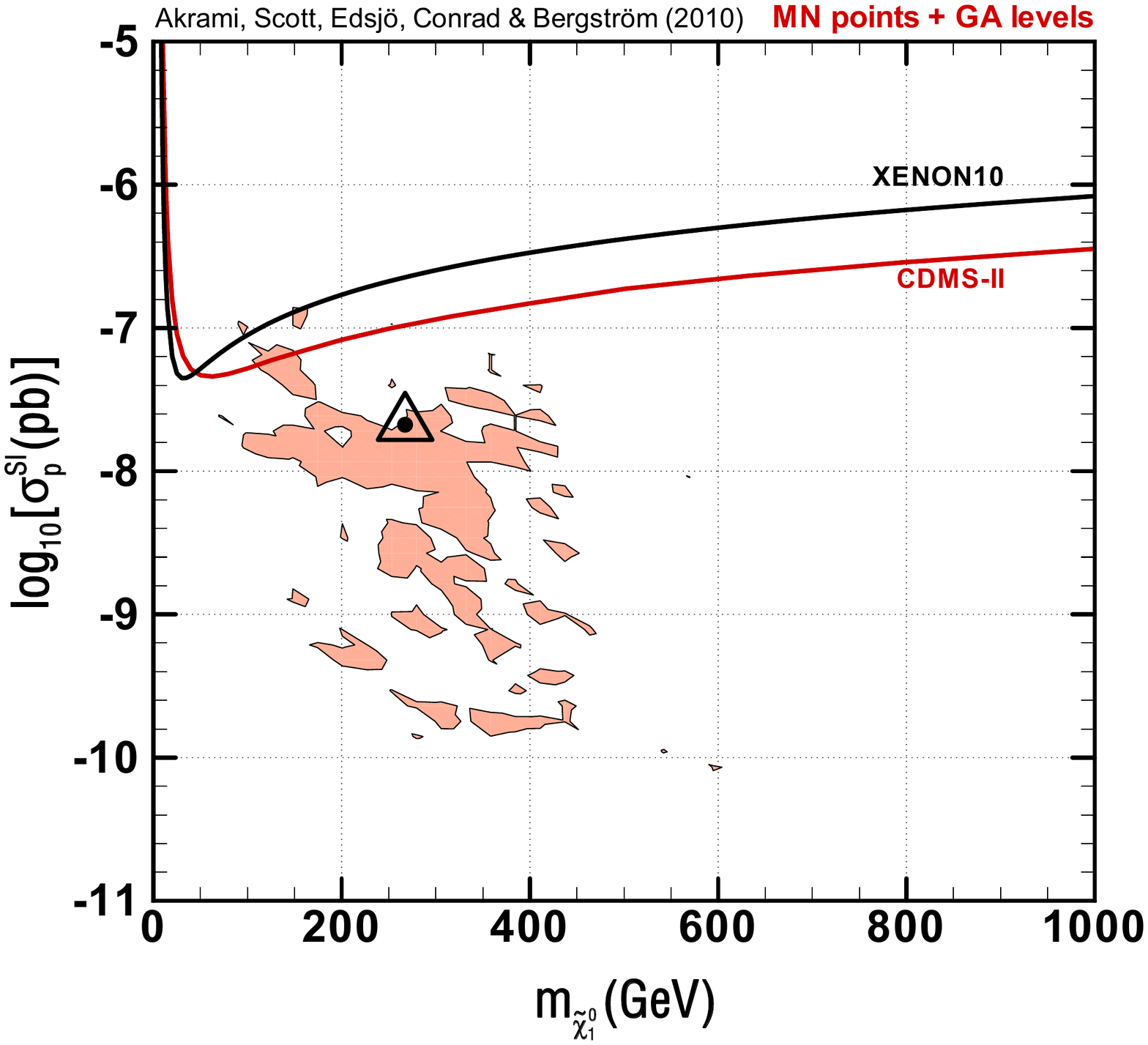}} \\
\subfigure[][]{\label{sigmaSImChi:c}\includegraphics[width=0.49\linewidth, trim = 70 0 70 50, clip=true]{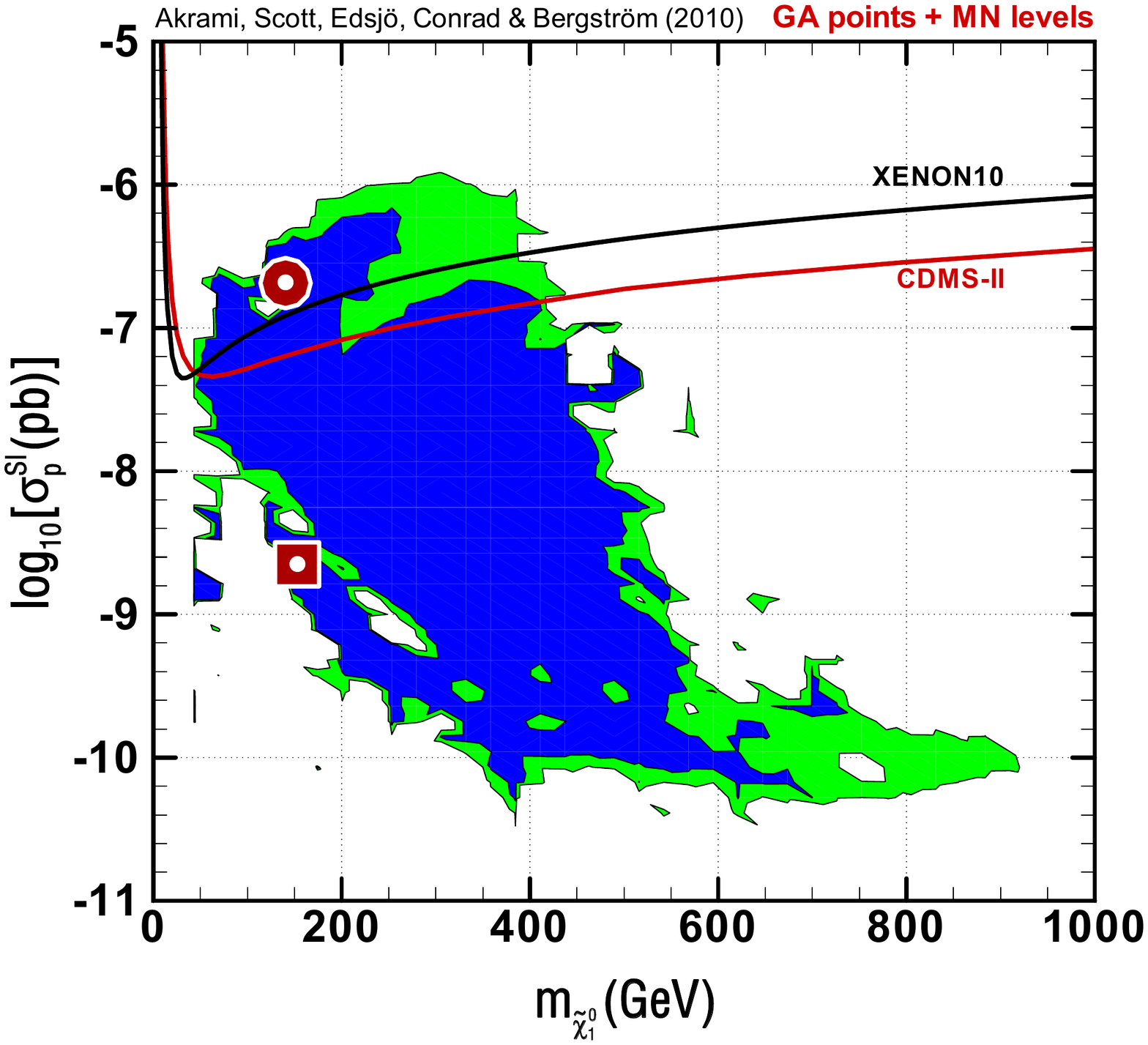}}
\subfigure[][]{\label{sigmaSImChi:d}\includegraphics[width=0.49\linewidth, trim = 70 0 70 50, clip=true]{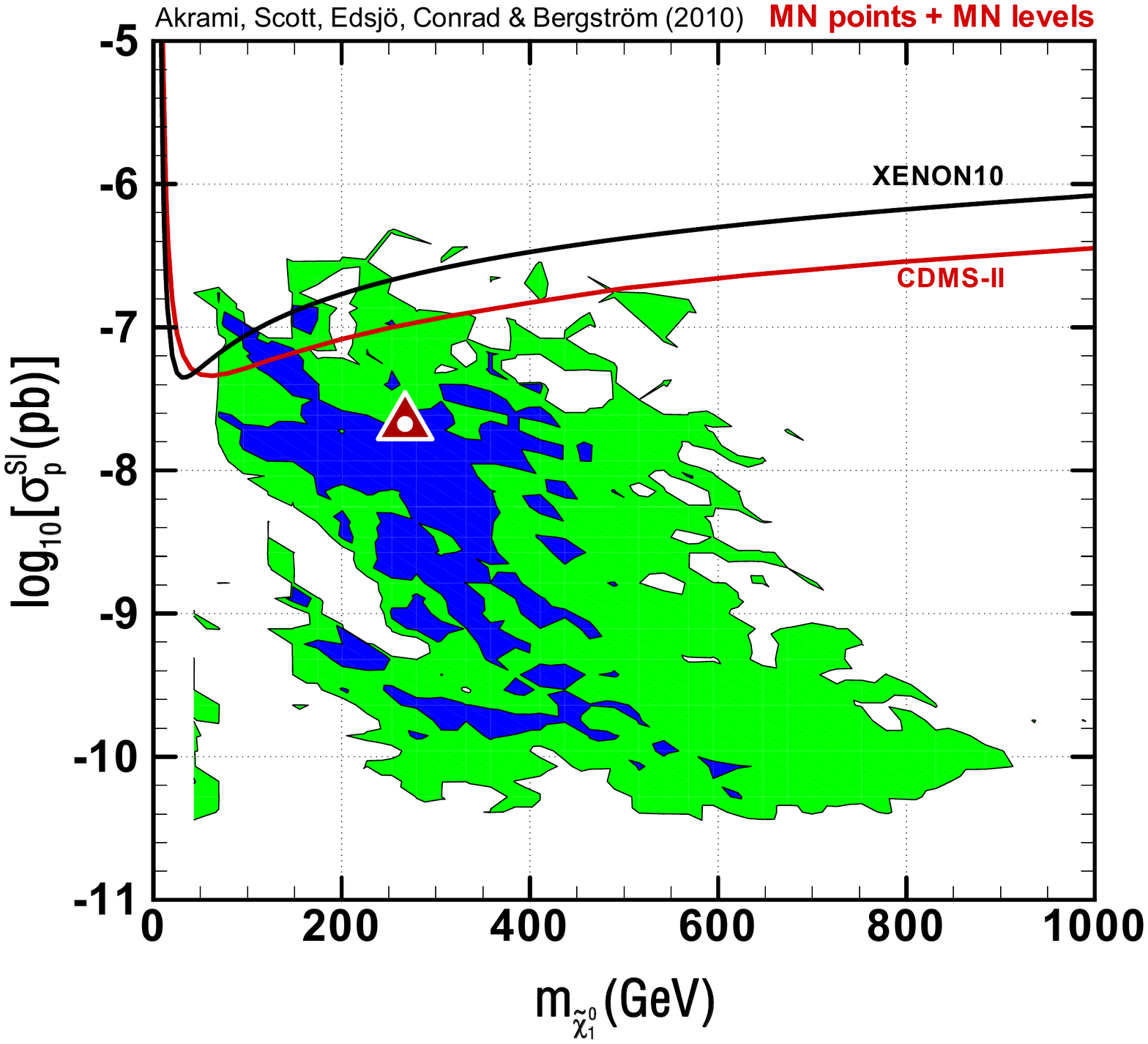}}
\caption[aa]{\footnotesize{As in \fig{fig:m0mhf} and \fig{fig:A0tanbeta}, but representing the best-fit points and high-likelihood regions for the spin-independent scattering cross-section of the neutralino and a proton $\sigma^{SI}_p$ versus the neutralino mass $m_{\tilde\chi^0_1}$.  Panels (a) and (d) show the statistically-consistent results of the GA and \MN~scans, respectively.  Panels (b) and (c) are given for comparative purposes only.  The latest experimental limits from CDMS-II~\cite{cdmsII} and XENON10~\cite{xenon10} are also shown, plotted as red and black curves respectively.  These curves are exclusion limits at the $90\%$ confidence level under the assumption of a standard local halo configuration.}} \label{fig:sigmaSImChi}
\end{center}
\end{figure}

As a natural continuation of our discussion of the consequences of our results for present and upcoming experiments, we turn now to dark matter, beginning with direct detection (DD) experiments.  One interesting quantity for these experiments is the spin-independent scattering cross-section $\sigma^{SI}_p$ of the neutralino and a proton.  This cross-section is often plotted against the neutralino mass $m_{\tilde\chi^0_1}$ when comparing limits from different direct detection experiments.  Predictions are given in this plane from both the \MN~and GA scans in \fig{fig:sigmaSImChi}, drawn similarly to \figs{fig:m0mhf}{fig:A0tanbeta}.  $\sigma^{SI}_p$ is shown in units of \pb~(i.e. $10^{-36} \mbox{cm}^2$).  Contours shown in the upper (lower) panels are generated according to the GA (\MN) best-fit point, and the points with highest likelihoods are marked as before.  Although no constraints from direct detection measurements have been used in forming the model likelihood in this paper (mainly in order to work with the same set of quantities and constraints as employed in Ref.~\citealp{Trotta:08093792}), we have also included the current best DD limits for comparison.  These are limits at the $90\%$ confidence level from CDMS-II~\cite{cdmsII} and XENON10~\cite{xenon10}.

Looking at \fig{fig:sigmaSImChi}, we first notice that all the conclusions we made earlier are reconfirmed here: there are many important points in the parameter space that have appeared by the use of GAs, having a strong impact on the statistical conclusions.  For example, instead of a rather spread and sparse $1\sigma$ confidence region produced by \MN~(\fig{fig:sigmaSImChi}d), GAs (\fig{fig:sigmaSImChi}a) reveal a more compact region, sharply peaked around the best-fit points.  It is interesting to see that in the latter case, most of the $1\sigma$ FP region around the dotted circle, including the point itself, is already excluded by CDMS-II and XENON10 under standard halo assumptions.  The global best-fit point has quite a large cross-section ($\sigma^{SI}_p \sim 2\times 10^{-7}\pb$) compared to the \MN~global best-fit point ($\sigma^{SI}_p \sim 1.7\times 10^{-8}\pb$), making it much more easily probed by direct detection (\tab{tab:DM}).  On the contrary, the best-fit COA point has a much lower $\sigma^{SI}_p$ ($\sim 2.2\times 10^{-9}\pb$), and is still well below these experimental limits.  With future experiments planned to reach cross-sections as low as $10^{-10}\pb$, this point will eventually be tested as well.  Even if we do not consider the highest-likelihood point found by the GA, and just compare the two lower panels in \fig{fig:sigmaSImChi}, \MN~has obviously only explored a small fraction of its $1\sigma$ high-likelihood region in the parameter space.  It has also missed most of its $1\sigma$ and $2\sigma$ points in the region $\sigma^{SI}_p > 10^{-7}\pb$.  This is a particularly interesting area, being within the reach of current dark matter DD experiments.

In \tab{tab:DM}, we also give the calculated values for the spin-dependent scattering cross-sections of the neutralino with a proton ($\sigma^{SD}_p$) and a neutron ($\sigma^{SD}_n$), for both the FP and COA best-fit points.

\TABLE[t]{
\centering
{\footnotesize
\begin{tabular}{|l | l | l|} \hline
\multicolumn{3}{|c|}{direct detection} \\ \hline\hline
 & GA global BFP  & \multicolumn{1}{c|}{GA BFP in COA region} \\ \hline \hline
$\sigma^{SI}_p$ &  $2.057\times 10^{-7}\pb$    & $2.236\times 10^{-9}\pb$ \\ \hline
$\sigma^{SD}_p$ &  $2.435\times 10^{-6}\pb$    & $4.231\times 10^{-6}\pb$ \\ \hline
$\sigma^{SD}_n$ &  $1.644\times 10^{-6}\pb$    & $3.142\times 10^{-6}\pb$ \\ \hline \hline
\multicolumn{3}{|c|}{indirect detection} \\ \hline\hline
 & GA global BFP & \multicolumn{1}{c|}{GA BFP in COA region} \\ \hline
$\left\langle \sigma v\right\rangle$ & $2.260\times 10^{-26} \mbox{cm}^3 \mbox{s}^{-1}$ & $5.385\times 10^{-28} \mbox{cm}^3 \mbox{s}^{-1}$ \\ \hline
\end{tabular}
} \caption[aa]{\footnotesize{Dark matter direct and indirect detection observables.  These are the spin-independent and spin-dependent scattering cross-sections of the neutralino with nucleons in \pb~($10^{-36} \mbox{cm}^2$), and the velocity-averaged neutralino self-annihilation cross-section.  These are calculated at the global best-fit point (BFP) located in the focus point (FP) region, and at the best-fit point in the stau co-annihilation (COA) region.}} \label{tab:DM}
}

\begin{figure}
\begin{center}
\subfigure[][]{\label{sigmavmChi:a}\includegraphics[width=0.49\linewidth, trim = 70 0 70 50, clip=true]{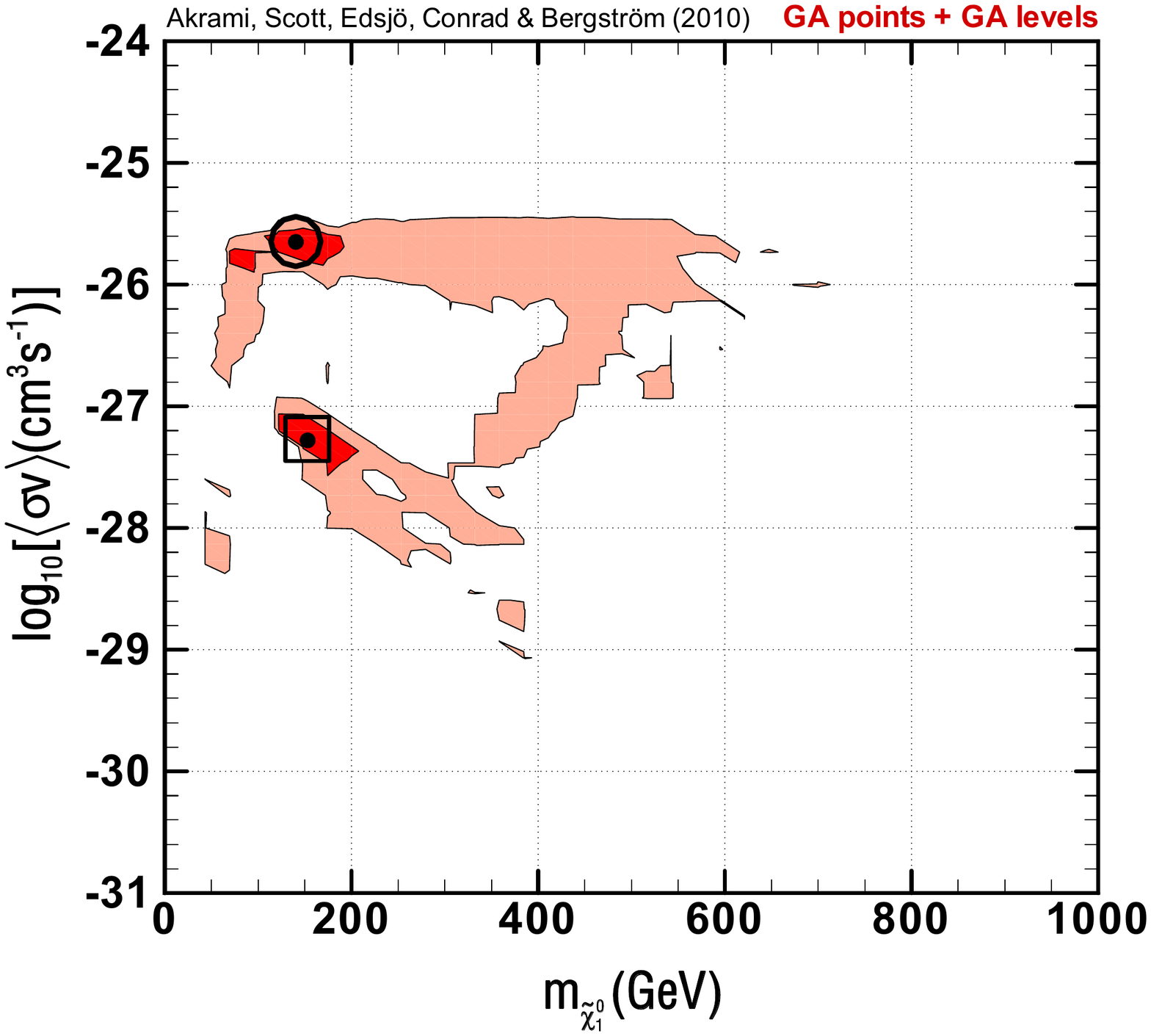}}
\subfigure[][]{\label{sigmavmChi:b}\includegraphics[width=0.49\linewidth, trim = 70 0 70 50, clip=true]{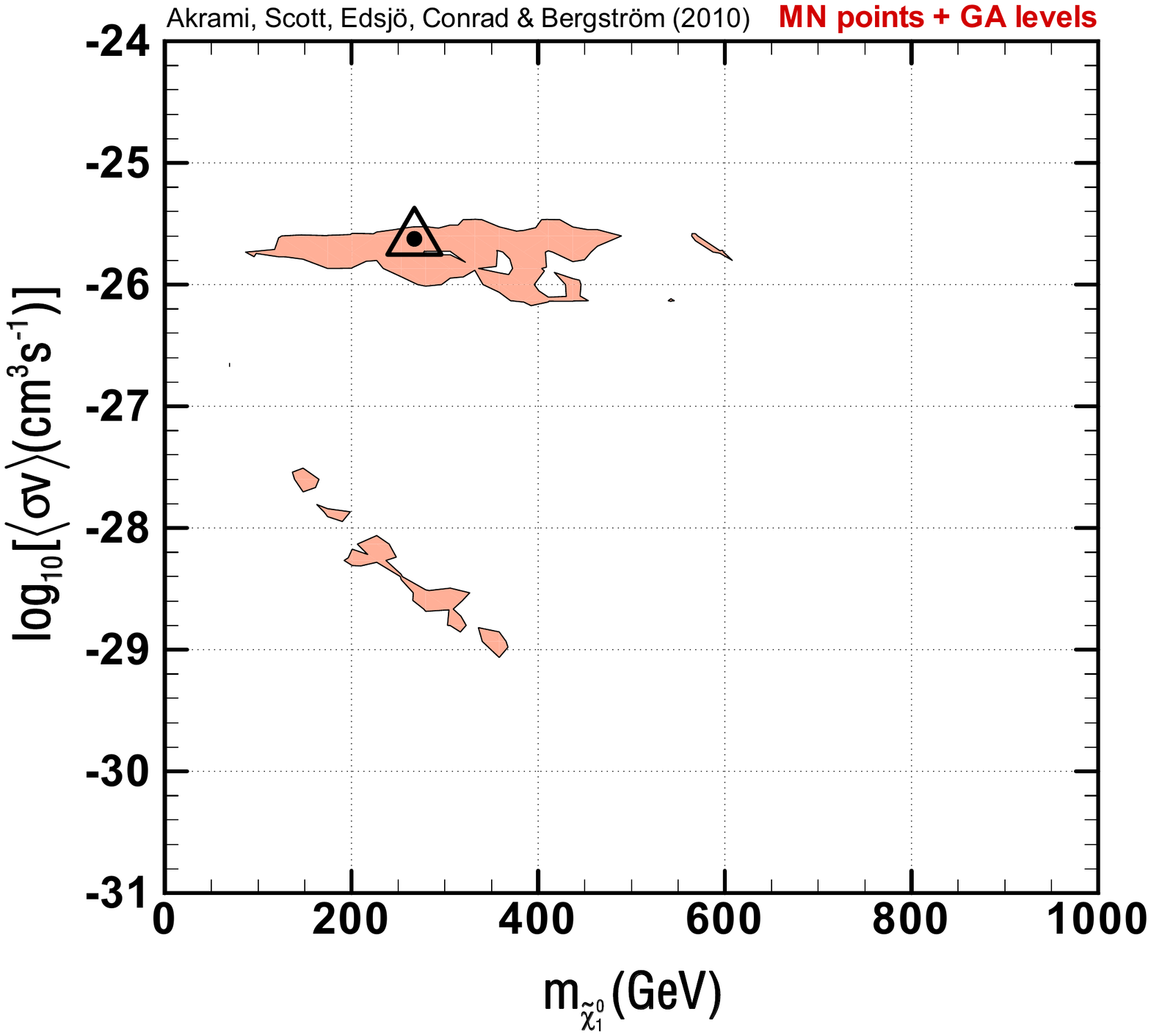}} \\
\subfigure[][]{\label{sigmavmChi:c}\includegraphics[width=0.49\linewidth, trim = 70 0 70 50, clip=true]{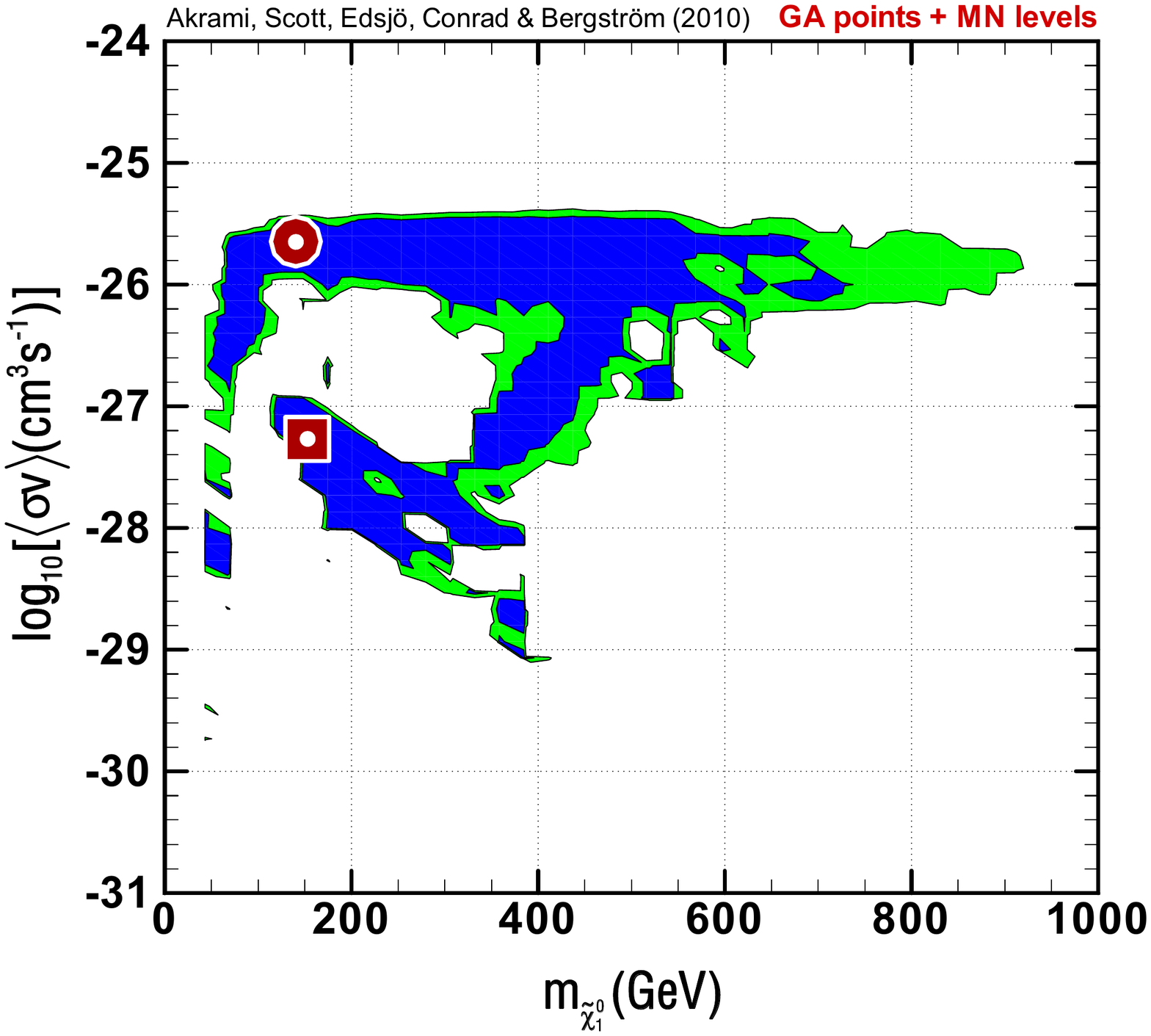}}
\subfigure[][]{\label{sigmavmChi:d}\includegraphics[width=0.49\linewidth, trim = 70 0 70 50, clip=true]{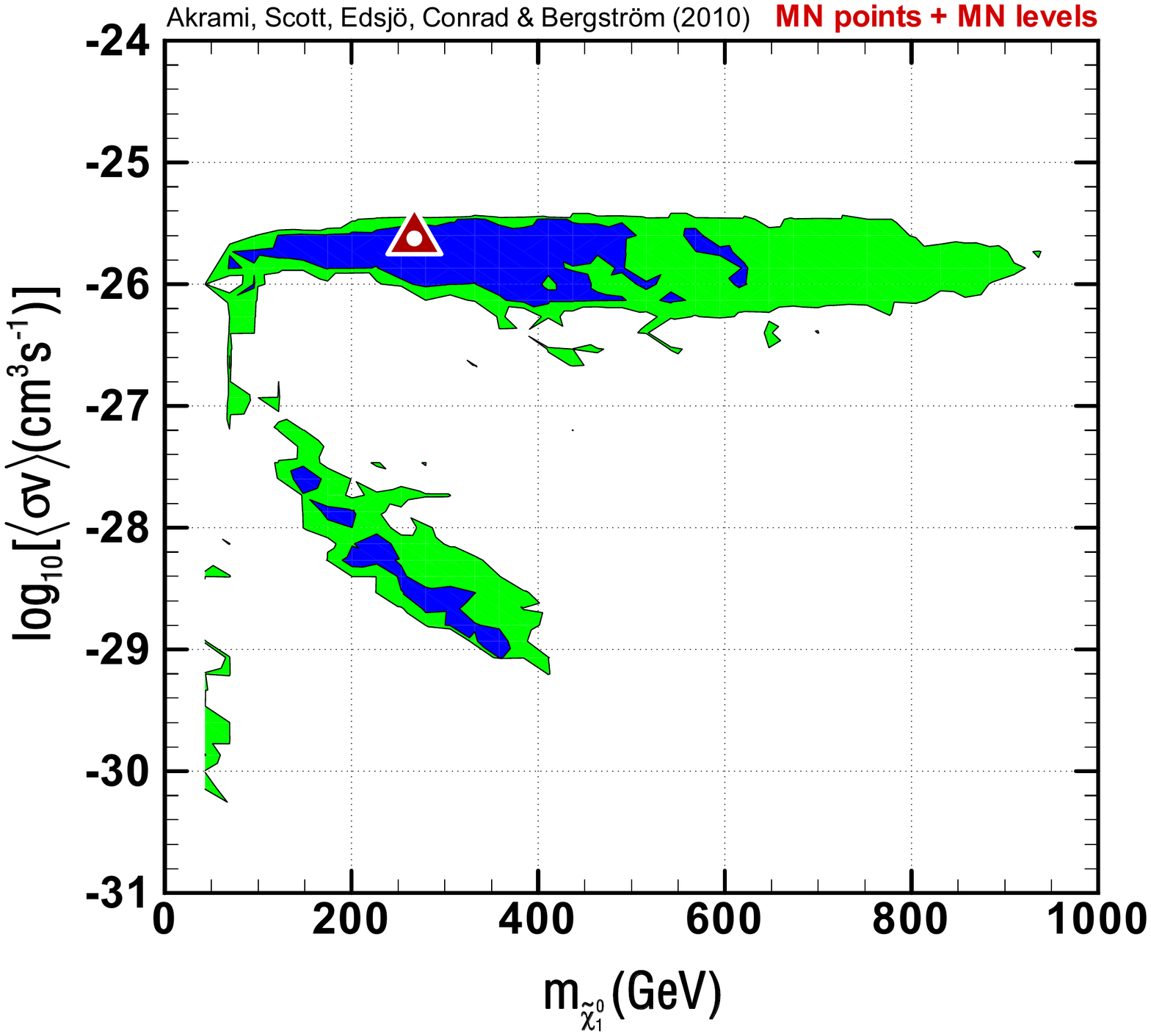}}
\caption[aa]{\footnotesize{As in \fig{fig:m0mhf}, \fig{fig:A0tanbeta} and \fig{fig:sigmaSImChi} but representing the best-fit points and high-likelihood regions for the velocity-averaged neutralino self-annihilation cross-section $\left\langle \sigma v\right\rangle$ versus the neutralino mass $m_{\tilde\chi^0_1}$.  Again, panels (a) and (d) show the statistically-consistent GA and \MN~results, respectively, and panels (b) and (c) are given for comparative purposes only.}} \label{fig:sigmavmChi}
\end{center}
\end{figure}

Considering implications for indirect detection (ID) of dark matter, one is often interested in a plot very similar to the one presented for the DD case, but now for the velocity-averaged neutralino self-annihilation cross-section $\left\langle \sigma v\right\rangle$ (instead of the scattering cross-section in the previous case), again versus the neutralino mass $m_{\tilde\chi^0_1}$.  Such plots are shown in \fig{fig:sigmavmChi} for all different cases in the same style as in \fig{fig:sigmaSImChi}.  Their general characteristics resemble very much those we enumerated for the DD case.

We first notice the strong correlation between the DD and ID plots, such as the generic similarities between the corresponding high-likelihood regions and the best-fit points.  One interesting feature visible in these plots is the existence of a new, considerably large, high-likelihood region spanned by the approximate ranges of $350\gev < m_{\tilde\chi^0_1} < 500\gev$ and $-27.5<\log_{10}(\left\langle \sigma v\right\rangle)<-26.5$ and almost completely missed by \MN.  To our knowledge, this region has not been introduced so far by any other Bayesian or frequentist analysis.  Further investigations show that these points are located in the stau co-annihilation region, but with very high $\mzero$ (up to $\sim 1750\gev$).  The very existence of such a high mass COA region compatible with all data indicates again that one should be very careful in drawing any conclusion that low masses in the parameter space are favoured over high masses by existing data.  This point was emphasised earlier, in \sec{sec:BFP}, with regards to the high-mass, high-likelihood points in the FP region.

Looking again at the high-likelihood regions and the best-fit points in \fig{fig:sigmavmChi}, it is interesting to realise how likely these different points and regions are to be tested by the current and upcoming ID instruments.  For example, depending on how much its sensitivity improves in future years, it might be possible for the Large Area Telescope (LAT)~\cite{Atwood:0902.1089} aboard the \emph{Fermi} gamma-ray space telescope to cover part of the high-likelihood FP region, including the global best-fit point with $\left\langle \sigma v\right\rangle \sim 2.3\times 10^{-26} \mbox{cm}^3 \mbox{s}^{-1}$ (\tab{tab:DM}).  It is in fact expected from pre-launch estimates~\cite{Baltz:08062911} that the instrument will cover some fraction of the parameter space below $\left\langle \sigma v\right\rangle \sim 10^{-26} \mbox{cm}^3 \mbox{s}^{-1}$, depending on the neutralino mass.  A detailed \MN~global fit of the CMSSM parameter space using 9 months of \emph{Fermi} data has already been performed~\cite{Scott:FermiLAT}, using the dwarf spheroidal galaxy Segue 1 as a target; constraints are already quite close to becoming interesting.  A similar analysis can also be done using GAs.  The other best-fit (COA) point, however, has $\left\langle \sigma v\right\rangle \sim 5.4\times 10^{-28} \mbox{cm}^3 \mbox{s}^{-1}$, which is well below what is realistically detectable by \emph{Fermi}.

\begin{figure}
\begin{center}
\includegraphics[width=0.49\linewidth, trim = 0 330 160 10, clip=true]{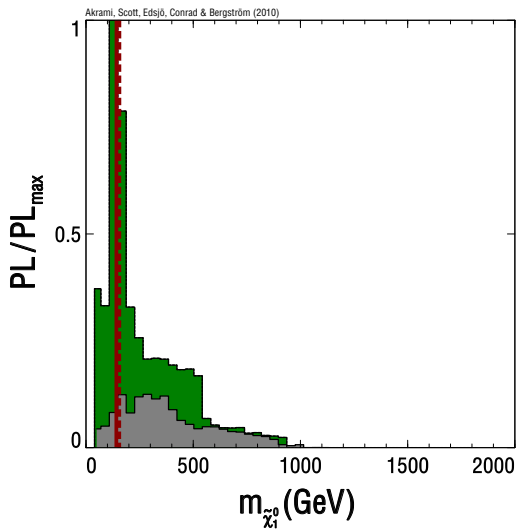}
\caption[aa]{\footnotesize{As in \fig{fig:params1D} and \fig{fig:LHC1D}, but for the neutralino mass $m_{\tilde\chi^0_1}$.}}
\label{fig:mChi}
\end{center}
\end{figure}

Finally, we have shown in \fig{fig:mChi} the 1D profile likelihood for the neutralino mass $m_{\tilde\chi^0_1}$.  The plot is produced in the same way as \figs{fig:params1D}{fig:LHC1D}, and compares the results of both GA and \MN~scans.  It is again important to notice the higher-likelihood points found by the GA almost everywhere in the interesting mass range.  We also observe that both the FP and COA best-fit points have quite similar and low neutralino masses ($\sim 150\gev$), similar to what was seen for $m_{\tilde g}$ in \fig{fig:LHC1D}.

\subsection{Technical comparison with nested sampling} \label{sec:TechComp}

\subsubsection{Dependence on priors and parameterisation}

Throughout the previous sections, we compared our GA results mostly with those of the linear-prior \MN~scan of the CMSSM, especially when we discussed the resultant 1D and 2D profile likelihoods.  We mentioned several times that \MN~implemented with log priors gives a better value of $11.90$ for the best-fit $\chi^2$, compared to the best fit when implemented with linear priors (13.51).  It is true that one can achieve better fits in certain regions of the parameter space by utilising e.g. a logarithmic prior in the search algorithm (see e.g. Ref.~\citealp{Trotta:08093792} and references therein for a discussion of the effects of priors on best-fit points and high-likelihood regions, as well as Bayesian posterior means and high-probability regions).  However, there are good reasons not to use the log-prior \MN~results for the main performance comparisons in this paper.  

Firstly, in order to make any comparison of the two algorithms reasonable, one should put both on the same footing.  On the one hand, the likelihood function, as defined in terms of the original model parameters, is the only statistical measure employed in any frequentist study.  Our genetic analysis is no exception.  Our GA scans the parameter space according to the likelihood, as a function of the original model parameters.  On the other hand, \MN, similarly to every other sampling technique optimised for Bayesian scans, performs the scan based on the posterior PDF (i.e. likelihood times prior) rather than the likelihood alone.  Consequently, a very natural way of comparing the two is to make the latter sampling algorithm also proceed according to the likelihood function only.  This can be achieved by choosing a flat prior in this case.

Imposing any other nontrivial prior (or equivalently, changing the scanning metric), although entirely justified in the Bayesian framework, is a very ad hoc approach in a frequentist framework.  In the Bayesian case, this simply means that the algorithm samples the regions containing larger prior volumes better, producing more sample points in these regions.  This is exactly what one requires for a Bayesian scan in which the density of samples reflects the posterior density at different points in the parameter space.  In the profile likelihood analysis however, we are interested in having reasonable maps of the likelihood function in terms of the given model parameters.  Imposing any prior in this case means nothing but giving different scanning weights to different parts of the parameter space, i.e. forcing the algorithm to scan some regions with higher resolutions than the others; this can make the algorithm miss important points in some regions.

In the frequentist language, the effect of imposing a non-flat prior is the same as reparameterising the model.  This for example means that, in the case of the log-prior scan, the likelihood function is redefined in terms of the logarithmically-scaled parameters rather than the original model parameters.  Results of a profile likelihood analysis should in principle be independent of the specific parameterisation of the model;  it should not matter if one works with e.g. one set of coordinates or another.  This statement is however correct only if one has perfect knowledge of the likelihood function.  No numerical scanning algorithm provides this perfect knowledge, as its resolution is always limited.  This means that different parameterisations of the model do give different results until the limit of `perfect sampling' has been reached.  Any specific choice should then be justified `a priori'.  One can for example argue that a specific scaling is theoretically better justified compared to others (e.g. that a log prior is geometrically preferred to a flat one).  In principle this is a Bayesian statement, as it places an implicit measure on the parameter space, but it does have a practical impact upon frequentist profile likelihood scans.  If one wanted to explore the effects of such reparameterisations, it would be entirely possible to do this by way of a GA, implemented in terms of genomes encoding the rescaled parameters.  We suspect that by using a logarithmically-encoded genome, we would for example find the funnel region properly and probably even some better-fitting points than our current best-fit.  Since our primary intention in the current work has been to look at the CMSSM model as it is, we have therefore adhered to the likelihood function defined in terms of the original parameters (and thus employed a linearly-encoded genome).  We leave the investigation of logarithmically-encoded genomes for future work, where we intend to compare results with those of log-prior \MN~scans.  

\subsubsection{Speed and convergence}

The results presented in this work are based on 3 million sample points in total, corresponding to the same number of likelihood evaluations.  The samples have been generated through 10 separate runs with different initial populations and $3000$ generations each.  The resultant samples were then combined to obtain the final set of points.  Compared to a typical number of likelihood evaluations required in a \MN~scan (around $500,000$), the computational effort here is larger by a factor of 6.  We have however chosen the number of runs and generations entirely by hand, not by any advanced convergence criteria; it may be possible to achieve similar (or better) results with less likelihood evaluations, using a more carefully tuned GA and/or a suitable stopping criterion.

There is in fact no way of checking whether or not our present implementation of the algorithm, although giving better results compared to \MN, has globally converged; there are indeed several reasons making us believe that it probably has not.

\begin{figure}[h!]
\begin{center}
\subfigure{\label{m0mhf_1:a}\includegraphics[width=0.29\linewidth, trim = 100 20 100 60, clip=true]{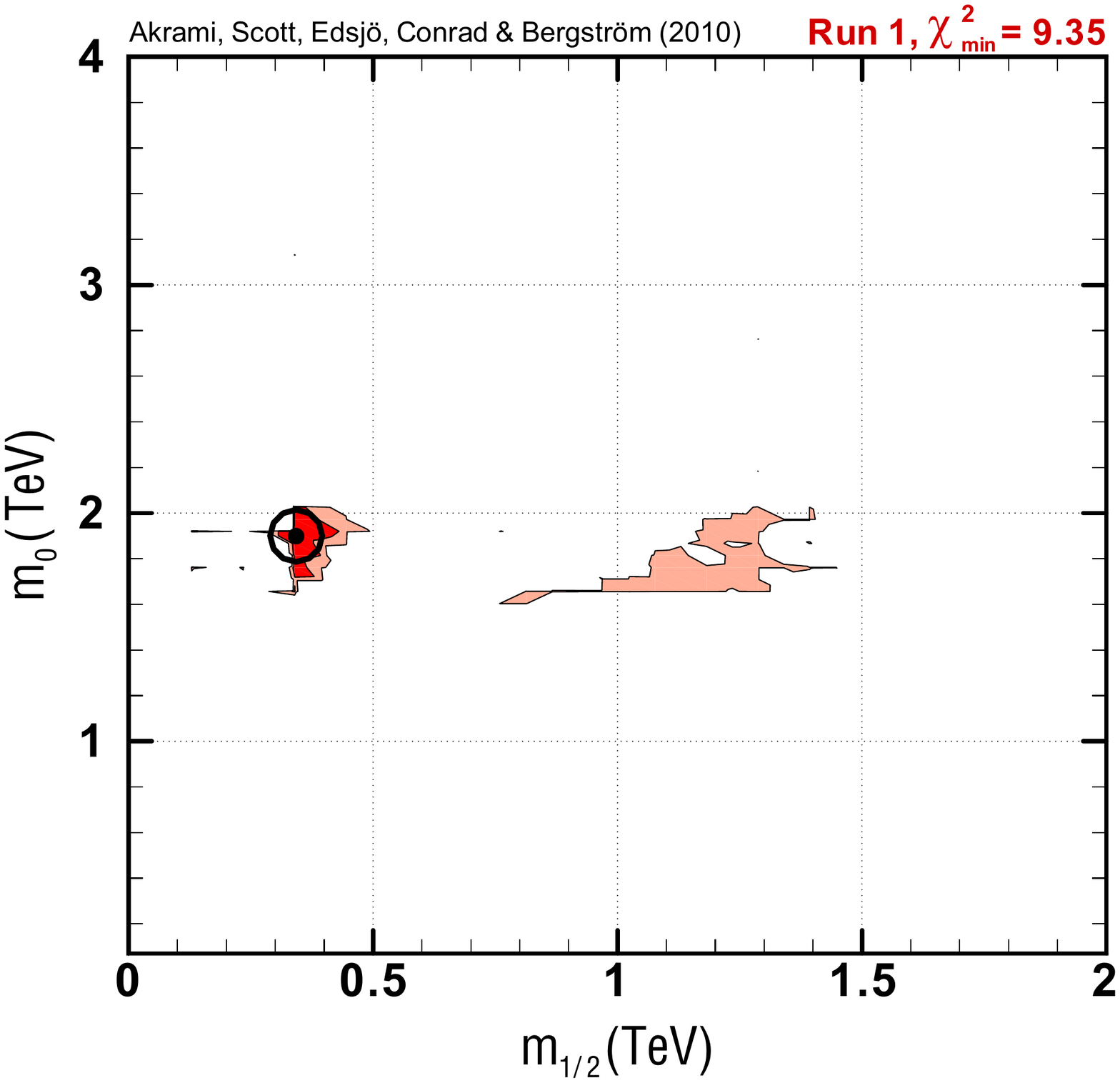}}
\subfigure{\label{m0mhf_2:b}\includegraphics[width=0.29\linewidth, trim = 100 20 100 60, clip=true]{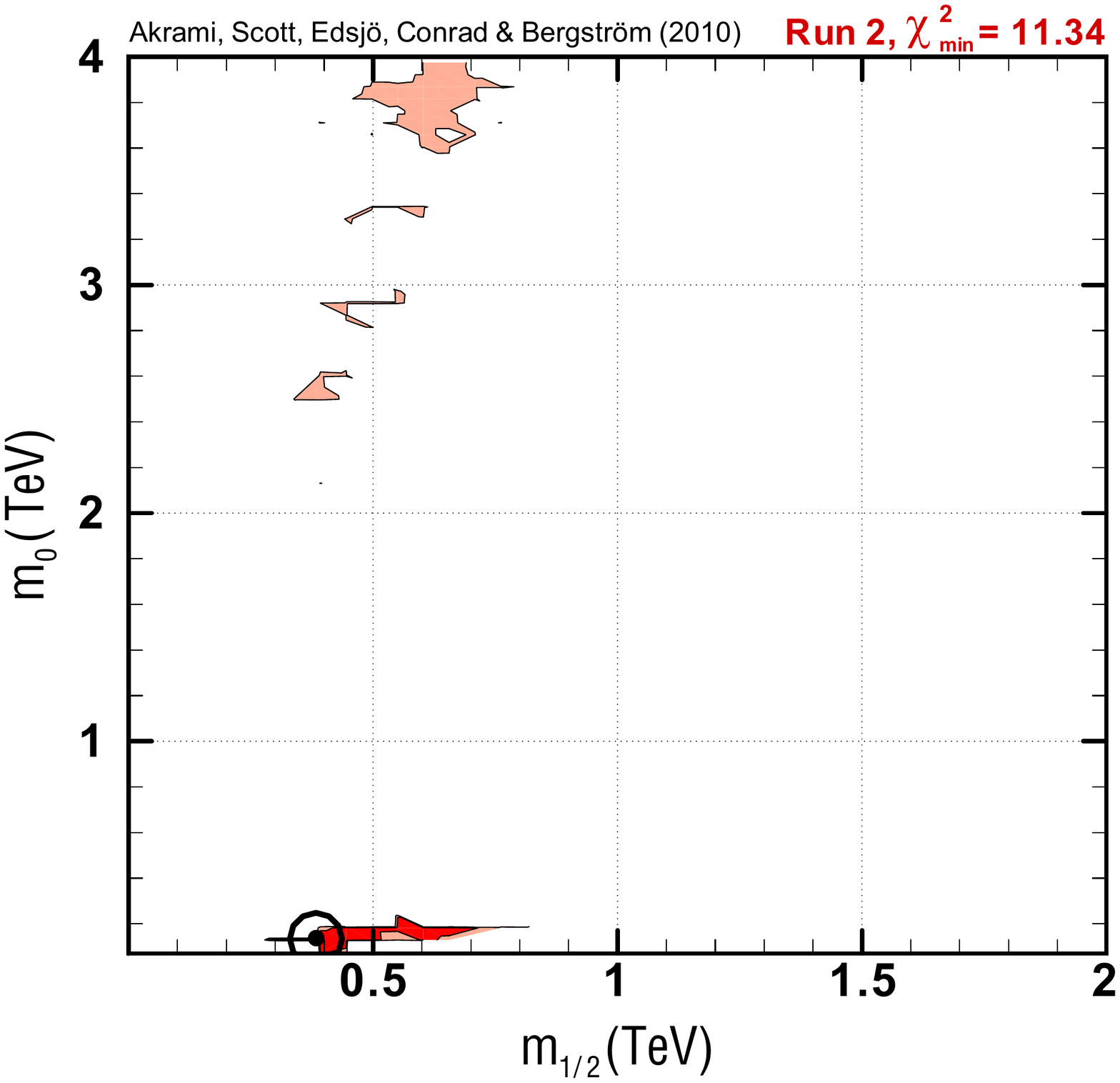}} 
\subfigure{\label{m0mhf_3:c}\includegraphics[width=0.29\linewidth, trim = 100 20 100 60, clip=true]{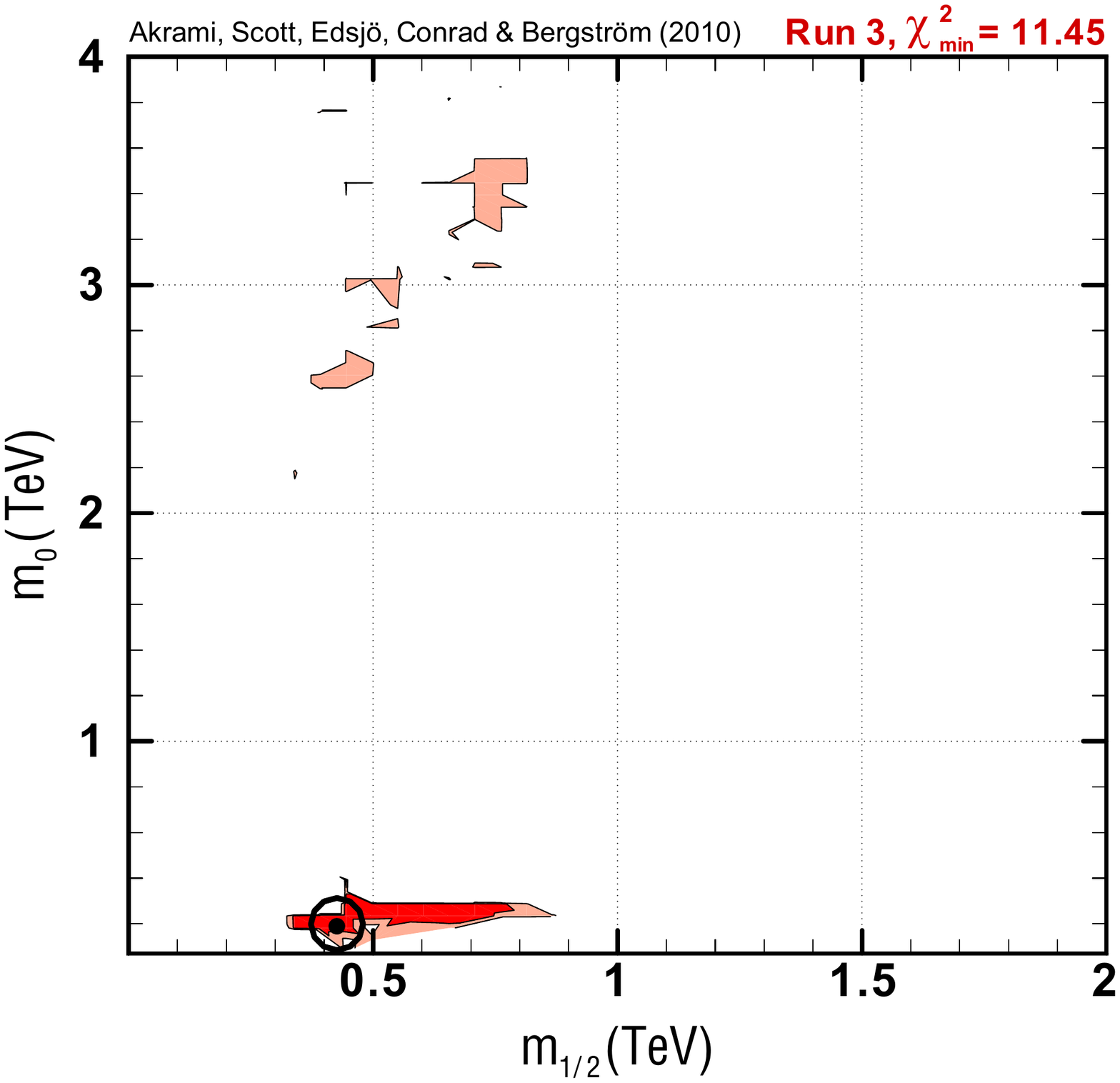}} \\
\subfigure{\label{m0mhf_4:d}\includegraphics[width=0.29\linewidth, trim = 100 20 100 60, clip=true]{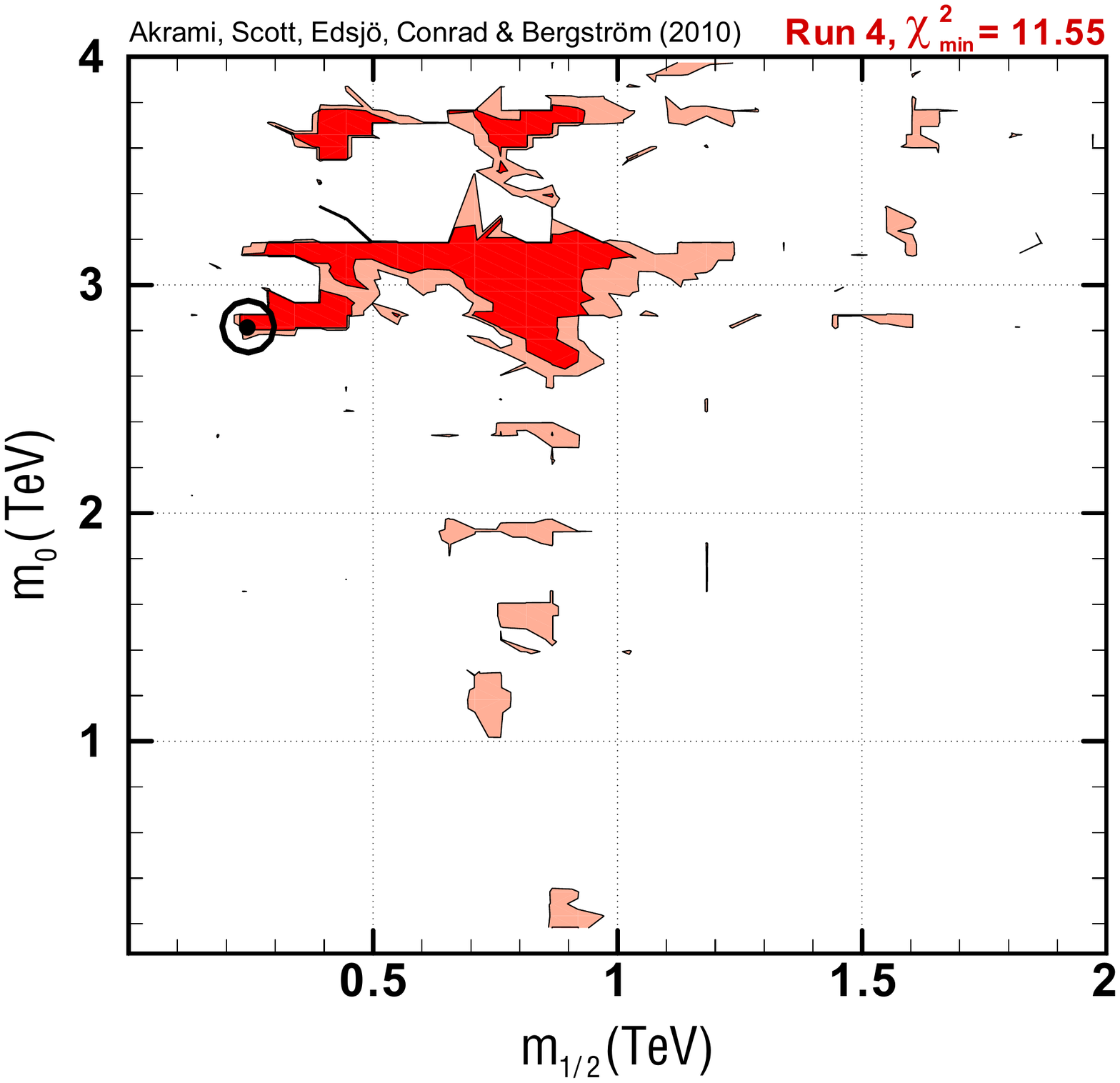}}
\subfigure{\label{m0mhf_5:a}\includegraphics[width=0.29\linewidth, trim = 100 20 100 60, clip=true]{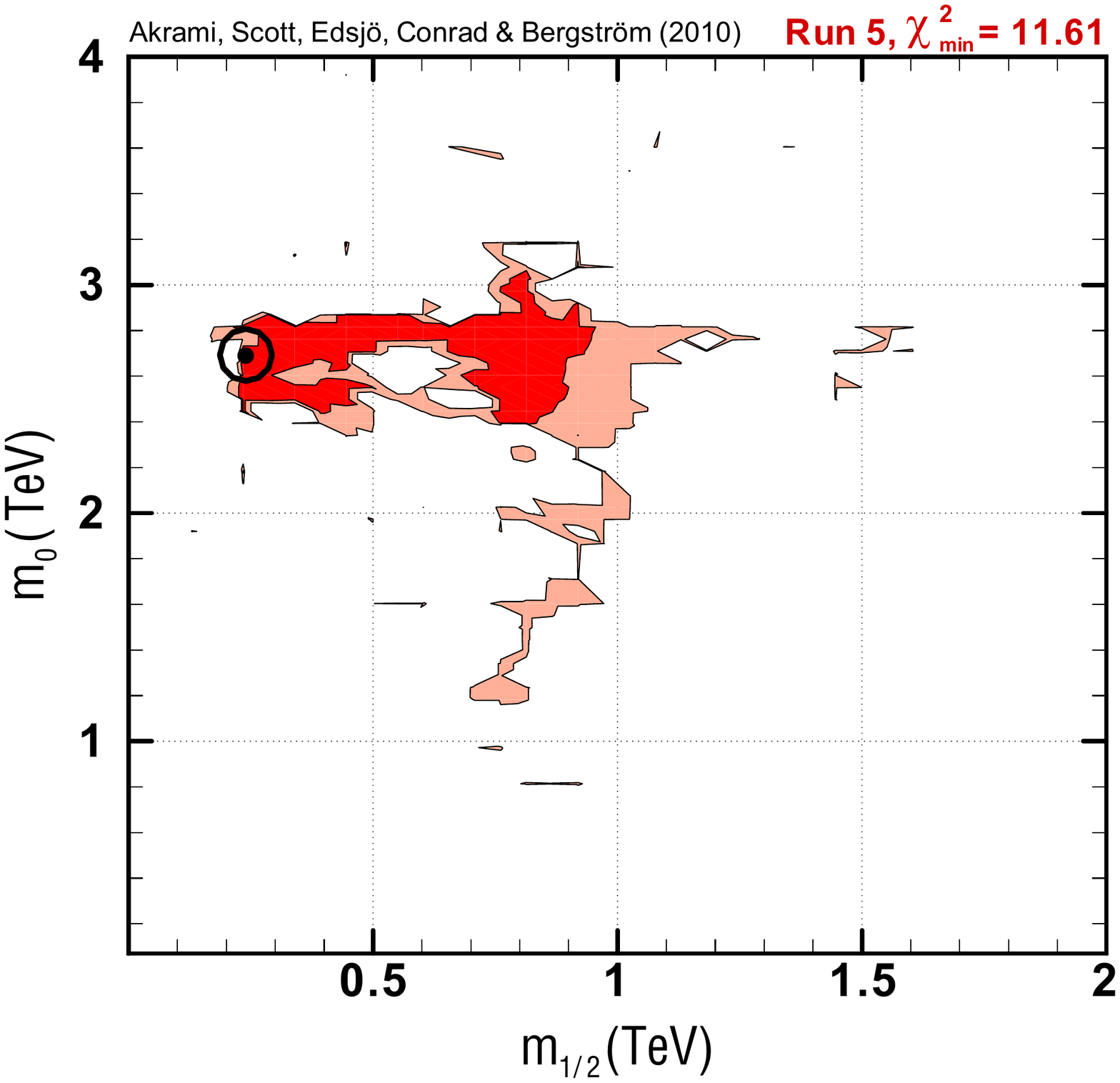}}
\subfigure{\label{m0mhf_6:b}\includegraphics[width=0.29\linewidth, trim = 100 20 100 60, clip=true]{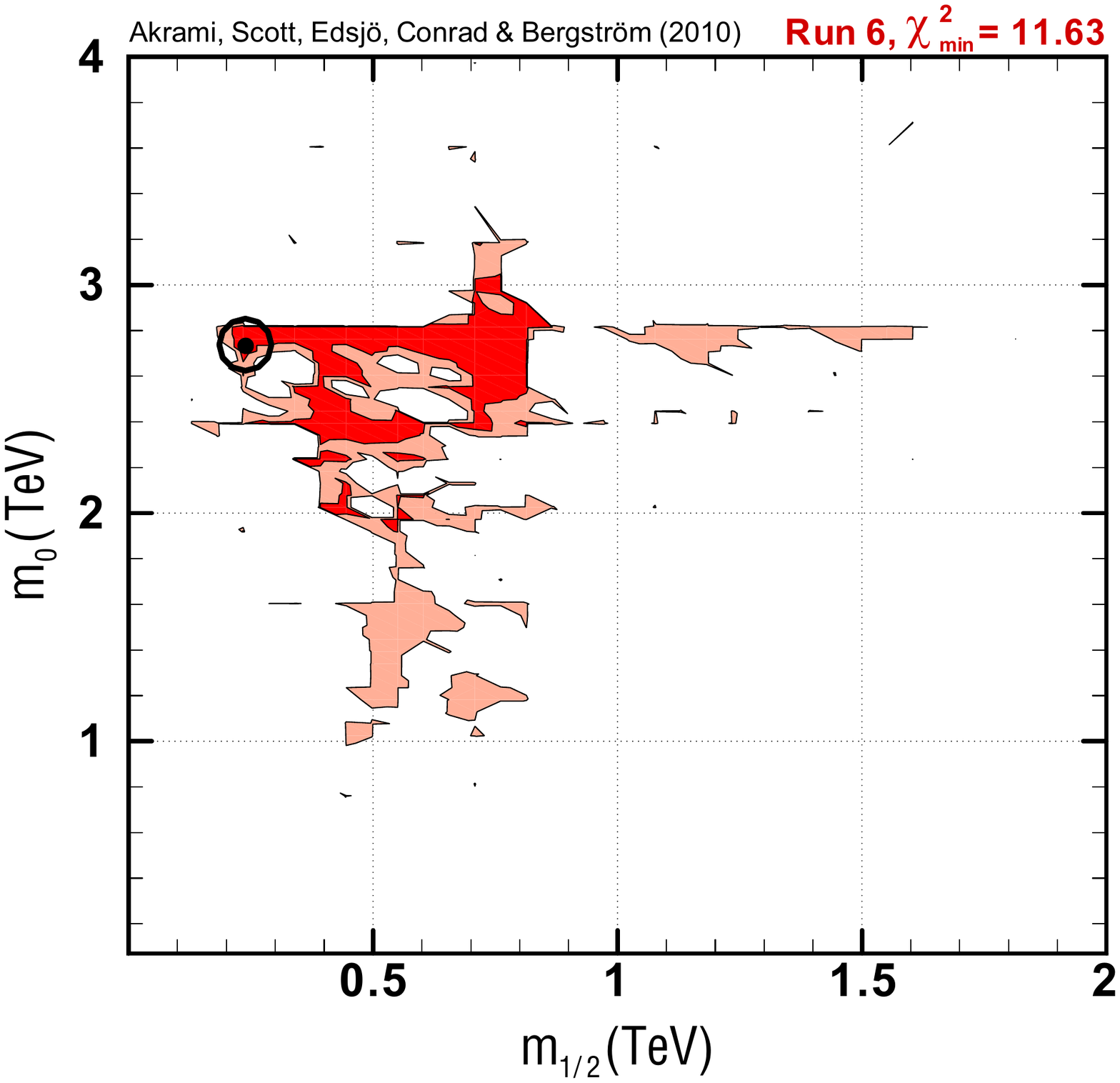}} \\
\subfigure{\label{m0mhf_7:c}\includegraphics[width=0.29\linewidth, trim = 100 20 100 60, clip=true]{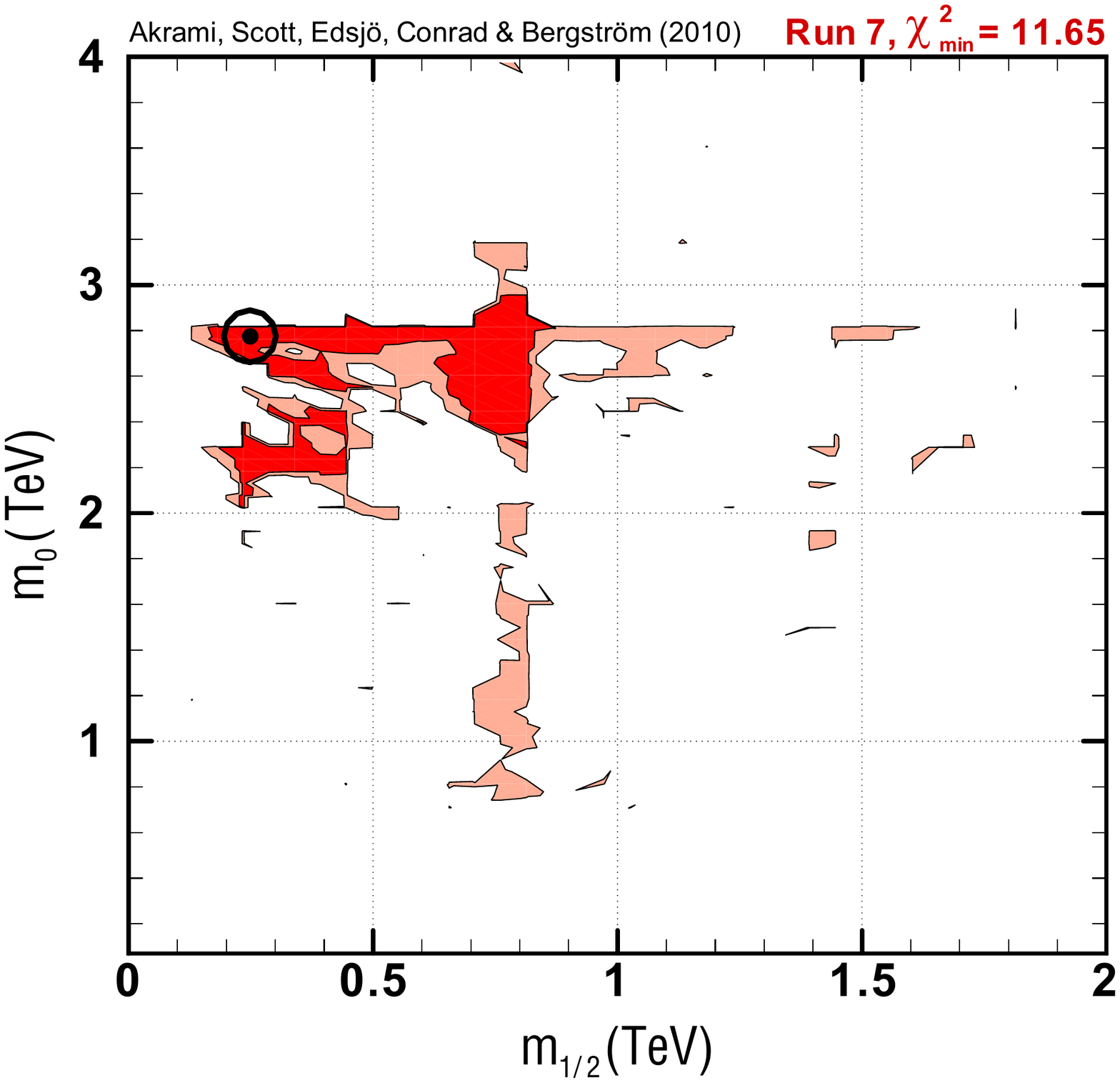}} 
\subfigure{\label{m0mhf_8:d}\includegraphics[width=0.29\linewidth, trim = 100 20 100 60, clip=true]{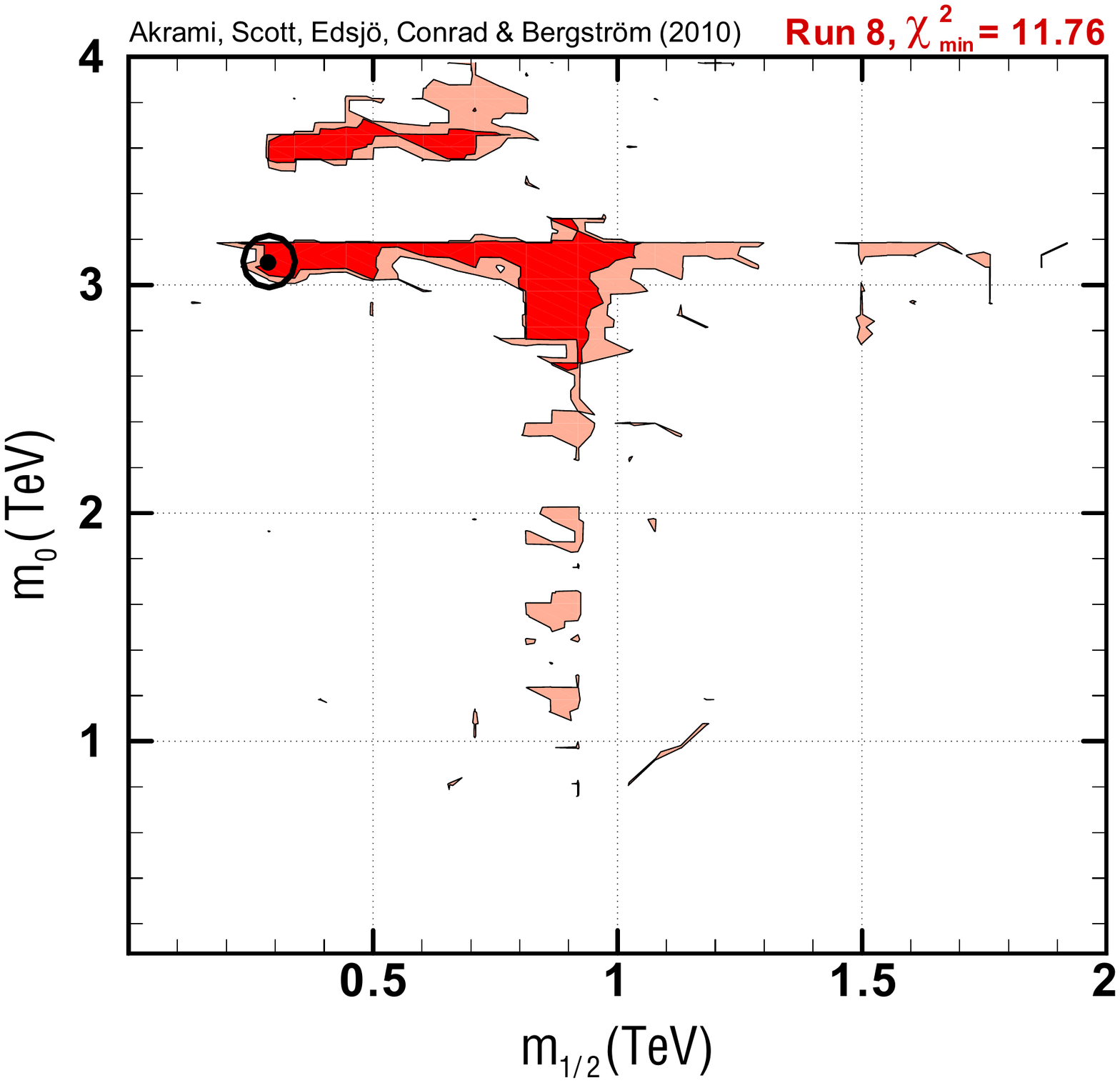}} 
\subfigure{\label{m0mhf_9:c}\includegraphics[width=0.29\linewidth, trim = 100 20 100 60, clip=true]{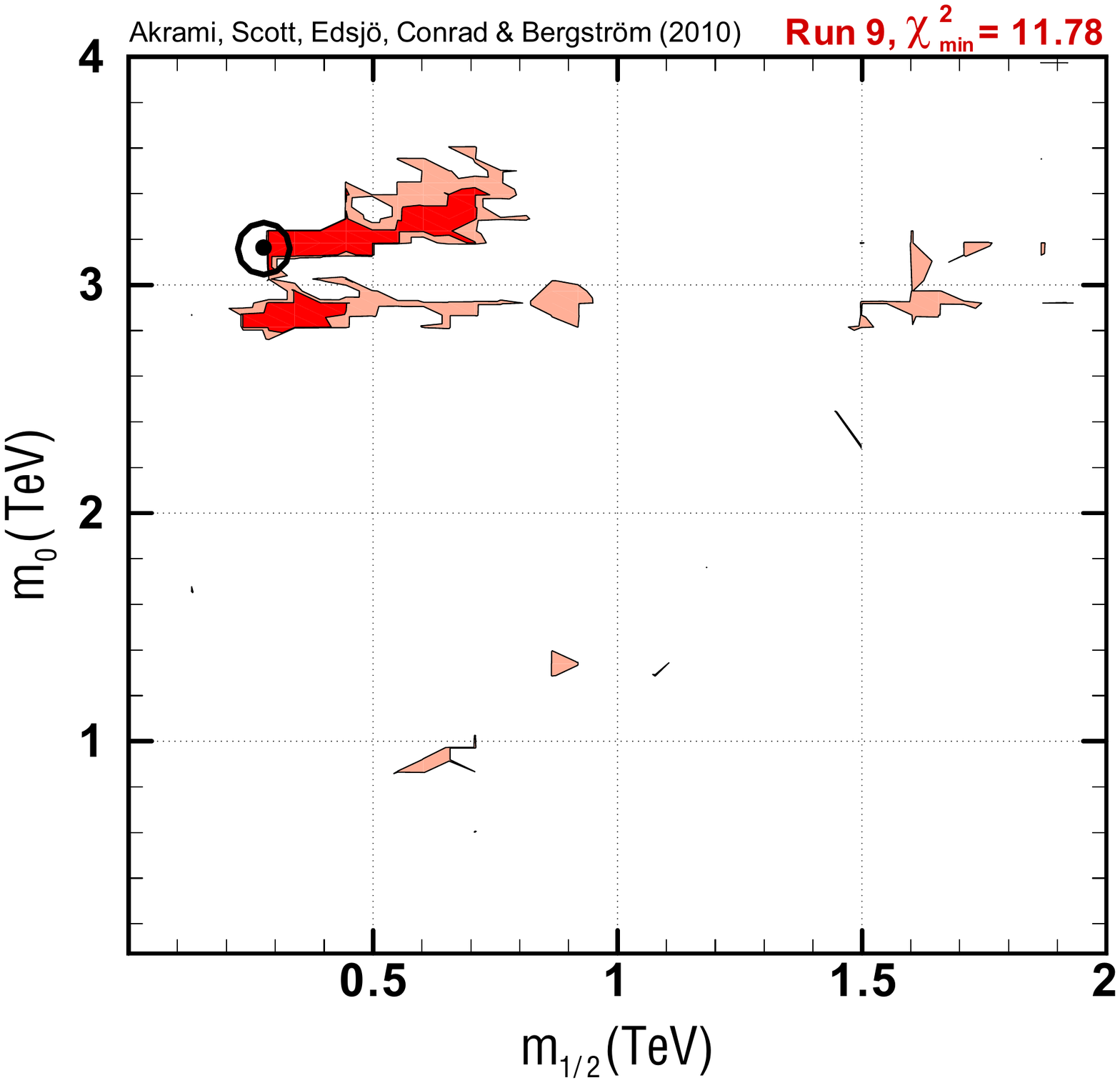}} \\
\subfigure{\label{m0mhf_10:d}\includegraphics[width=0.29\linewidth, trim = 100 20 100 60, clip=true]{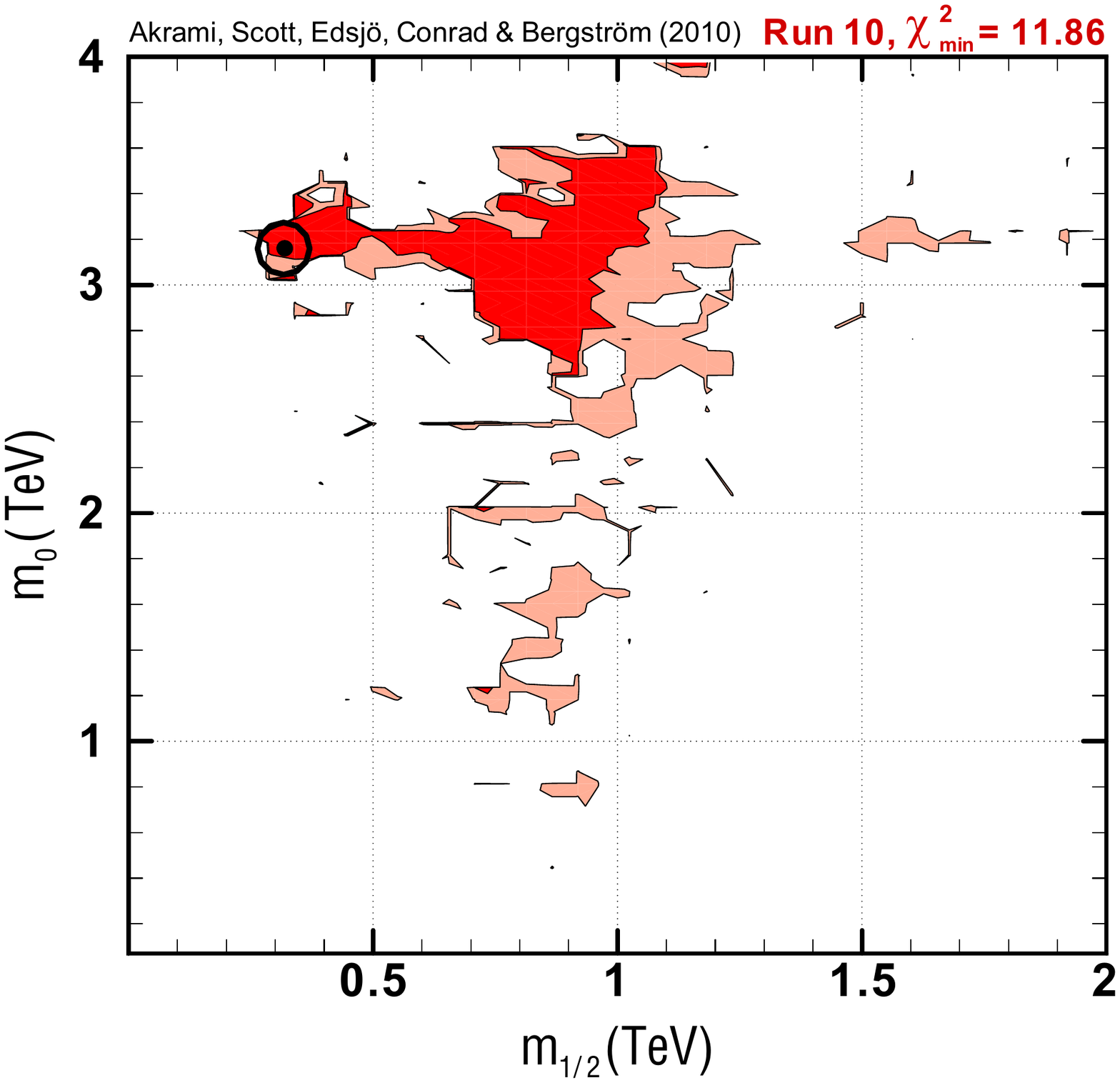}}
\caption[aa]{\footnotesize{Individual two-dimensional profile likelihoods in the $m_0$-$m_{1/2}$ plane for each of the 10 runs employed in our analysis.  Each panel shows the statistically-consistent results of each scan based on the best-fit point found in that scan.  As in previous figures, the inner and outer contours represent $68.3\%$ ($1\sigma$) and $95.4\%$ ($2\sigma$) confidence regions, respectively.  The dotted circles show the best-fit points of each run.  The sample points have been divided into $75\times75$ bins in all plots.  Panels are sorted according to the values for the best-fit $\chi^2$.  The final two-dimensional profile likelihood in Fig.~\ref{fig:m0mhf}a is obtained by combining all 10 scans and drawing iso-likelihood contours relative to the global best fit, i.e. $\chi^2=9.35$.}} \label{fig:m0mhfRuns}
\end{center}
\end{figure}

One way of seeing this is to look at the results for each of the 10 runs separately.  To illustrate this, we have given in \fig{fig:m0mhfRuns} the corresponding two-dimensional profile likelihoods in the $m_0$-$m_{1/2}$ plane.\footnote{Other two- and also one-dimensional profile likelihood plots for the parameters and observables exhibit similar features, so add little to the discussion.}  The iso-likelihood contours in each panel show the statistically-consistent $1\sigma$ and $2\sigma$ confidence regions, based on the best-fit point found in that specific scan.  Panels are sorted according to the best-fit $\chi^2$ values.  The values of all model and nuisance parameters at the best-fit points for all 10 runs are also given in \tab{tab:ParamsRuns}, together with the best-fit $\chi^2$ values.

\TABLE[t]{
\centering
{\scriptsize
\begin{tabular}{|c||c|c|c|c|c|c|c|c||c|c|} \hline
 & $m_0$ & $m_{1/2}$ & $A_0$ & $\tan\beta$ & $\mtpole$ & $m_b (m_b)^{\overline{MS}}$ & $\alphas$ & $1/\alphaemmz$ & $\chi^2_{min}$ \\
 & $\mbox{(GeV)}$ & $\mbox{(GeV)}$ & $\mbox{(GeV)}$ & & $\mbox{(GeV)}$ & $\mbox{(GeV)}$ & & & \\ \hline \hline
Run 1 & $1900.5$ & $342.8$ & $1873.9$  & $55.0$ &  $172.9$ & $4.20$ & $0.1172$ & $127.955$ & $\textbf{9.35}$ \\ \hline
Run 2 & $133.9$ & $383.1$ & $840.6$ & $17.9$ & $173.3$ & $4.20$ & $0.1183$ & $127.955$ & $\textbf{11.34}$ \\ \hline
Run 3 & $198.8$ & $426.3$ & $1059.4$ & $22.6$ & $173.3$ & $4.20$ & $0.1183$ & $127.955$ & $\textbf{11.45}$ \\ \hline
Run 4 & $2817.3$ & $244.6$ & $1712.9$ & $51.2$ & $172.9$ & $4.20$ & $0.1179$ & $127.954$ & $\textbf{11.55}$ \\ \hline
Run 5 & $2693.1$ & $240.0$ & $1706.0$ & $51.0$ & $172.6$ & $4.20$ & $0.1179$ & $127.954$ & $\textbf{11.61}$ \\ \hline
Run 6 & $2737.8$ & $238.7$ & $1692.6$ & $51.0$ & $172.6$ & $4.20$ & $0.1176$ & $127.954$ & $\textbf{11.63}$ \\ \hline
Run 7 & $2775.1$ & $248.2$ & $1780.2$ & $51.1$ & $172.6$ & $4.20$ & $0.1179$ & $127.954$ & $\textbf{11.65}$ \\ \hline
Run 8 & $3102.3$ & $287.3$ & $1976.8$ & $51.2$ & $172.8$ & $4.20$ & $0.1176$ & $127.954$ & $\textbf{11.76}$ \\ \hline
Run 9 & $3158.9$ & $276.6$ & $1901.2$ & $51.1$ & $173.1$ & $4.21$ & $0.1174$ & $127.955$ & $\textbf{11.78}$ \\ \hline
Run 10 & $3159.3$ & $317.0$ & $2136.8$ & $51.2$ & $172.7$ & $4.20$ & $0.1180$ & $127.955$ & $\textbf{11.86}$ \\ \hline
\end{tabular}
} \caption[aa]{\footnotesize{Parameter and $\chi^2$ values at the best-fit points found by Genetic Algorithms in each of 10 runs.  The final inference is based on all points found in all runs.}} \label{tab:ParamsRuns}
}

Our global best-fit point ($\chi^2=9.35$) is found only in one run.  The other 9 runs have not found best-fit points better than $\chi^2=11.34$.  However, all these other best-fit values ($\chi^2=11.34$, $11.45$, $11.55$, $11.61$, $11.63$, $11.65$, $11.76$, $11.78$, and $11.86$) are also significantly better than the one found by \MN~with linear priors ($\chi^2=13.51$).  Although some of these best fits in individual runs are very similar to the value found by \MN~with logarithmic priors ($\chi^2=11.90$), they are still at least slightly better in every single case.

The plots in \fig{fig:m0mhfRuns} indicate that none of our individual GA runs have been able to cover all the interesting high-likelihood regions on their own.  It seems for example that in runs 2 and 3, the algorithm has become trapped in local minima in the COA region.  In runs 4-10, the same has happened at high masses, including the focus point region.  However, if the individual GAs continued to run, they may have escaped these minima.  In runs 6 and 7 for example, although the $1\sigma$ region does not include the global best-fit region of run 1, is very likely to extend and eventually uncover that region if the run continues.  The other difficulty is that if the current version of the algorithm finds the global best-fit point (or some high-likelihood points in its vicinity) relatively quickly, the chance that it will then scan other points with lower likelihoods within the interesting confidence regions is dramatically reduced.  This indicates a drawback of the algorithm in correctly mapping confidence intervals around the global best-fit point.  This problem is not unexpected, as the primary purpose of GAs is to find global maxima or minima for a specific function as quickly as possible; if they succeed in this, there is no reason for them to start mapping the other less important surrounding points.  This issue obviously becomes more serious when spike-like best-fit regions exist, a characteristic which appears to be the case for the CMSSM.  For example, the existence of relatively large high-likelihood regions in panels 4,5,6,7,8 and 10 is probably due to the fact that there are many points with likelihood values very close to the highest one, distributed in a large area; this is not the case for runs 1,2, or 3.  This means that if the GAs continue to run in those cases and it so happens that the global best-fit point of run 1 is found at some stage, their mapping of the interesting regions will be much better than the present case in run 1.  This is again an indication that there is a trade-off between how quickly we want the algorithm to find the actual global best-fit point and how accurately it is supposed to map the confidence regions; employing more advanced operators and strategies in the algorithms might improve the situation.

As far as our current implementation of GAs is concerned, all the above imply that the algorithms have not converged properly in every individual run.  Firstly, they have not been able to find the actual global best-fit point in all (or even necessarily any) of the scans, and secondly, in the run with the highest-likelihood best-fit point (run 1), the mapping of the confidence regions around that point is rather unsatisfactory.  On the other hand, most of the unwanted features discussed here are alleviated by combining the results of all 10 runs.  This demonstrates the important role that parallelisation could play in improving the efficiency of GAs.  Although our full set of sample points seems to provide better results than \MN, we are still not sure that this parallel version of the algorithm has converged either.  This is again because the number of separate runs employed in our analysis is chosen rather arbitrarily.  The arbitrariness in both the number of runs and the termination criteria in each run means that no result-driven convergence condition exists in our GAs.  One could possibly utilise more sophisticated criteria in the termination condition, giving rise to better estimates of the convergence in each run, but to our knowledge no problem-independent such alternatives exist; we leave the investigation of such possibilities for future work.

As discussed in~\sec{sec:BFP}, even though such a convergence condition does exist for the \MN~scans, making the algorithm terminate after less total likelihood evaluations than our GAs, it still misses many points that are important in the frequentist framework.  The convergence criterion for \MN~is defined in terms of the Bayesian evidence; even if a run is properly converged in terms of the evidence, this convergence makes sense only in the context of the Bayesian posterior PDF, not a frequentist profile likelihood analysis.  That is, many isolated spikes might have been missed, and consequently the likelihood might not have been mapped effectively.  Even tuning the convergence parameters (such as the tolerance) may not help.  This is because if the \MN~algorithm does not find a high-likelihood point on its first approach to a region, it is given no chance to go back and find it at a later stage.  This means that even if the convergence parameter for \MN~is tuned in such a way that the algorithm runs for the same number of likelihood evaluations as a GA (i.e.~3 million here), this does not help in finding better-fit points.  One should thus be very careful in introducing convergence criteria to any scanning techniques (including GAs and \MN) dealing with a complex model such as the CMSSM;  the criteria should be carefully defined depending on which statistical measure is employed.

We also point out that the isolated likelihood spikes missed by Bayesian algorithms will only fail to affect the posterior mass if a limited number of them exist.  If there are a significant number of spikes, they could add up to a large portion of the posterior mass, and affect even the Bayesian inference.  Our GA results indicate that many such missed points actually exist, hinting that \MN~might not have even mapped the entire posterior PDF correctly.  Given the frequentist framework of this paper, this must unfortunately remain mere speculation, as our results do not allow any good estimate to be made of the posterior mass contained in the extra points.

We conclude this section by emphasising the role of parallelisation in terms of required computational power.  Not only does the parallelisation enhance the scanning efficiency of GAs by reducing the probability of premature convergence and trapping in local maxima, it also increases the speed significantly.  This can be done in different ways.  Firstly, GAs work with a population of points instead of a single individual, providing extensive opportunity for treating different individuals in parallel.  As an immediate consequence, the required time in a typical simple GA run with full generational replacement (the scheme we have used) can in principle decrease by a factor of $n_p$, the number of individuals in each population ($100$ in our case).  The other way to parallelise GAs is to have separate populations evolve in parallel.  By employing more advanced genetic operators and strategies such as `immigration', individuals in different populations can even interact with each other.  Although we have not employed such advanced parallel versions of the algorithm in our analysis, the samples have been generated in a parallel manner, i.e. through 10 separate runs with $3000$ generations each.

\section{Summary and conclusions} \label{sec:concl}

Constraining the parameter space of the MSSM using existing data is under no circumstances an easy or straightforward task.  Even in the case of the CMSSM, a highly simplified and economical version of the model, the present data are not sufficient to constrain the parameters in a way completely independent of computational and statistical techniques.

There have been several efforts to study properties and predictions of different versions of the MSSM.  Many recent activities in this field have used scanning methods optimised for calculating the Bayesian evidence and posterior PDF.  Those analyses have been highly successful in revealing the complex structure of SUSY models, demonstrating that some patience will be required before we can place any strong constraints on their parameters.  The same Bayesian scanning methods have also been employed for frequentist analyses of the problem, particularly in the framework of the profile likelihood.  These methods are not optimised for such frequentist analyses, so care should be taken in applying them to such tasks.

We have employed a completely new scanning algorithm in this paper, based on Genetic Algorithms (GAs).  We have shown GAs to be a powerful tool for frequentist approaches to the problem of scanning the CMSSM parameter space.  We compared the outcomes of GA scans directly with those of the state-of-the-art Bayesian algorithm \MN, in the framework of the CMSSM.  For this comparison, we mostly considered \MN~scans with flat priors, but kept in mind that e.g. logarithmic priors give rise to higher-likelihood points at low masses in the CMSSM parameter space;  we justified this choice of priors.

Our results are very promising and quite surprising.  We found many new high-likelihood CMSSM points, which have a strong impact on the final statistical conclusions of the study.  These not only influence considerably the inferred high-likelihood regions and confidence levels on the parameter values, but also indicate that the applicability of the conventional Bayesian scanning techniques is highly questionable in a frequentist context.  Although our initial motivation in using GAs was to gain a correct estimate of the likelihood at the global best-fit point, which is crucial in a profile likelihood analysis, we also realised that they can find many new and interesting points in almost all the relevant regions of parameter space.  These points strongly affect the inferred confidence regions around the best-fit point.  Even though we cannot be confident of exactly how completely our algorithm is really mapping these high-likelihood regions, it has certainly covered large parts of them better than any previous algorithm.

We think that by improving the different ingredients of GAs, such as the crossover and mutation schemes, this ability might even be enhanced further.  We largely employed the standard, simplest versions of the genetic operators in our analysis, as well as very typical genetic parameters.  These turned out to work sufficiently well for our purposes.  Although we believe that tuning the algorithm might produce even more interesting results, it is good news that satisfactory results can be produced even with a very generic version.  This likely means that one can apply the method to more complicated SUSY models without extensive fine-tuning.

One interesting outcome of our scan is that the global best-fit point is found to be located in the focus point region, with a likelihood significantly larger than the best-fit point in the stau co-annihilation region (which in turn actually still has a higher likelihood than the global best-fit value obtained with \MN, even with logarithmic priors).  The focus point region is favoured in our analysis over the co-annihilation region, in contrast to findings from some recent MCMC studies~\cite{Buchmueller:08084128,Buchmueller:09075568}, where the opposite was strongly claimed.  We also found a rather large part of the stau co-annihilation region, consistent with all experimental data, located at high $\mzero$.  This part of the co-annihilation region seems to have been missed in other recent scans.  All these results show that, at least in our particular setup, high masses, corresponding either to the FP or the COA regions, are by no means disfavoured by current data (except perhaps direct detection of dark matter).  The discrepancy between this finding and those of some other authors that the FP is disfavoured might originate in the different scanning algorithms employed, or in the different physics and likelihood calculations performed in each analysis.  We have however shown, by comparing our results with others produced using exactly the same setup except for the scanning algorithm, that one should not be at all confident that all the relevant points for a frequentist analysis can be found by scanning techniques optimised for Bayesian statistics, such as nested sampling and MCMCs.

We have also calculated some of the quantities most interesting in searches for SUSY at the LHC, and in direct and indirect searches for dark matter.  We showed that GAs found much better points compared to \MN~almost everywhere in the interesting mass ranges of the lightest Higgs boson, gluino and neutralino.  We confirmed previous conclusions that the LHC is in principle able to investigate a large fraction of the high-likelihood points in the CMSSM parameter space if it explores sparticle masses up to around $3\tev$.  As far as the Higgs mass is concerned, there are many points with rather low masses that, although sitting just below the low mass limit given by LEP, are globally very well fit to the experimental data.  In the context of dark matter searches, we noticed that the global best-fit point and much of the surrounding $1\sigma$ confidence level region at high cross-sections are actually already mostly ruled out by direct detection limits, if one assumes the standard halo model to be accurate.  We also argued that some of these points may be tested by upcoming indirect detection experiments, in particular the \emph{Fermi} gamma-ray space telescope.  Finally, we realised that the high-likelihood stau co-annihilation region at large $\mzero$ introduces a new allowed region in the combination of the neutralino mass and self-annihilation cross-section, which (to our knowledge) has not been observed previously.

We also compared our algorithm with \MN~in terms of speed and convergence, and argued that GAs are no worse than \MN~in this respect.  GAs have a large potential for parallelisation, reducing considerably the time required for a typical run.  This property, as well as the fact that the computational effort scales linearly (i.e. as $kN$ for an $N$-dimensional parameter space), also makes GAs an excellent method for the frequentist exploration of higher-dimensional SUSY parameter spaces.

Finally, perhaps the bottom line of the present work is that we once again see that even the CMSSM, despite its simplicity, possesses a highly complex and poorly-understood structure, with many small, fine-tuned regions.  This makes investigation of the model parameter space very difficult and still very challenging for modern statistical scanning techniques.  Although the method proposed in this paper seems to outperform the usual Bayesian techniques in a frequentist analysis, it is important to remember that it may by no means be the final word in this direction.  Dependence of the results on the chosen statistical framework, measure and method calls for caution in drawing strong conclusions based on such scans.  The situation will of course improve significantly with additional constraints provided by forthcoming data.

\acknowledgments{} The authors are grateful to the Swedish Research Council (VR) for financial support.  YA was also supported by the Helge Axelsson Johnson foundation.  JC is a Royal Swedish Academy of Sciences Research Fellow supported by a grant from the Knut and Alice Wallenberg Foundation.  We thank Roberto Trotta for helpful discussions.  We are also thankful to the authors of Ref.~\citealp{Buchmueller:08084128} for useful comments on a previous version of the manuscript.






\end{document}